\documentclass[12pt]{article}
\pdfoutput=1
\usepackage{amsmath,amsfonts,amssymb,amsthm,amssymb,bbm,bm}
\usepackage{pdfsync}
\usepackage{graphicx,import}
\usepackage[latin1]{inputenc}
\usepackage{hyperref}
\usepackage{empheq}
\usepackage[all]{xy}
\usepackage{stmaryrd}
\usepackage{rotating}
\usepackage{color}  
\usepackage{slashed}
\usepackage{cancel}
\usepackage{array, makecell} %
\usepackage{comment}
\usepackage{tikz-cd}
\usepackage{hhline}
\usepackage{comment}
\usepackage{cancel}
\makeatletter

\DeclareFontFamily{U}{matha}{\hyphenchar\font45}
\DeclareFontShape{U}{matha}{m}{n}{
      <5> <6> <7> <8> <9> <10> gen * matha
      <10.95> matha10 <12> <14.4> <17.28> <20.74> <24.88> matha12
      }{}
\DeclareSymbolFont{matha}{U}{matha}{m}{n}
\DeclareMathSymbol{\oright}       {2}{matha}{"69}

\newcommand{\doublehat}[1]{%
\begingroup%
  \let\macc@kerna\z@%
  \let\macc@kernb\z@%
  \let\macc@nucleus\@empty%
  \hat{\raisebox{.55ex}{\vphantom{\ensuremath{#1}}}\smash{\hat{#1}}}%
\endgroup%
}

\newcommand{\p}{\partial}

\newcommand{\bit}{\begin{itemize}}
\newcommand{\eit}{\end{itemize}}
\newcommand{\bd}{\begin{description}}
\newcommand{\ed}{\end{description}}

\newcommand{\bc}{\begin{center}}
\newcommand{\ec}{\end{center}}

\newcommand{\R}{{\mathbb R}}

\newcommand{\cM}{{\cal M}}
\newcommand{\cP}{{\cal P}}
\newcommand{\cJ}{{\cal J}}
\newcommand{\cN}{{\cal N}}
\newcommand{\cT}{{\cal T}}

\newcommand{\B}{{b}}

\newcommand{\bmw}{BMSW }
\newcommand{\bmsw}{\textsf{bmsw }}

\newcommand{\cL}{{\mathcal L}}

\newcommand{\cC}{{\mathcal C}}
\newcommand{\cS}{{\mathcal S}}
\newcommand{\cE}{{\mathcal E}}

\newcommand{\D}{{\mathrm{D}}}

\renewcommand{\sl}{{\mathfrak{sl}}}

\def\be#1\ee{\begin{align}#1\end{align}}
\newcommand{\bea}{\begin{eqnarray}}
\newcommand{\eea}{\end{eqnarray}}
\newcommand{\bs}{\begin{subequations}}
\newcommand{\es}{\end{subequations}}
\newcommand{\nn}{\nonumber}
\newcommand{\la}{\label}

\newcommand{\f}{\frac}

\def\p{\partial}
\def\bD{D}
\def\bC{C}
\def\bP{P}
\def\bq{q}
\def\bU{U}
\def\bF{F}

\def\bxi{\bar\xi}

\def\d{\delta}

\def\m{\mu}
\def\n{\nu}

\def\s{\sigma}
\def\t{\tau}

\def\D{\Delta}

\def\rd{\mathrm{d}}
\def\pa{\partial }

\newcommand{\tC}{{\tilde C}}
\newcommand{\tcJ}{{\tilde \cJ }}
\newcommand{\tcM}{{\tilde{\mathcal M}}}

\newcommand{\tc}{{\tilde c }}
\newcommand{\tn}{{\tilde n }}

\newcommand{\scri}{\cal I}

\begin{document}

\title{\Large{\bf Gravity from duality symmetry} \\}
\title{\Large{\bf Gravity from symmetry and its duality}\\}
\title{\Large{\bf Gravity from symmetry and impulsive waves}}
\title{\Large{\bf Gravity from symmetry: \\
Duality  and impulsive waves}}

\author{ Laurent Freidel$^1$\thanks{lfreidel@perimeterinstitute.ca} , 
Daniele Pranzetti$^{1,2}$\thanks{dpranzetti@perimeterinstitute.ca} 
}
\date{\small{\textit{
$^1$Perimeter Institute for Theoretical Physics,\\ 31 Caroline Street North, Waterloo, Ontario, Canada N2L 2Y5\\ \smallskip
$^2$ Universit\`a degli Studi di Udine,
via Palladio 8,  I-33100 Udine, Italy
}}}

\maketitle
\begin{abstract}
We show that we can derive the asymptotic Einstein's equations that arises at order $1/r$ in asymptotically flat gravity purely from symmetry considerations.  This is achieved by studying the transformation properties of functionals of the metric and the stress-energy tensor under the action of the Weyl BMS group, a recently introduced asymptotic symmetry group that includes arbitrary diffeomorphisms and local conformal transformations of the metric on the 2-sphere. Our derivation, which  {encompasses} the  inclusion of  matter sources,
leads to the identification of covariant observables that provide a definition of conserved charges parametrizing the non-radiative corner phase space.     These observables, related to the Weyl scalars, reveal a duality symmetry and a spin-$2$ generator which allow us to recast the asymptotic evolution equations  in a simple and elegant form as  conservation equations for  a null fluid living at null infinity.
 Finally we identify non-linear gravitational  impulse waves that describe transitions among gravitational vacua and are non-perturbative solutions of the asymptotic Einstein's  equations. {This provides a new picture of  quantization of the asymptotic phase space,  where gravitational vacua are representations of the asymptotic symmetry group and impulsive waves are encoded in their couplings.} 
\end{abstract}

%

\newpage
\tableofcontents

\section{Introduction}

The program of {\it local holography} is grounded in the fundamental role played by symmetries. It aims to provide a new description of quantum geometry in terms of the representation theory of the gravitational symmetries associated to the codimension-2 surface bounding a general finite region in spacetime, the {\it corner} \cite{DonnellyFreidel, Freidel:2015gpa, Freidel:2016bxd, Freidel:2018pvm, Freidel:2019ees, Freidel:2019ofr, Freidel:2020xyx, Freidel:2020svx, Freidel:2020ayo, Donnelly:2020xgu}. 
Since the seminal work of Emmy Noether \cite{Noether:1918zz}, the notion of symmetry has represented a very helpful and effective tool to unravel the correct description of the fundamental forces of Nature, both at the classical and, in the case of the Standard Model and Condensed Matter,  in the quantum regime. We believe that this invaluable tool will ultimately prove itself crucial also to guide us through the ultimate and most impervious stretch of this discovery journey, leading to the quantization of gravity. 

From this perspective, it is fundamental to understand the pivotal role of symmetries in describing the properties of a gravitational system 
in a finite bounded region of spacetime. 
The full power of the Noether theorem for local symmetries implies that the symmetry charges are supported by codimension two surfaces lying at the corner of the spacetime region under consideration \cite{DonnellyFreidel}.
This symmetry group lying at the corner  can naturally be split into `kinematical' symmetries that carries no symplectic flux and are readily quantizable and `dynamical' symmetries that include supertranslations along the null normals  and which carry fluxes.\footnote{ Recent developments appeared shortly after the first version of this manuscript   have allowed us to also include supertranslation as canonical transformation into the gravitational phase space by extending this with a dressing field \cite{Freidel:2021dxw,Ciambelli:2021nmv}.}
The study of the kinematical gravitational symmetries of a finite bounded region of spacetime 
 has been performed originally in \cite{DonnellyFreidel} for the Einstein--Hilbert formulation of gravity and then further extended to other first order formulations in \cite{Freidel:2020xyx, Freidel:2020svx}. 

 These analyses have led to the notion of {\it corner symmetry group}, which is the kinematical subgroup  generated by internal gauge transformations  and the residual diffeomorphisms which vanish at the corner. In the Einstein--Hilbert formulation, the corner symmetry algebra $\mathfrak{g}_S $ has been shown  \cite{DonnellyFreidel}  to have the semi-direct sum structure 
\be\la{gS-intro}
\mathfrak{g}_S = \mathrm{diff}(S)\oright  \mathfrak{sl(2,\mathbb{R})}^S\,,
\ee
where $ \mathrm{diff}(S)$ corresponds to the Lie algebra generated by diffeomorphisms tangent to the corner $S$ and  $\mathfrak{sl(2,\mathbb{R})}^S$ the Lie algebra generated by the surface boosts that linearly transform the normal plane of $S$ in a position-dependent way. 

The inclusion of normal  supertranslations that move the corner has led to the notion of {\it extended corner symmetry group} in  \cite{Ciambelli:2021vnn, Freidel:2021cbc} and given by
\be\label{gext-intro}
\mathfrak{g}^{\rm ext}_S = \left(  \mathrm{diff}(S)\oright\mathfrak{sl(2,\mathbb{R})}^S \right) \oright (\R^2)^S \,,
\ee
where the second semi-direct sum involves the two normal time translations.  

At the same time, it has been shown in \cite{Chandrasekaran:2018aop} that a similar semi-direct sum structure \eqref{gext-intro} captures the symmetries  of a general non-stationary null surface at finite distance equipped with a thermal structure.
This group was dubbed  Weyl BMS, or BMSW for short, in \cite{Freidel:2021yqe} and shown two satisfy two key properties.
On the one hand, it is a subgroup of the extended corner symmetry group: The subgroup that preserves, up to scale, the canonical null generator of the null surface. On the other hand, it is also the symmetry group of null infinity. 

More precisely, the  Lie algebra of  BMSW  possesses a semi-direct sum structure
\be\la{bmsw-intro}
\bmsw:= \left(  \mathrm{diff}(S) \oright \mathbb{R}_W^S\right)  \oright \mathbb{R}_T^S\,,
\ee
where $\R_W^S$ denotes the Weyl transformations labeled by functions $W$ on the sphere,
while $\R_T^S$ denotes the super-translations labelled by weight $1/2$ densities  $T$  on the sphere.
This algebra contains all the known extensions  of the BMS algebra  \cite{Bondi:1960jsa, BMS, Sachs62} that have been recently introduced  as candidates for the   gravitational symmetries of null infinity in \cite{Barnich:2009se, Barnich:2011mi, Campiglia:2014yka, Flanagan:2015pxa, Compere:2018ylh}. Besides  super-translation transformations,  it includes  arbitrary sphere diffeomorphisms and local Weyl rescalings of the 2D sphere metric at $\scri$. 
 Importantly, it has been shown in \cite{ Freidel:2021cbc} that the extended corner symmetry algebra \eqref{gext-intro} reduces to the $\bmsw$ Lie algebra in the limit $r\rightarrow\infty$, with the Weyl rescaling corresponding to the $\mathfrak{sl(2,\mathbb{R})}^S$ generator  preserving the null generator of $\scri$, while the $\R_T^S$ contribution  corresponds to the   normal super-translation along $\scri$. This result provides clear evidence that the local holography program can be equally well applied to null infinity and this is what we concentrate on in this manuscript.

 From this perspective, it is fundamental to understand the full power of symmetries in describing the properties of a gravitational system, 
 and investigate the role of the kinematical subgroup of BMSW as well as the role of supertranslations.
 Therefore,  our goal is to understand how far the symmetry principle can take us in the description of a gravitational system and its asymptotic dynamics, and from there to the quantum realm of gravity. We aim to establish that the  dominant asymptotic Einstein's equations can be  recovered purely from a symmetry argument. A first indication that this is indeed possible, in the context of null infinity, comes from the analysis of  \cite{Freidel:2021yqe}, where a new charge bracket generalizing a previous proposal of Barnich and Troessaert  \cite{Barnich:2011mi}, and derived from first principles in \cite{Freidel:2021cbc}, was introduced to represent the $\bmsw$ Lie algebra in terms of the Noether charges associated to it. It was shown that the demand that the BMSW Lie algebra being represented  at all times along $\scri$ without any  2-cocycle extension is {\it equivalent } to imposing the asymptotic Einstein's equations at null infinity.
On a similar vein, evidence that the asymptotic symmetry group is strong enough to reconstruct the MHV sector of S-matrix amplitudes has been given by Banerjee et al. \cite{ Banerjee:2020zlg, Banerjee:2020vnt, Banerjee:2021cly, Banerjee:2021dlm}. Providing evidence that symmetry can be strong enough to significantly constraint S-matrix amplitudes is one  of the corner-stones of the program of local holography \cite{Strominger:2017zoo, Pasterski:2021rjz, Raclariu:2021zjz}. 
Our work can be viewed as a classical and group-theoretical analog of this quest.
 It provides new evidence  that symmetry might be strong enough to determine the dynamics. Let us point out that an approach  similar in spirit has been applied in \cite{Ciambelli:2018ojf} to relate the study of dynamics in Carrollian geometries to the analysis of symmetries at null infinity; in this case though the group of interest is the boundary group of Carrollian diffeormorphism, not the corner symmetry group.

Here, we exploit the BMSW group structure in order to derive the asymptotic Einstein's equations  at null infinity in a more direct way, making the symmetry argument even more explicit. 
More precisely, in Section \ref{sec:Ano}, after deriving the symmetry transformations of the asymptotic metric components, we first identify a set of  semi-covariant observables  $(\cN^{AB},\cJ^A,\cM,\tcM,\cP_A,\cT_{AB})$.
They are defined as  Bondi metric functionals that do not possess quadratic anomalies, under the BMSW transformations.
They also transform homogeneously (i.e. tensorially) under  the non-extended BMSW group, namely when time super-translations are not included. In Section \ref{sec:Weyl-scalars} we show that these are in direct relation with the five asymptotic Weyl scalars \cite{NP62, Newman:1962cia, Adamo:2009vu} at null infinity.
We then look for combinations containing  time derivatives of the semi-covariant observables, which transform homogeneously, that is with no anomalies at all, under the  BMSW group.
This singles out five relations which express the asymptotic Einstein's evolution equations  {at leading order in the large-$r$ expansion around null infinity}
in an elegant and simple form. 
As derived in Section \ref{sec:EOM}, these read
\begin{subequations}\la{ceom-intro}
\be
 \dot \cJ^A &= \f12  \bD_B {\cN}^{AB}\,,\\ 
\dot \cM &= \f12  \bD_A \cJ^A  +  \f18  C_{AB} {\cN}^{AB}\,,\\
\dot \tcM &= \f12  \bD_A \tilde \cJ^A  +  \f18  C_{AB} {\tilde \cN}^{AB}\,,\\
\dot\cP_A &=  D_A \cM + \tilde{D}_A\tcM  + C_{AB} \cJ^B\,, \\
\dot \cT_{AB}&= \bD_{\langle A} \cP_{B\rangle}+\f32\left( C_{AB}\cM+\tilde C_{AB} \tcM\right)\,.
\ee
\end{subequations}
In the expressions above ${\cN}^{AB}=\dot N^{AB}$ is the time derivative of the news tensor; the tilde denotes a notion of duality in the gravitational phase space at null infinity that we introduce in Section \ref{sec:Ano} and it encompasses the notion of dual gravitational charges introduced in  \cite{Godazgar:2018qpq,Godazgar:2018dvh,Godazgar:2019dkh} and further studied in \cite{Godazgar:2020kqd,Godazgar:2020gqd,Kol:2019nkc,Kol:2020vet,Oliveri:2020xls}. The Einstein's evolution equations recast as in \eqref{ceom-intro} exhibit a manifest invariance under this duality transformation. Moreover, we show how the symmetry argument can be applied in the presence of matter as well, allowing us to derive the correct combination of stress-energy tensor  components (and their derivatives) that source the Einstein's evolution equations.

{While our derivation of the asymptotic  Einstein's evolution equations  is totally independent from the Newman--Penrose formalism, 
the final form of the equations agree with their  central results \cite{Newman:1962cia, NP62, Adamo:2009vu}.
This can be seen by  exploiting the explicit relation between the semi-covariant observables and the asymptotic Weyl scalars, summarized in Section   \ref{sec:Weyl-scalars},
and showing that \eqref{ceom-intro} agrees with the Newman--Penrose derivation of the time evolution of the asymptotic Weyl scalars.
The advantage and novelty of our approach is  the explicit derivation of the symmetry transformations for these observables  from the Bondi gauge variables (see also \cite{Freidel:2021yqe, Grant:2021sxk})  and the  emphasis that invariance under the   BMSW asymptotic symmetry group is enough to ensure the derivation of the equations of motion. This represents the first main result of the paper.
It gives a posteriori a justification for the success of the Penrose-Newman formalism, by showing that  it is naturally adapted  to the concept of asymptotic symmetries.
}
This sets the stage for  the rest of our analysis and it opens the way towards a quantum analysis.
 

More precisely, in Section \ref{sec:CO} we focus our attention on the non-radiative phase space. 
The no radiation condition is defined by the vanishing of the time derivative of the news tensor, that is $\dot{N}^{AB}=0$, which provides a more relaxed definition of non-radiative phase space than the usual condition ${N}^{AB}=0$, and corresponds to the case where no outgoing radiation is registered at $\scri$.
We can, under the no radiation condition, integrate the evolution equations
and we construct a new set  of \emph{conserved} charges  $(j_A, m,\tilde{m},p_A,t_{AB})$ defined in terms of the covariant ones. These charges parametrize the non-radiative corner phase space on $\scri$ and their transformation properties are obtained to be
\begin{subequations}\la{ddc-intro}
\be
\d_{(T,W,Y)} j^A &= \left[\cL_Y + 4W \right] j^A\,, \\
 \d_{(T,W,Y)} m &=
\left[\cL_Y + 3 W \right]m 
+  j^A \pa_AT   +  
\f{T}2 \bD_A j^A\, , \\
\delta_{(T,W,Y)}  \tilde{m} &=
[ {\cL}_Y +3W] \tilde{m} 
+  \tilde{j}^A \pa_A {T} +  
\f{T}2  \bD_A\tilde j^A\,, \\
\d_{(T,W,Y)} p_A &= \left[ \cL_{Y} + 2 W \right] p_A + 
\f32 (m \pa_A T + \tilde{m} \tilde\pa_AT)
 + \f T2 \left(\p_A m +\tilde{\p}_A \tilde{m} +  c_{AB} j^B \right)\,, \\
\d_{(T,W,Y)} t_{AB} &= \left[ \cL_{Y} +  W \right] t_{AB}  +
\f83  p_{\langle A}  \p_{B \rangle} T 
 + T\left(\f23  \bD_{\langle A}p_{B\rangle} +   \f12 c_{AB}m+ \f12 \tilde c_{AB} \tilde{m}\right)\,,
\ee
\end{subequations}
{where $T,W,Y$ are transformation parameters which are functions of the coordinates on the  celestial sphere and label respectively supertranslations, Weyl rescalings and tangent diffeomorphisms.
This set of transformations represents the second main result of the paper as it generalizes \footnote{ In \cite{Barnich:2011ty, Barnich:2016lyg, Barnich:2019vzx}  the metric is restricted to be conformally spherical and the diffeomorphisms are restricted to be local Killing vector fields. Our  derivation relax this restriction  and includes the full group of sphere diffeomorphism.}  to the BMSW group the   one identified  by Barnich et al. \cite{Barnich:2011ty, Barnich:2016lyg, Barnich:2019vzx} in the Penrose--Newman formalism for the extended BMS group \cite{Barnich:2009se, Barnich:2011mi}.
Our derivation provides a more direct and independent derivation of these transformation laws from the Bondi formalism.
}
These transformations  constitute the starting point for the construction 
in \cite{Barnich:2021dta} of the moment map between the non-radiative corner phase space of null infinity and the dual Lie algebra of its full symmetry group. 
Indeed the charge conservation and the closure of symmetry transformations are two indicators that the charges can be understood as moment maps representing the action of an extended symmetry group 
on the asymptotic gravity phase space.  Even if the explicit  action of the new spin-2 charge $t_{AB}$ on the gravity phase space has now been  revealed in \cite{Freidel:2021dfs, Freidel:2021ytz},\footnote{These two references have also appeared on the arXiv shortly after the first version of this manuscript.  } establishing that these conserved charges, including the dual mass and the spin two charge aspect, define a moment map for a generalization of the BMS group still needs to be carried out. 

In order to understand the relationship between the charge aspects we  revealed and the radiation, we
investigate in Section \ref{sec:IW} how an initial vacuum state of the non-radiative phase space is changed by an impulsive gravitational wave localized at $u=0$, transitioning into a new vacuum. 
To do so, we find the non-linear impulsive solutions that describe this transition by integrating the evolution equations \eqref{ceom-intro}   in the case where the Weyl tensor component $\cN_{AB}$ is proportional to a delta function $\d(u)$ through an impulse news function on the 2-sphere. By demanding continuity of the induced metric, we are able to integrate all the evolution equations without encountering distributional singularities. Quite surprisingly, we find that all the Weyl scalars are activated by the gravitational impulse. This is in contrast with the usual solution for an impulsive gravitational wave where only one Weyl scalar is non-vanishing  \cite{Aichelburg:1966su, Szekeres:1970vg, Khan:1971vh, Penrose-72, Hogan:1993xj,Aliev:2000jp, Podolsky:2002sa, Luk:2012hi,Luk:2013zr}---although other Weyl scalars can be activated from collisions---.
 More precisely, the remaining covariant charges are of the form
\begin{subequations}\la{IWsol-intro}
\be
\cJ_{A}(u)&= \cJ_{A}^ {\mathrm{NR}}+\cJ_{A}^{ \mathrm{R}1}\,,\\
\cM(u)&=\cM^{\mathrm{NR}}+\cM^{\mathrm{R1}}\,,\\
\tcM(u)&=\tcM^{\mathrm{NR}}+\tcM^{\mathrm{R1}}\,,\\
\cP_{A}(u)&= \cP_{A}^{\mathrm{NR}}+\cP_{A}^{\mathrm{R}1}+\cP_{A}^{\mathrm{R}2}\,,\\
\cT_{AB}(u)&= \cT_{AB}^{\mathrm{NR}}+\cT_{AB}^{\mathrm{R}1}+\cT_{AB}^{\mathrm{R}2}\,,
\ee
\end{subequations}
where the label $\mathrm{NR}$ denotes the non-radiative expressions given in  \eqref{C-exp}, $\mathrm{R1}$  a distributional radiative component   linear in the impulse news and  $\mathrm{R2}$  a secular radiative component quadratic in the impulse news. The explicit expressions for the impulsive wave transition represent the third main result of the paper and they are given in Section \ref{sec:IW-sol}.

We present our conclusions in Section \ref{sec:conc} and many technical derivations of various relations used in the main text in a series of Appendices \ref{App:action}, \ref{sec:Variations}, \ref{App:Peom}, \ref{sec:SET}.

\paragraph{Notation}
We use units in which $8\pi G=1$ and $c=1$. 
Greek letters are used for spacetime indices and uppercase Latin letters $\{A, B, C, \dots\}$ for coordinates over the 2D sphere. 
The symbol $\stackrel{\scri}=$ is used when the right-hand side is evaluated at future null infinity $\scri$. We denote the symmetric, trace-free part of a tensor $T_{AB}$ with the brackets $\langle \cdot \rangle$, namely
\be
T_{\langle A  B  \rangle}=\frac12\left(T_{A  B }+T_{BA} -\bq_{AB} \bq^{CD}T_{CD}\right)\,,
\ee
{where $q_{AB}$ is the   asymptotic metric on the two-sphere.}

\section{Future null infinity}\la{sec:BMSW}

Let us introduce Bondi coordinates $x^\mu=(u,r,\sigma^A)$, where $u$ labels null outgoing geodesic congruences which intersect infinity along 2d spheres, $r$ is a parameter along these geodesics   measuring the  sphere's radius ($r$ is the luminosity distance) and 
 $\sigma^A$ denotes coordinates  on the celestial sphere.
In these coordinates, the metric is given by \cite{Bondi:1960jsa, BMS, Sachs:1962wk}
\be \label{eq:BondiMetric}
\rd s^2 = -2e^{2\beta} \rd u \left( \rd r  + \Phi \rd u\right) + r^2 \gamma_{AB} \left(\rd \sigma^A -\f{\Upsilon^A}{r^2}\rd u\right)\left(\rd \sigma^B -\f{\Upsilon^B}{r^2}\rd u\right).
\ee
This metric satisfy the Bondi \emph{gauge conditions} given by  
\be\label{Bondig}
g_{rr}=0,\qquad g_{rA}=0,\qquad \pa_r \sqrt{\gamma}=0. 
\ee
In addition to the gauge condition we 
impose \emph{extended}\footnote{Note that we do not require $R(q)=2$ which is why we call our boundary conditions \emph{extended}.} Bondi \emph{asymptotic boundary conditions} \cite{Barnich:2009se,Barnich:2016lyg,Barnich:2010eb}, which are given  by  
\be\label{Bondias}
g_{ur}\stackrel{\scri}= -1,\qquad  \frac1rg_{uu}\stackrel{\scri}= 0,\qquad 
\frac1rg_{uA}\stackrel{\scri}= 0, 
\qquad 
\pa_u q_{AB} \stackrel{\scri}= 0. 
\ee
In this work we assume the usual Bondi like  asymptotic  boundary conditions  which imply that 
the metric components $(\Phi, \beta, \gamma_{AB}, \Upsilon^A)$ have the following fall-off behavior \footnote{ This is a restrcition of our analysis: One could be more general by allowing $b$ to  be order $1$ and by allowing $\Phi$ to admit a term growing linearly in $r$.
We follow here the  original treatment \cite{Bondi:1960jsa, BMS, Sachs62}   and all the  subsequent extensions \cite{Barnich:2009se, Barnich:2011mi, Campiglia:2014yka, Flanagan:2015pxa, Compere:2018ylh}   of asymptotic symmetries of null infinity, we have chosen fall-off conditions that do not  include these terms from the beginning in order to simplify the rest of the analysis.
We come back to this point at the beginning of Section \ref{sec:EOM}.
}
\begin{subequations}\la{eq:FallOff}
\begin{align}
\Phi&= F({u}, \s^A) - \frac{  M(u, \s^A)}{r}+o(r^{-1})\,,\\
\beta&=\frac{b(u, \s^A)}{r^2}+o(r^{-2})\,,\\
\Upsilon^A&={U^A(u, \s^A)}- 
\frac{2 {q}^{AB}}{ 3 r } 
( P_B+ \bC_{BC} U^C +\pa_B  \B   )(u,\s^A)  +o(r^{-1})\,,\label{PA}\\
\gamma_{AB}&=  {q}_{AB}( {u}, \s^A) + \frac{C_{AB} (u, \s^A)}{r}+ \frac{1}{r^2}\left(D_{AB}+\f14 q_{AB}C_{CD}C^{CD} \right)(u, \s^A)
+ \f{E_{AB}(u, \s^A)}{r^3}  +  o(r^{-3})\,.
\la{gamma}
\end{align}
\end{subequations}
The expansions of the different coefficients are  needed to obtain the 
expansion of the metric $ g_{\mu\nu} \rd x^\mu \rd x^\nu$ 
to order\footnote{Since $\rd r$ is of order $O(r)$, $g_{ur}$ needs to be expanded to order $O(r^{-2})$, since $g_{AB}= r^2 \gamma_{AB}$,  $\gamma_{AB}$ needs to be expanded to order $O(r^{-3})$ and since $g_{uA} = \gamma_{AB} \Upsilon^A$,  $\Upsilon^A$ needs to be expanded to order $O(r^{-1})$,  to achieved  $O(r^{-1})$ for the expansion of  the metric $g_{\m\n} \rd x^\m \rd x^\n$.} $O(r^{-1})$.
%
Here $M$ is the Bondi mass aspect, $\bU^A$ is the asymptotic velocity,    $\bC_{AB}$ is {twice} the asymptotic shear.
If one restricts ${q}_{AB}$ to be the round sphere metric $\mathring{q}_{AB}$ with $R(\mathring{q})=2$, one recovers the restricted Bondi \emph{boundary} conditions.
We will avoid doing that in the following and keep the asymptotic conditions
just stated.  Because of the Bondi determinant gauge condition the symmetric tensors $C_{AB}, D_{AB}, E_{AB}$ are all \emph{traceless} when contracted with the inverse asymptotic metric $q^{AB}$.
The  $O(r^{-2})$ factor in the metric expansion is uniquely  determined by the Bondi gauge condition;  the demand that logarithmic anomalies vanish requires $D_{AB}=0$ \cite{Winicour16} and we assume this in the following.
When evaluating the asymptotic, we use the metric ${q}_{AB}$  to lower and raise the indices $\{A, B, \dots\}$ on the 2-sphere. 
%

The leading asymptotic Einstein's equations (EEs) give a first relation
\be\la{asym-EE}
\pa_u {q}_{AB}=0\,,
\ee
which can be understood as a boundary condition, implying the $u$-independence of the leading order of the metric component   $g_{AB}$. {While we derive from symmetry all the other EEs, this one is assumed and taken as a boundary condition from now on.}
There is then a second set of asymptotic vacuum Einstein's equations given by the relations \footnote{{The boundary condition \eqref{asym-EE} together with the Einstein's equation \eqref{EEF} clearly imply that the leading order of $g_{uu}$ is time independent, namely $\pa_u F=0$. }}
\be
 \textsf E_F & :=F- \frac{R({q})}4=0\,,\la{EEF}\\
 \textsf E_{U}^A&:=  {U}^A+ \frac12 {D}_B \bC^{AB}=0\,,\la{EEU}\\
 \textsf E_{\B}&:=  \B  + \frac{1}{32 }  \bC_{AB}\bC^{AB}=0\,,\la{EEB}
\ee
where ${D_A}$ is the covariant derivative associated with $q_{AB}$. These can be understood as constraints between phase space data {and we will show below how they can be obtained simply from the BMSW transformation properties and the  requirement of covariance under it}. 
The  next two asymptotic equations are 
\begin{align}
 \textsf E^{ M}  &:=\dot{M} - \f14 \bD_A\bD_{B} N^{AB} -\f18 {\Delta}R + \f18 N_{AB}N^{AB} =0\, ,\label{EEM}\\
 \textsf E^P_{A} &:= \dot{\bP}_A -\bD_A M - \f18 \bD_A \left( C^{BC}N_{CB}\right) - \f14 C_{AB} \pa^B R \cr
&\quad -\f14 \bD_C \left( \bD_A \bD_B C^{BC} - \bD^C\bD^{B}C_{AB}\right)\cr
&\quad -\f14 \bD_B \left( N^{BC}C_{AC} - C^{BC}N_{AC}\right) + \f14 N^{BC}\bD_A C_{BC}=0\,,\la{EEP}
\end{align}
and they correspond to evolution equations for the energy aspect $M $ and the momentum aspect $P_A$. 
At the order we are working in, the last asymptotic equation is an evolution equation for the spin-$2$ tensor $E_{AB}$. 
It was written explicitly in the gauge we are adopting here by Nichols\footnote{ Nichols $N_A$ is equal to
\be 
N_A= P_A + \f14 C_{A  D}  \bD_C C^{CD}
+\f1{16}\p_{A} (C_{CD}C^{CD})\,.
\ee It is the same as what we call the covariant momentum later.} in \cite{Nichols:2018qac}
\be\la{EEE}
 \textsf E^E_{AB } &:=
\dot E_{AB}
-\f12C_{AB} M-\f13 \bD_{\langle A} P_{B\rangle} 
+ \frac{1}{8}\epsilon_A{}^{C}C_{CB}\epsilon_{D}{}^E D_E D_C C^{CD}
-\f14  C_{AB} N_{CD} C^{CD}
\cr
&-\f1{3} \bD_{\langle A}  \left(\f14 C_{B\rangle D}  \bD_C C^{CD}
+\f1{16}\p_{B\rangle} (C_{CD}C^{CD})
\right)\,.
\ee
These three evolution equations are  derived as well using our symmetry argument.

 \subsection{\bmw vector fields}
 
The infinitesimal \bmw diffeomorphisms introduced in \cite{Freidel:2021yqe} are 
spacetime diffeomorphisms  preserving the \emph{boundary} conditions above. These are labelled by a vector field $Y^A$ on $S$  representing  asymptotic diffeomorphisms of the celestial sphere, a super-translation parameter $T$ and a Weyl transformation parameter $W$,
which are all  independent of $u$ and $r$.
The \bmw vector fields can be conveniently  written, as $\xi_{(\tau,Y)}$,  in terms of  the parameter  $\tau=\tau(T,W)$ given by
\be
\tau :=T+ u W\,,\qquad
\dot{\tau} = W\,,
\qquad \ddot{\tau}=0.
\ee
The \bmw vector fields $\xi_{(\tau,Y)} $ are characterized as the bulk vector fields preserving  the Bondi gauge and asymptotic conditions (\ref{Bondig}), (\ref{Bondias})
which evaluate on $\scri$ to  
$\xi_{(\tau,Y)}\stackrel{\scri}= \bxi_{(\tau,Y)}$,
where the  asymptotic \bmw vector fields are  
\be
\bxi_{(\tau,Y)}&:=\tau \pa_u + Y^A\pa_A -\dot{\tau}\, r\pa_r,\cr
&= T \pa_u + W(u\pa_u - r\pa_r) + Y^A \pa_A.
\la{xib}
\ee
We see that with the first parametrization $\tau$ labels time translations, while
$\dot{\tau}$ labels conformal rescaling.
In the second parametrization $T$ labels time translations, while $W$ labels asymptotic boosts.

To write down explicitly the bulk extension $\xi_{(\tau,Y)} $, it is convenient to define
\be
 I^{AB}:= \left(\int_r^\infty  \f{\rd r'}{r^{'2}} e^{2\beta}  \gamma^{AB}\right),
 \ee
 which is such that  $I^{AB}\stackrel{\scri}=0$ and 
$ r\pa_r I^{AB}= -\frac1{r} e^{2\beta}  \gamma^{AB}$.
The corresponding vector fields are of the form  \footnote{ We have used that, thanks to the Bondi gauge, we have  
$D^\gamma_A Z^A=\frac1{\sqrt{\gamma}}\pa_A ( \sqrt{\gamma}Z^A) = 
\frac1{\sqrt{q}}\pa_A ( \sqrt{q} Z^A)= {D}_AZ^A$, for a generic vector $Z$ on the sphere, with $D^\gamma_A$ the covariant derivative associated to $\gamma _{AB}$.}
\begin{subequations}\la{vu}
\be
\xi^u_{(\tau,Y)}&=\tau\,,\\
\xi^A_{(\tau,Y)}&=Y^A -I^{AB} \p_B \tau\,, \label{vA}\\
 \xi^r_{(\tau,Y)}&= r\left( \f{1}{2}D_A(I^{AB} \p_B \tau) + \f{1}{2r^2} \Upsilon^A\p_A \tau   - \dot\tau  \right) \label{vr}\,.
\ee
\end{subequations}

 We can  check that  $W(\sigma^A)$ induces a  Weyl rescaling of the celestial sphere, as
 \be\la{dxiq}
{\cal L}_{\xi} \sqrt q= ( D_AY^A-2 W )\sqrt q\,,
\ee
 where $q:=\det(q_{AB})$. 
 The generalized BMS group proposed in \cite{Campiglia:2014yka, Compere:2018ylh} is recovered by setting $W=\f12 D_AY^A$, so that the condition $\delta \sqrt{q}=0$ is preserved by the symmetry transformations.

\subsection{\bmw symmetry transformations}

The boundary \bmw symmetry group  is asymptotically generated by the vector fields $\xi_{(\tau,Y)}$.
Its Lie algebra is isomorphic to the double semi-direct sum \cite{Freidel:2021yqe}
\be\la{bmsw}
\bmsw:= \left(  \mathrm{diff}(S) \oright \mathbb{R}_W^S\right)  \oright \mathbb{R}_T^S .
\ee
The first factor $\R_W^S$ denotes the Weyl transformations labeled by functions $W$ on the sphere,
while the second factor $\R_T^S$ denotes the super-translations labelled by functions $T$ on the sphere of density weight $1/2$.
The commutators are given by 
\be
[\bxi_{(\tau_1, Y_1)}, \bxi_{(\tau_2, Y_2)} ]
= 
\bxi_{\left(\tau_{12}, Y_{12}\right) }.
\ee
Here we parametrized the $(\R_W^S \oright \R_T^S)$ by functions on the sphere $\tau = T +u W$ which are linear in time and we have denoted
\be
\tau_{12}=  \tau_1\dot{\tau}_2- \tau_2\dot{\tau}_1
+ Y_1[\tau_2] -Y_2[\tau_1]\,,\quad
Y_{12}^A = [Y_1,Y_2]_{\mathrm{Lie}}^A\,, 
\ee
where $Y[\tau]:=Y^A\pa_A\tau$. 
%

%

The quantities of physical interests such as $( q_{AB}, C_{AB}, F, M, b, U_A, \bP_A)$ are 
functionals $\Phi^i(g_{\m\n})$ of the metric. The transformations
 of these functionals  are given by the chain rule
$
\delta_{(\t, Y)} \Phi^i = \int \frac{\delta \Phi^i}{\delta g_{\m\n}} \cL_{\xi_{(\t, Y)}}g_{\m\n}.
$
In practice, to evaluate the variations  $\delta_{(\t, Y)} \Phi^i$ we use that $g_{\m\n}$ is determined by $\Phi_i$ and we evaluate 
the condition 
\be
{\cal L}_{\xi_{(\t,Y)}} g_{\mu\nu}[\Phi^i]=\left.\frac{\pa}{\pa \epsilon} g_{\mu\nu}[\Phi^i+ \epsilon\d_{(\tau,Y)} \Phi^i] \right|_{\epsilon=0}\,.
\la{Ld}
\ee
The explicit derivations are given in Appendix \ref{App:action}. 

By focusing on the metric component $g_{AB}$, one can easily derive the transformations
\begin{subequations}\la{d}
\be
\d_{(\tau,Y)} \bq_{AB}
&=\left[ \cL_Y - 2 \dot{\tau}\right]  \bq_{AB}\,,
\la{dqAB}
\\
\d_{(\tau,Y)} C_{AB}
&=\left[ \tau \pa_u  + \cL_Y - \dot{\tau} \right]  C_{AB} - {2\bD_{\langle A} \bD_{B\rangle} \tau }\,
\la{dC}.
\ee
Taking the trace  of $q$  implies that 
\be 
\d_{(\tau, Y)}\sqrt{\bq}&=\left[\bD_{A} Y^A - 2\dot{\tau}\right]  \sqrt{\bq}\,.\la{dq}
\ee
Taking the time derivative $N_{AB} =\dot{C}_{AB}$ and the divergence of the shear $(\bD\!\cdot\! C)^B:= D_AC^{AB}$ implies that 
\be 
\d_{(\tau,Y)} N_{AB}
&=\left[ \tau \pa_u  + \cL_Y \right]  N_{AB} - {2\bD_{\langle A} \pa_{B\rangle} \dot\tau }\,,
\label{dN}\\
\d_{(\tau,Y)} (\bD\!\cdot\! C)^B &= [\tau\pa_u + \cL_Y + 3\dot\tau ](\bD\!\cdot\! C)^B
+ (N^{BA}\pa_A \tau- C^{BA} \pa_A \dot\tau) 
-(R(q)\pa^B\tau +\pa^B\D \tau)\,.
\la{dDC}
\ee 
The second equality is established in Appendix \ref{sec:Variations}.

Next, focusing on the $g_{uu}$ component, one can derive the two transformations
\be
\d_{(\tau,Y)} \bF&=\left[\cL_Y + 2 \dot{\tau} \right] \bF  + \f12  \Delta \dot{\tau},\label{dF}\\
\d_{(\tau,Y)} M&=\left[\tau \pa_u +\cL_Y + 3 \dot{\tau} \right]M
+ \left( \f12 \bD_B N^{AB} +\pa^A F \right)\pa_A \tau\cr
&+
\f14 N^{AB} \bD_A\pa_B\tau 
+\f14  C^{AB} \bD_A \p_B \dot\tau
+  {\f12 ({\textsf E_U^A \pa_A \dot\tau}-  \dot{ \textsf E}_U^A \pa_A \tau)} \,,
\la{dM}
\ee
where $\textsf E_{U}^A$ is the asymptotic Einstein's equation \eqref{EEU}.
The $g_{ur}$ component gives the transformation
\be
\d_{(\tau,Y)}  \B  &=\left[\tau \pa_u +\cL_Y + 2 \dot{\tau} \right]  \B   
+\f1{8} C^{AB} D_A\pa_B \tau
+\f14  \textsf E_{U}^A \p_A \t
\,.
\la{dB}
\ee
From the Lie derivative of the $g_{uA}$ component we can read off the transformation  of the   two functionals
\be
\d_{(\t,Y)} \bU_A &=   \left[\tau\pa_u + \cL_{Y} + \dot\tau  \right]\bU_A 
   + \f12 (4 \bF  \pa_A\tau +\pa_A \Delta\tau)  +\f12( C_{A}{}^{B} \pa_B\dot\tau - N_{A}{}^{B}\pa_B\tau)\,,\la{dUA}\\
 \d_{(\tau,Y)} P_A &\,\hat{=}\, [\tau \pa_u +\cL_Y + 2\dot\tau ] P_A  
 +3M\p_A \t
-\f18 C_{BC} N^{BC} \p_A\t
+\f12 C_{AB}  N^{BC}\p_C \tau
  \cr
 &+ \f34 (\bD_A\bD_C C_B{}^{C}- \bD_B\bD_C C_A{}^{C}) \pa^B \tau
 + \f14 \p_A ( C^{BC} \bD_B\bD_C \tau) \cr
 &+ \f12 \bD_{\langle A } \bD_{B\rangle}\tau \bD_C C^{BC} 
 +  C_{AB} \left( F \pa^B\tau + \f14\pa^B \Delta \tau \right)
 -2\dot{ \textsf E}_{\B} \pa_A \tau\,.
 \la{dP}
\ee
\end{subequations}
The hatted equality refers to the fact that 
we have used the asymptotic equations $\textsf E_{U}^A\,\hat=\,0$ to simplify the RHS of  expression \eqref{dP}.

Finally,  the most extensive calculation concerns the variation of the traceless component $E_{AB}$. One finds in Appendix \ref{App:Eano} that 
\be
\d_{(\tau,Y)}E_{AB}&\,\hat{=}\, [\tau \pa_u  + \cL_Y  +\dot{\tau}]  E_{AB}\cr
&+\frac{4 }{ 3 } \left(\bP_{\langle A}+\f14 D_D C^{DC}  C_{C\langle A} 
-8\pa_{ \langle A}\B \right)  \p_{B\rangle } \tau \cr
&+\f1{2}\left( 
C^{CD}\bD_C C_{AB}
- C_{AB} {\bD}_CC^{CD}
\right)\p_D \tau- \bD_{\langle A} C_{B \rangle  C}  C^{CD}\p_D \tau
  \cr
 &+ 4 \B \bD_{\langle A} \p_{B\rangle } \tau  - \f14 C_{AB} C^{CD}{\bD}_C \p_D\tau
 -\f{16}3\textsf E_\B  \bD_{\langle A} \p_{B\rangle} \tau+\f{32}3\p_{ \langle A}  \textsf E_\B  \p_{B \rangle} \tau\,.\label{dE}
\ee

\subsection{Relation to the extended corner symmetry group}\la{sec:gext}

As shown in \cite{DonnellyFreidel}, the \emph{ corner symmetry} group is the  gravitational symmetry associated to a generic codimension-2 surface called corner. This surface can be thought of as  bordering  a bounded region of space. 
The  \emph{corner symmetry algebra} is simply the subalgebra of diffeomorphisms that do not change the position of the corner surface.  It is given by the semi-direct sum of the surface diffeomorphism and surface boosts. Explicitly, we have
\be\la{gS}
\mathfrak{g}_S = \mathrm{diff}(S)\oright  \mathfrak{sl(2,\mathbb{R})}^S\,.
\ee
Its extension to include time translations normal to the surface yields  the notion of the 
 \emph{extended corner symmetry algebra}, as revealed in \cite{Ciambelli:2021vnn, Freidel:2021cbc}, which includes also two copies of $\mathbb{R}$ corresponding to the surface translations along the two normal directions. Its explicit structure is given by
\be\label{gext}
\mathfrak{g}^{\rm ext}_S = \left(  \mathrm{diff}(S)\oright\mathfrak{sl(2,\mathbb{R})}^S \right) \oright (\R^2)^S \,.
\ee
As shown in \cite{Freidel:2021cbc}, the \textsf{bmsw} Lie algebra \eqref{bmsw} corresponds to  a subalgebra of the  extended corner symmetry algebra $\mathfrak{g}^{\rm ext}_S$ in the bulk, where the 
$\R_W^S$ contribution is given by one of the  $\sl(2, \R)$  generators (namely, the one  preserving the null generator of $\scri$), while the $\R_T^S$ contribution  corresponds to one of the two  normal super-translations (namely, the one along $\scri$). We can thus understand the BMSW group as the $r\rightarrow \infty$ limit (a contraction) of the extended corner symmetry group.

\section{Anomalies}\la{sec:Ano}
The anomaly operator $\Delta_{\tau}$ associated with 
a functional $\cal{O}$ of conformal dimension $s$ is given by 
\be
\Delta_{\tau} {\cal{O}} := \delta_{(\tau,Y)}{\cal{O}}- (\tau\pa_u +\cL_Y + s \dot\tau) {\cal{O}}.
\ee
This anomaly measures the difference  between the natural action of the BMSW group on  $\cal{O}$   and its field space action. By construction the anomaly only depends on $\tau$.
The transformation rules reported in the previous section have the general structure
\be
\delta_{(\tau,Y)} {\cal{O}}= [\tau \pa_u +\cL_Y + s \dot\tau ] {\cal{O}}  + L_{\cal{O}}^A\pa_A\tau +\bar{L}_{\cal{O}}^A \pa_A\dot\tau +
Q_{\cal{O}}^{AB} \bD_A\pa_B \tau +  
\bar{Q}_{\cal{O}}^{AB} \bD_A\pa_B \dot\tau.
\ee
The first term is the homogeneous transformation that involves the scale weight\footnote{$s$ can also be understood as a boost weight.} $s$  of the functional ${\cal{O}}$.  All scale weights of the different functionals can be found by assigning scale weight $s(\rd s^2)=0$, while $s(r)=s(\rd r)=+1$ and $s(u)=s(\rd u)=-1$ in the metric expansion, hence the scale weight of $\pa_u $ is $+1$. 
Functionals that transform homogeneously are sections of the scale bundle $P$. \footnote{{We call {\it scale bundle} a   line bundle $P\to \scri $ over $\scri$ whose  automorphism group includes the asymptotic BMSW vector fields \eqref{xib}}.} 
The inhomogeneous terms are of two types: $(L_{\cal{O}}^A , \bar{L}_{\cal{O}}^A)$ which we call \emph{linear anomalies} and terms $(Q_{\cal{O}}^{AB},  \bar{Q}_{\cal{O}}^{AB})$ which are the \emph{quadratic anomalies}.
An example of anomaly is 
\be
\Delta_\tau C_{AB}  = -2 D_{\langle A}\pa_{B\rangle} \tau\,.
\ee
 The functional $\cal{O}$ is said to be tensorial when both linear and quadratic anomalies vanish.
  The first examples of tensorial combinations are  the quantities $ q_{AB}, \dot N_{AB}$ which satisfy
\be
\Delta_{\tau} q_{AB}=0,\qquad \Delta_\tau \dot N_{AB}= 0\,.
\ee
They are operators of scale weight $(-2, +1)$ respectively. 
The main theme of our paper is that we can recover the equations of motion by identifying the tensorial combinations. For instance, we can easily see  from the previous expressions that the asymptotic equations of motion \eqref{asym-EE}, \eqref{EEU}, \eqref{EEB} all
transform tensorially: $\textsf E_F:= R-\f14 F$ as a section of weight $2$, $\textsf E_\B$ as a section of weight $2$, $\textsf E_U^A$
as a vector of weight $3$.

Another class of operators which will be of interest to us is the {\it pseudo-tensors} of  weight $s$. These are characterized by the fact that  the quadratic anomaly  and  the anomaly $\bar{L}_{\cal{O}}^A$ linear in $\dot \tau$ vanish while    the linear anomaly $ {L}_{\cal{O}}^A$ does not.
The pseudo-tensors can be understood as tensorial for the subgroup of symmetry that does not include super-translations.



The next example we want to study involves the Liouville stress tensor \cite{Compere:2018ylh}. 
Given a metric $q_{AB}$ we can define its Liouville stress tensor\footnote{ The conserved energy momentum tensor of Liouville  is $\tau_{AB}: = T_{AB} + \f12 q_{AB} R(q)$. Its trace is 
$q^{AB}\tau_{AB}= R(q)$. The tensor $T_{AB}(q)$ is also called the Geroch tensor
when $q= e^{\varphi} \mathring{q}$ \cite{Geroch:1977jn}. }
to be the symmetric traceless tensor $T_{AB}(q)$  such that 
\be\la{Tc}
D_A T^{AB} +\f12 \nabla^B R =0.
\ee
The fact that this tensor can be uniquely determined follows first from the fact that 
the equation $\mathring{D}_A T^{AB}=0$ implies, when $S$ is a sphere, that 
$T^{AB}(\mathring{q})=0$.  Second, from the  following covariance
properties under Weyl transformation with parameter $W$ and diffeomorphism $\varphi: S\to S$
\be\la{CovT}
T_{AB}(e^{2 W} {q}) = T_{AB}(q) - 2( D_{\langle  A}W D_{B\rangle} W +D_{\langle A} D_{B\rangle } W), 
\qquad
\varphi^*(T_{AB}(q))= T_{AB}(\varphi^*(q)).  
\ee
Indeed, by the uniformization theorem, any metric on the sphere can be written as 
$q = e^{W} \varphi^*(\mathring{q})$. So the transformation properties \eqref{CovT} allow one to determine $T_{AB}(q)$.
This means that the combination 
\be\la{bN}
\bar N_{AB}:=  N_{AB}   - T_{ AB }(q)
\ee
possesses no  anomaly $\Delta_\tau  \bar N_{AB}=0$. 
It is a tensor operator of scale weight $0$.
From this we can construct a covariant current
\be
\cJ^A := \f12 \bD_B\bar N^{BA}=  \f12 \bD_B N^{AB} + \f14 \pa^A R\,, \la{cJ}
\ee
where the second equality follows from \eqref{Tc}.
The covariant current yields  the Weyl scalar $\Psi_3$, it possesses no quadratic anomaly and it is  a pseudo-tensor of dimension $4$
\be\la{dcJ}
 \d_{(\tau,Y)} \cJ^A = \left[\tau \pa_u +\cL_Y + 4\dot{\tau} \right] \cJ^A + 
 \f12 \dot{N}^{AB} \pa_B\tau\, .
\ee
This can be seen by taking the time derivative of \eqref{dDC} (an alternative derivation is given in Appendix \ref{sec:Variations}).
The explicit relation between all the covariant observables and the Weyl scalars is shown in Section \ref{sec:Weyl-scalars}.
Since  $\dot{N}^{AB}$ is a tensor and $\cJ^A$ is a  pseudo-tensors whose anomaly vanish when $\dot{N}^{AB}=0$, we can define the non-radiative vacua   to be such that  $\cJ^A=0=\dot{N}_{AB}$.
The non-radiative vacua are transformed into each other  by the symmetry transformations.

\subsection{Covariant mass}


We are  interested in combinations of the physical quantities parametrizing the Bondi metric \eqref{eq:BondiMetric}  that transform as pseudo-tensors, with no quadratic anomaly and no Weyl linear anomaly.
To this aim, we introduce the notion of {\it covariant mass}
\be\boxed{\,\,
{\cal M}:= M + \f18  C_{AB}N^{AB}\,.\,\,
}
\la{cM}
\ee

The justification for this name comes from the fact that, by means of \eqref{dM}, \eqref{dC}, \eqref{dN},  the quantity above transforms\footnote{We recall that the hatted equality refers to the fact that 
we use the asymptotic equations $\textsf E_{U}^A=0$.} as
\be \label{dcM}
\d_{(\tau,Y)} {\cal M} \,\hat{=}\,
\left[\tau \pa_u +\cL_Y + 3 \dot{\tau} \right]{\cal M}
+ \cJ^A \pa_A \tau\,.
\ee
 We thus see that only a linear anomaly term appears in the transformation of $\cM$ and moreover that  the linear anomaly depends only on $\tau$, not $\dot\tau$. 
 A nontrivial consistency check for this formula comes from the fact that the 
 variation of $\cJ^A$ does not contain any quadratic anomaly terms, as shown by  \eqref{dcJ}.

This  indicates also that if the non-radiative structure $\cJ^A=0=\dot{N}_{AB}$ is satisfied,  then
the covariant mass aspect $\cM$ transforms homogeneously.
Moreover, flat vacua can be defined by the conditions  $\cM=0=\cJ^A= \dot{N}^{AB}$.
These were parametrized and studied in \cite{Compere:2018ylh, Compere:2016jwb}.

\subsection{Duality and covariant mass}
In this section we show that it is possible to construct 
from $C_{AB},N_{AB}$ and their derivative another scalar of dimension $3$ 
that possesses no quadratic anomaly: The \emph{dual covariant mass}.
To describe its construction, let us introduce the volume form on $S$ denoted $\epsilon_{AB}$ and given by 
$\frac12 \epsilon_{AB} \rd \s^A \wedge \rd \s^B=\sqrt{q} \rd^2\s$.
Raising one of its indices with the metric, one gets the complex structure
\be
\epsilon_A{}^B := \epsilon_{AC} q^{CB},
\qquad 
\epsilon_A{}^B\epsilon_B{}^C =-\delta_A^C. 
\ee
The complex structure is a  tensor of weight $0$
\be
\delta_{(\t,Y)}  \epsilon_A{}^B = (\tau\pa_u + \cL_{Y})\epsilon_A{}^B.
\ee
We can use this complex structure to define a duality transform for  the traceless tensors
$C,N$ and the derivatives. We introduce the notation
\be
\tilde{C}_{AB}:= \epsilon_{A}{}^C C_{C B}= \epsilon_{B}{}^C C_{AC}\,,  
\qquad 
\tilde{N}_{AB}:= \epsilon_{A}{}^{C} N_{CB}=\epsilon_{B}{}^C N_{AC}\,,
\qquad
\tilde{V}_A :=\epsilon_A{}^B V_B\,.
\ee
Note that the equality $\epsilon_{A}{}^C C_{C B}= \epsilon_{B}{}^C C_{AC}$ is only valid for symmetric traceless tensors.

The tilde operation is  a duality $\tilde{\tilde{N}}_{AB}=- {N}_{AB}$,
and we have the properties
\be
\tilde{V}_A W^A = - V_A \tilde{W}^A,\qquad
\tilde{N}_{AB} V^B = - N_{AB} \tilde{V}^B,
\ee
where we denoted $\tilde{V}^A = q^{AB} \tilde{V}_B$.
In particular, this means that 
\be
\widetilde{(D\!\cdot\! N)}_A= \epsilon_{A}{}^B D^C N_{CB} 
= (D\!\cdot\! \tilde{N})_A=- (\tilde{D}\!\cdot\! {N})_A\,.
\ee
The tensor $\epsilon_{AB}$ can be used to convert 2-forms on the sphere into (pseudo)-scalars.
In particular, given $J_{AB}=J_{[AB]}$  a 2-form on $S$ this can be written as 
\be
J_{AB}=\frac12 \tilde{J} \epsilon_{AB},\qquad \tilde{J}= \epsilon^{AB} J_{AB}. 
\ee
An identity that we will repeatedly use in the following derivations is the condition that 
\be\label{key-id}
D_{[A} N_{B]}{}^C  \pa_C\tau =  D_{C} N_{[A}{}^C \pa_{B]}\tau,
\ee
which  follows from the Fierz identity $ D_{[A} N_{B}{}^C  \pa_{C]}\tau=0$ and the fact that $N$ is traceless.
It will also be useful to simplify some tensors using the  identity
\be
\epsilon_{AB}\epsilon^{CD} = \delta _A^C \delta_B^D -\delta_A^D\delta_B^C. 
\ee
Given these preliminaries we can now present the construction of the dual covariant mass.
From the transformation \eqref{dDC}  of $ (D\!\cdot\! C)_B$ we conclude in Appendix \ref{sec:Variations} that 
\be\la{DDtC}
\delta_{(\tau,Y)} (\bD_{A} ( \bD\!\cdot\!\tilde{C})^{A}) &=
[\tau\pa_u + \cL_Y +3\dot\tau ] (\bD_{A} ( \bD\!\cdot\!\tilde{C})^{A}) +  4 \tilde{\cJ}^A \pa_A\tau + ( \tilde{N}^{BC} {D}_{B}\pa_C \tau + C^{BC} \tilde{D}_{B}\pa_C \dot{\tau} ).
\ee
This means that the following combination
\be\la{dualM}
\boxed{\,\,
\tcM:= \frac14 (\bD_{A} ( \bD\!\cdot\!\tilde{C})^{A}) + \f18 C_{AB} \tilde{N}^{AB}\,,\,\,}
\ee
called the {\it covariant dual mass}, 
possesses no quadratic anomalies.
The explicit transformation follows from  the transformations \eqref{DDtC} and \eqref{dC}, \eqref{dN} and 
it is given by 
\be
\delta_{(\tau,Y)}  {\tcM} =
[\tau \pa_u+ {\cL}_Y +3\dot{\tau}] {\tcM} 
+  \tilde\cJ^A \pa_A {\tau}\,.
\la{ddcM}
\ee
This is in absolute parallel with the mass transformation formula \eqref{dM}.
Note that the role of the mass aspect $M$ is played here by the ``vorticity'' of the fluid with velocity $U_A$:
\be 
\tilde{M}:= -\frac12 \epsilon^{AB} D_A U_B  = \frac14 (\bD_{A} ( \bD\!\cdot\!\tilde{C})^{A}) .
\ee
The  covariant mass $\cM$ and the dual covariant mass $\tilde \cM $ determine respectively the real and the imaginary part of the Weyl scalar $\Psi_2$ at $\scri $. 


\subsection{Covariant  momentum}
We now focus on the construction of the covariant momentum.
The transformation of the momentum\footnote{To compare these transformations with the one of \cite{Barnich:2011mi, Compere:2018ylh} one needs to use that $N_A= P_A + \pa_A\B $.} is given in \eqref{dP}.
In order to analyze and simplify this equation, one can  follow the same strategy 
as the one that led to the definition of the covariant mass and look for counter-terms that cancel all the quadratic anomaly terms proportional to $D_A\pa_B \tau$ and $\Delta \tau$.
To do so  one first establishes, using \eqref{dDC} again, that 
\be
\d_{(\tau,Y)} (\bD_C C^{CB} C_{BA})&= [\tau\pa_u + \cL_Y +2\dot\tau ]
(\bD_C C^{CB} C_{BA}) 
+C_{AB} (N^{BC}\pa_C \tau- C^{BC} \pa_C \dot\tau) \cr
&-C_{AB} (R\pa^B\tau +\pa^B\D \tau) 
- 2 \bD_{\langle A}\bD_{B\rangle} \tau \bD_C C^{CB}\,.
\ee


We also use that 
\be
\d_{(\tau,Y)} \pa_A\left( -\f1{32} C_{BC}C^{BC}\right) &\,\hat{=}\, [\tau\pa_u + \cL_Y +2\dot\tau ]
\pa_A\left( -\f1{32} C_{BC}C^{BC}\right) +\f18 \pa_A (C^{BC} D_B \pa_C\tau)\cr
&  -\f1{32}\p_u\left(  C_{BC}C^{BC}\right) \pa_A \tau  -\f1{16} C_{BC}C^{BC} \pa_A\dot\tau\,.
\ee
This means that the last three terms in the variation \eqref{dP} of $P_A$ can therefore be cancelled by the modification
\be\la{cP}
\boxed{\,\,{\cal P}_A :=P_A +\f14   (\bD_C C^{CB})C_{BA}
+\f1{16} \p_A (C_{BC}C^{BC})\,,\,\,}
\ee
which defines the \emph{covariant momentum} and  yields the Weyl scalar $\Psi_1$.
To compute explicitly the variation of this covariant momentum one uses that
\be
C_{AB}C^{BC} = \frac12 (C_{BD}C^{BD}) \delta_A^C,\qquad
C_{AB}N^{BC} = \frac12 (C_{BD}N^{BD}) \delta_A^C + \frac12  (C_{BD}\tilde{N}^{BD})\epsilon_{A}{}^C\,.
\ee
This means that the variation of the covariant momentum  drastically simplifies into 
\be
 \d_{(\tau,Y)}{\cal P}_A &\,\doublehat{=}\, [\tau \pa_u +\cL_Y + 2\dot\tau ] {\cal P}_A  +3 \cM_{AB} \pa^B\tau\,,
 \la{dcP1}
 \ee
 where the tensor $\cM_{AB}$ is given by 
 \be
 \cM_{AB} &:=  \cM q_{AB} + \tcM \epsilon_{AB}\,,
 \la{JABanti}
\ee
and  the double  hatted equality refers to the fact that 
we have used both asymptotic equations $\textsf E_{U}^A\,\doublehat{=}\,0\,\doublehat{=}\,\textsf E_b$.

The momentum transformation involves the mass and dual mass on a symmetric level.
It  is a clear  improvement from the cumbersome transformation \eqref{dP} and it exhibits, as anticipated, a  self-dual symmetry of the transformation rules.
This expression also provides a powerful and nontrivial consistency check of \eqref{dP}.
Indeed, since ${\cal P}_A$ does not contain quadratic anomaly, its variation should also be 
expressed only in terms of semi-covariant tensors that do not contain quadratic anomalies.
This is indeed the case since both  $\cal M$ and $ \tcM$ are semi-covariant.
To summarize,  the transformation property \eqref{dcP1} of the covariant momentum
is self-dual and given by 
\be\la{dcP}
\d_{(\t,Y)} {\cal P}_A\,\doublehat{=}\,  [\tau\pa_u + \cL_{Y} + 2 \dot\tau  ] {\cal P}_A 
 + 3 \left({\cal M}\pa_A\tau  +   \tcM \tilde\pa_A\tau\right) \,.
 \ee

%
%

\subsection{Covariant stress }
We finally focus on the construction of the covariant spin-$2$ observable.
It is easy to see that the first two terms in the last line of \eqref{dE} are cancelled in the following combination
\be\boxed{\,\,
{\cal T}_{AB}:={3}\left( E_{AB}-\f1{16}  C_{AB} C_{CD} C^{CD}\right)\,,
\,\,}
\ee
which defines the {\it covariant stress} and   yields the Weyl scalar $\Psi_0$.
Using the definition of the covariant momentum, we can write its  transformation as (see Appendix \ref{App:Eano})
\be
\d_{(\tau,Y)}\cT_{AB}&\, \hat{=}\, [\tau \pa_u  + \cL_Y  +\dot{\tau}]  \cT_{AB} +4 \cP_{\langle A}  \p_{B\rangle } \tau  \cr
&+\f3{2}\left( 
C^{CD}\bD_C C_{AB}
- C_{AB} {\bD}_CC^{CD}
\right)\p_D \tau
-3 \bD_{\langle A} C_{B \rangle  C}  C^{CD}\p_D \tau
+ \f34 \pa_{\langle A}(C_{CD}C^{CD}) \pa_{B\rangle}\t\cr
&-4\textsf E_\B  \bD_{\langle A} \p_{B\rangle} \tau\,.
 \ee
We can simplify this expression considerably using the identity
\be
\f1{2}\left( 
C_D{}^{C}\bD_C C_{AB}
- C_{AB} {\bD}_CC^{C}{}_D
\right)
&= \bD_{\langle A} C_{B \rangle  C}  C^{C}{}_D- \f14\p_{\langle A}(C_{CE}C^{CE}) q_{B \rangle D}\,,
\ee
that can be proven using complex coordinates. 
Finally, this means  that we simply have 
\be \la{dE2}
\d_{(\tau,Y)} \cT_{AB} &\,\doublehat{=}\, \left[\tau\pa_u + \cL_{Y} +  \dot\tau  \right] \cT_{AB}  +
 4 \cP_{\langle A}  \p_{B \rangle} \tau.
\ee
Again a drastic simplification from the original transformation \eqref{dE}.

\subsection{Covariant tensors and Weyl scalars}\la{sec:Weyl-scalars}

In order to elucidate the relation between the covariant observables introduced above and the Weyl scalars in the Newman--Penrose formalism \cite{NP62, Newman:1962cia}  at null infinity, let us introduce a  doubly-null
tetrad $(\ell, t, m, \bar m)$ adapted to the 2 + 2 foliation defined by {two null vectors  $\ell, t$  transverse to the sphere and a complex dyad $m, \bar m$ tangent to the sphere, with $q^{AB}=2m^{(A}\bar m^{B)}$. Explicitly, in the Bondi coordinates $(u,r)$ on $\scri$ these vectors are given by} 
\be
\ell =\p_r\,,\quad t=\p_u\,\quad m=m^A\p_A\,.
\ee
By contracting the Weyl tensor $W_{\mu \nu \rho \sigma}$ with the tetrad field above, we obtain the 5 Weyl scalars $(\Psi_4,\Psi_3,\Psi_2,\Psi_1,\Psi_0)$.  The  asymptotic values of the Weyl scalars, which are determined by the peeling theorem, are respectively given by $(\tfrac12  \dot N^{AB}, \cJ^A, \cM-i\tcM, \cP_A, \cT_{AB})$  (see Appendix D of \cite{Freidel:2021yqe})
\begin{subequations}
\be
\Psi_4  &:=-W_{t  \bar m t \bar m}   =\frac{1}{2r}\,  \dot N^{AB} \bar m_A \bar m_B + o(r^{-1})\,,\\
\Psi_3  &:=-W_{  t \ell  t\bar m} =  \frac{1}{r^2}\,  \cJ^{A} \bar m_A+ o(r^{-2})\,,  \\
\Psi_2  &:=-\f 12 \left(W_{\ell t  \ell t}+ W_{\ell t m \bar m}\right)  =\frac{1}{r^3}\,(\cM+i \tilde{\cM})+ o(r^{-3}),\\
\Psi_1  &:=-W_{\ell t \ell m} =\frac{1}{r^4}\, \cP_{A} m^A  + o(r^{-4})\,,\\
\Psi_0  &:=-W_{\ell m \ell m} =  \frac{1}{r^5}\, \cT_{AB}m^A m^B+ o(r^{-5})\,.
\ee
\end{subequations}
This means that the covariant observables are simply, and up to normalisation, the asymptotic Weyl scalars.

\section{EOM from symmetry}\la{sec:EOM}

In the previous section we have constructed the tensor $\dot{N}^{AB}$ and the pseudo-tensors \\ $(\cJ^A,\cM,\tcM,\cP_A,\cT_{AB})$. 
By design these are covariant under the kinematical  part of the $\bmsw$ algebra and they represent the metric data up to order $1/r$ in the metric expansion.
We now want to explore their covariance properties under supertranslations and derive from it the asymptotic evolution equations, that appear as restrictions on the free data.

The strategy that defines our symmetry argument  is as follows. For a given scale weight $s$ and a given spin, we first identify the combinations of free data that transform with that weight and are covariant under the action of the BMSW infinitesimal transformations. The associated asymptotic EEs are then obtained by setting those combinations to zero. 
To be more precise the first step of the argument requires identifying quantities denoted $\cE$ which are now tensorial under the full symmetry group including supertranslaions. One can then argue that in the absence of sources, that is for pure gravity, the only possible consistent equation is $\cE=0$.  
The reason we put the RHS of the equation equal to zero instead of $1$ say is that $\cE$ can be understood as transforming in the coadjoint representation of the BMSW group. It would be inconsistent to fixed $\cE=1$ as the transformed value $g\cE g^{-1}$, for $g$ an element of the BMSW group, would now be different from 1. The only option is to have an equality of the form $\cE= O$ where $O$ is an object that transforms under the coadjoint representation of BMSW. In the absence of matter no such object exists and the only admissible coadjoint orbit that can source the equation is $O=0$. Later in the section we identify, in the presence of matter, which combination of the energy-momentum tensor transforms in the same orbit as $\cE$.  This  leads to a proposal for the asymptotic equations of motion in the presence of matter.

Let us emphasize that the derivation of Einstein's equation from symmetry is only valid for the  asymptotic equations of motion that arises in a $1/r$ expansion of the metric. The rest of the equations that allow to reconstruct the bulk metric,  through the radial evolution,  are not derived in that way but they  are assumed to hold. This is  consistent with the holographic perspective where the boundary is assumed to have a unique bulk reconstruction.
Let us also emphasize that going from the identification of a tensor $\cE$ to the imposition $\cE=0$ as an equation of motion is not new. It is the same strategy that Einstein used to derive the  equation $G_{\mu\nu}=0$ from a {\it gauge} symmetry argument \cite{Einstein:1916vd}.

Before focusing on the derivation of the set of Einstein's equations,  we need to distinguish the equations that are derived from a symmetry argument form the ones that are imposed as boundary conditions. 
The only  equations that we impose as boundary conditions are listed in \eqref{Bondias}, which  in particular contain the equation \eqref{asym-EE}.
It is possible to relax these boundary conditions and perform a more full fledge analysis where the equation \eqref{asym-EE} is also derived from symmetry, but we do not do this here.

We now look systematically at the metric components that transform homogeneously under the full BMSW group. 
One starts by the scalar data of weight $s=1$. At this weight there is only one datum  of weight $s=1$ that  transforms homogeneously under BMSW transformations. It is the  scalar given by  $b_1$ which labels the  term of order $b_1/r$ in the expansion of $\beta$, which transforms as $\d_{(\tau,Y)} b_1=[\tau \pa_u +Y^A \pa_A + \dot{\tau} ] b_1$. Since this is the only datum of $s=1$ that transforms homogeneously, the covariant relation $b_1=0$  falls in the set of covariant relations to impose in the absence of external sources and it  is  thus included in the set of asymptotic EEs derived using exclusively BMSW transformation properties.

%

For the next steps we analyze the two scalar invariants of weight $s=2$ and the spin one invariant of weight $s=3$,
representing evolution equations for covariant observables. This allows us to apply our symmetry argument 
for an immediate derivation of the  Einstein's equations \eqref{EEF}, \eqref{EEU}, \eqref{EEB}. 
More precisely,
from the transformations of the 2d Ricci scalar  under the metric
rescaling $q_{AB} \to e^{-2\dot \t} q_{AB}$ and   \eqref{dF} it is immediate to see that the combination 
\be
\textsf E_F:=F- \frac{R({q})}4
\ee
transforms homogeneously as a scalar of weight $s=2$, namely
\be
\d_{(\tau,Y)}  \textsf E_F= \left[\cL_Y + 2 \dot{\tau} \right] \textsf E_F\,.
\ee
Hence the EE \eqref{EEF} is recovered as the covariant expression  $\textsf E_F = 0$.
From the transformations \eqref{dDC}, \eqref{dUA}, it is straightforward to show that the combination 
\be
\textsf E_{U}^A&:=  {U}^A+ \frac12 {D}_B \bC^{AB}
\ee
transforms homogeneously as a vector of weight $s=3$, namely
\be
\d_{(\tau,Y)}  \textsf E_{U}^A =   [\tau\pa_u + \cL_Y + 3\dot\tau ] \textsf E_{U}^A\,.
\ee
This shows that the EE  \eqref{EEU} is also recovered as the covariant expression  $\textsf E_{U}^A = 0$.


Similarly to the case for $\textsf E_{U}^A$, from the transformations \eqref{dC},  \eqref{dB} we see right away that the scalar combination
\be
\textsf E_{\B}&:=  \B  + \frac{1}{32 }  \bC_{AB}\bC^{AB}
\ee
transforms as 
\be
\d_{(\tau,Y)} \textsf E_{\B}=
\left[\tau \pa_u +\cL_Y + 2 \dot{\tau} \right] \textsf E_{\B}
+\f14  \textsf E_{U}^A \p_A \t\,.
\ee
Therefore, on-shell of the previously just derived EE  $\textsf E_{U}^A\, \hat {=} \,0$, we recover  \eqref{EEB} as well from the requirement of covariance  under the BMSW group action.

Now that we have shown that our symmetry argument can be applied to derive both Einstein's equations \eqref{EEU}, \eqref{EEB}, in the following we will at times go on-shell of these two equations in order to simplify some of the expressions. We recall that imposition of  \eqref{EEU} alone is denoted by a single hat while impositions of  \eqref{EEB} as well by a double hat. In particular, we point out already that, when including matter sources in our analysis, we will refrain from imposing  \eqref{EEB} as the combination $ \B  + \frac{1}{32 }  \bC_{AB}\bC^{AB}$ picks up a stress-energy tensor contribution in the Einstein's equations and this needs to be taken properly into account when studying covariance properties of matter terms as well. This means that some of the transformations derived in Sections \ref{sec:M}, \ref{sec:P}, \ref{sec:T} will need to be generalized to include terms proportional to $\textsf E_b$; this is done in Appendix \ref{App:action}.

\subsection{Mass evolution from symmetry}\la{sec:M}

The goal of this section is to show that, quite remarkably, the symmetry transformation of $\cM$ completely determines its equation of motion. To see this, one evaluates the transformation of the covariant mass time derivative and the current divergence
\be
\d_{(\tau,Y)} \dot{\cal M} &\,\hat{=}\, 
\left[\tau \pa_u +\cL_Y + 4 \dot{\tau} \right]\dot{\cal M}
+ \pa_u(\cJ^A \pa_A \tau)\,,\cr
\delta_{(\tau,Y)} \bD_A \cJ^A &=  
\left[\tau \pa_u +\cL_Y + 4 \dot{\tau} \right] \bD_A \cJ^A +2 \pa_u (\cJ^A \bD_A \tau)  +  \f12 \dot{N}^{AB} \bD_A\pa_B\tau\,,
\ee
where we used that $\f12  \bD_A\dot{N}^{AB} = \dot \cJ^B $.
This means that the quantity
\be\label{covariantE}\boxed{\,\,
\cE:= \dot \cM-\f12  \bD_A \cJ^A  - \f18 \dot{N}^{AB} C_{AB}\,\,}
\ee
transforms homogeneously under the symmetry transformation.
Therefore, the covariant conservation equation is $\cE=0$ or 
\be
 \dot M -\f12   \bD_A \cJ^A  = - \f18 N_{AB} N^{AB}\,,\la{EE}
\ee
when written in terms of the original variables. 

This is one of the Einstein's equation which is derived purely from symmetry principle.
If one uses a fluid analogy where $\cM$ plays the role of the energy density, we see that 
 $- \cJ^A$ is the energy transport current while 
$\f18 N_{AB}N^{AB}$ plays the role of  the entropy production.
The expression \eqref{covariantE} in terms of the covariant quantities 
shows that when no radiation is present, which corresponds to the condition $\dot{N}_{AB}=0$,
then the covariant mass is conserved
\be
\dot\cM-\f12\bD_A \cJ^A=0.
\ee
The same can be followed for the dual mass, using that 
\be
\delta_{(\tau,Y)}  \dot{\tcM} &=
[\tau \pa_u+ {\cL}_Y +4\dot{\tau}] {\tcM} 
+ \pa_u( \tilde{\cJ}^A \pa_A {\tau}),\cr
\delta_{(\tau,Y)} \bD_A \tilde\cJ^A &=  
\left[\tau \pa_u +\cL_Y + 4 \dot{\tau} \right] \bD_A \tilde\cJ^A +2 \pa_u (\tilde\cJ^A \bD_A \tau)  +  \f12 \dot{\tilde{N}}^{AB} \bD_A\pa_B\tau\,,
\ee
from which we deduce  that the quantity
\be\label{covariantdE}
\tilde\cE:= \dot \tcM-\f12  \bD_A \tcJ^A  - \f18 \dot{\tilde{N}}^{AB} C_{AB}
\ee
transforms homogeneously under the symmetry transformation.
Therefore, the covariant conservation equation for the dual mass is $\tilde\cE=0$.

\subsection{Momentum evolution from symmetry}\la{sec:P}

We can establish in a similar manner that  the asymptotic evolution equation for the  momentum can be written as ${\cal E}_A=0$, with (see also details  in Appendix \ref{App:Peom})
\be\label{Peom}
{\cal E}_A:=  
\dot{\cal P}_A - \pa_A {\cal M} -\tilde \pa_A \tilde {\cal M} - C_{AB} \cJ^B\,,
\ee
a vector that transforms homogeneously under the BMSW group.
 The goal is show that the anomaly of ${\cal E}_A$ vanishes when $\cE$ and $\tilde \cE$ vanish, namely
\be
\Delta_\tau {\cal E}_A\,\doublehat{=}\, 0\,.
\ee
We rely on the duality  between the covariant mass and dual mass to simplify the proof.

Let us first focus on the quadratic anomalies, which we denote $\Delta^{(2)}_\tau$. From the transformation properties derived in Appendix \ref{sec:Variations}, we can write
\be
\Delta^{(2)}_\tau \p_A {\cal M} &\,\hat{=}\,  \cJ^B \bD_A \pa_B \tau= \cJ^B \bD_{\langle A} \p_{B\rangle} \tau + \f12 \cJ_A \Delta \tau\,, \cr
\Delta^{(2)}_\tau \tilde \p_A \tilde{\cal M}
&=
 \tcJ^B\tilde \bD_A \pa_B \tau=- \cJ^B\tilde \bD_{\langle A} \tilde\pa_{B\rangle} \tau
 +\f12 \tcJ^B \epsilon_{AB} \Delta \tau\cr
 &= \cJ^B \bD_{\langle A} \p_{B\rangle} \tau - \f12 \cJ_A \Delta \tau\,,\cr
\Delta^{(2)}_\tau  C_{AB} \cJ^B &= - 2\cJ^B  \bD_{\langle A} \p_{B\rangle} \tau\,,
\ee
from which it is straightforward to see that
\be
\Delta^{(2)}_\tau {\cal E}_A\,\doublehat{=}\, 0\,,
\ee
since $\Delta^{(2)}_\tau \dot{\cP}_A\,\doublehat{=}\, 0$ as immediate from \eqref{dcP}.

Let us now look at the linear anomaly $\Delta^{(1)}_\tau$. Using again the results of Appendix \ref{sec:Variations}, we can write
\be
\Delta^{(1)}_\tau \dot{\cal P}_A  &\,\doublehat{=}\,
3 \dot {\cal M} \pa_A\tau 
+3 {\cal M} \pa_A\dot \tau  
+  3 \dot\tcM \tilde\pa_A\tau 
+  3 \tcM \tilde\pa_A\dot\tau\,,\\
\Delta^{(1)}_\tau \p_A {\cal M} &\,\hat{=}\, 
\p_A\tau  \dot {\cal M}+3\p_A\dot \tau {\cal M}+ D_A \cJ^B\pa_B\t,\\
\Delta^{(1)}_\tau \tilde \p_A \tilde{\cal M} 
&=
\tilde\p_A\tau  \dot {\tcM}+3\tilde\p_A\dot \tau {\tcM}+\tilde{D}_A \tcJ^B \pa_B\t \cr
\Delta^{(1)}_\tau   C_{AB} \cJ^B &=
  \f12 C_{AB}  \dot{N}^{BC} \pa_C\tau
= \f14 (C_{BC}  \dot{N}^{BC}) \pa_A\tau+ \f14 (C_{BC}  \dot{\tilde{N}}^{BC}) \tilde{\pa}_A\tau\,.
\ee
Moreover,  given the identity 
\be
{D}_A \cJ_B +\tilde{D}_A \tcJ_B =  q_{AB} (D_C\cJ^C)+ \epsilon_{AB} (\tilde{D}_C \tcJ^C)\,,
\ee
we  have
\be
\Delta^{(1)}_\tau {\cal E}_A&\,\doublehat{=}\, \left[ 2 \dot {\cal M}-(D_C\cJ^C) -\f14 (C_{BC}  \dot{N}^{BC})\right]  \pa_A\tau+ \left[ 2 \dot {\tcM}-(D_C\tcJ^C)-\f14 (C_{BC}  \dot{\tilde{N}}^{BC}) \right]\tilde{\pa}_A\t   \cr
&= 2\cE \pa_A \t + 2\tilde\cE \tilde{\pa}_A\t = 0\,,
\ee
on-shell of $\cE=0=\tilde \cE$.
We thus see that  also the equation of motion for the momentum, $\cE_A=0$, can be derived from purely symmetry principles.

\subsection{Stress tensor evolution from symmetry}\la{sec:T}

Let us show that the same strategy applies to the spin-2 equation of motion as well. 
For the spin-2  sphere metric component we have seen in \eqref{dE2} that the combination
$
{\cal T}_{AB}=3\left( E_{AB}-\f1{16}  C_{AB} C_{CD} C^{CD} \right)
$
has no quadratic anomaly.
 The goal is two show that the combination 
\be
\boxed{\cE_{AB}:= 
\dot \cT_{AB}- \bD_{\langle A} \cP_{B\rangle}-\f32\left( C_{AB}\cM+\tilde C_{AB} \tcM\right)
}
\la{Vireom}
\ee
is also free of anomaly on-shell of the momentum evolution equation $\cE_A=0$ and therefore it determines the tensorial equation of motion.
%
Since the  quadratic anomaly of  $\dot\cT_{AB}$ vanishes, 
one only needs to evaluate the   following quadratic anomalies 
\be\la{D2E}
\Delta^{(2)}_\t\left ( 
 \bD_{\langle A} \cP_{B\rangle}+\f32\left( C_{AB}\cM+\tilde C_{AB} \tcM\right)
\right)&\,\doublehat{=}\,3\Big( \tcM \bD_{\langle A}\tilde \pa_{B\rangle}\tau +  {\cal M}  \bD_{\langle A}\pa_{B\rangle}\tau\cr
&-\cM {\bD_{\langle A} \p_{B\rangle} \tau } -\tcM {\tilde \bD_{\langle A} \p_{B\rangle} \tau }\Big)\cr
&=0\,,
\ee
where we used the result in Appendix \ref{sec:Variations} for the  anomaly of the quantity $\bD_A \cP_B$, as well as \eqref{dC}.

We are thus left to show that the linear anomaly vanishes too. For the same quantities as in \eqref{D2E},  we have
\be
\Delta^{(1)}_\t\left ( 
 \bD_{\langle A} \cP_{B\rangle}\right)&\,\doublehat{=}\,
4\cP_{\langle A}\p_{B\rangle} \dot \tau
+  \dot \cP_{\langle A}\p_{B\rangle}\t
+  3\bD_{\langle A}  \tcM \tilde  \pa_{B\rangle}\tau +3\bD_{\langle A}  {\cal M} \pa_{B\rangle}\tau
\,,\\
\Delta^{(1)}_\t\left ( \f32
C_{AB}\cM
\right)&\,\hat{=}\,  \f32  C_{AB}  \cJ^C \p_C\t
\,,\\
\Delta^{(1)}_\t\left ( \f32
\tilde C_{AB} \tcM
\right)
&= \f32  \tC_{AB}  \tcJ^C \p_C\t\,.
\ee
Therefore, combining these together we have
\be
& \Delta^{(1)}_\t\left ( 
 \bD_{\langle A} \cP_{B\rangle}+\f32\left( C_{AB}\cM+\tilde C_{AB} \tcM\right)
\right)\cr
&\,\doublehat{=}\,4\cP_{\langle A}\p_{B\rangle} \dot \tau
+  \dot \cP_{\langle A}  \p_{B\rangle}\t  
+   3\bD_{\langle A}  \tcM \tilde  \pa_{B\rangle}\tau +3\bD_{\langle A}  {\cal M} \pa_{B\rangle}\tau
+\f32C_{AB}\cJ^C \p_C\t   + \f32 \tC_{AB}\tcJ^C \p_C\t\cr
&=4\p_u(\cP_{\langle A}\p_{B\rangle}  \tau) 
-3 \left(\dot \cP_{\langle A} -  \tilde\bD_{\langle A}  \tcM
+ \bD_{\langle A}  {\cal M}  \right) \p_{B\rangle}\t 
+ \f32C_{AB}\cJ^C \p_C\t   + \f32 \tC_{AB}\tcJ^C \p_C\t \cr
&= 4\p_u(\cP_{\langle A}\p_{B\rangle}  \tau) -3 \cE_{\langle A}  \p_{B\rangle}\t.
 \ee
This was simplified by using the identity 
\be
 C_{AB} \p_C\t  - \tC_{AB} \tilde\p_C\t=  
 C_{ C\langle A } \p_{B\rangle} \t  
 - 
  \tC_{C\langle A} \tilde\p_{B\rangle}  \tau = 2 C_{ C\langle A } \p_{B\rangle} \t  . 
\ee
We also used the definition of the momentum equation of motion \eqref{Peom}.
From this, taking the time derivative of \eqref{dE2}, we conclude that $\Delta^{(1)}_\tau \cE_{AB}\,\doublehat{=}\,0$ as well once we use $\cE_A=0$.

\subsection{Matter sources}

The previous results show that the multiplet $(\cE,\tilde\cE,\cE_A,\cE_{AB})$ 
of  evolution equations transforms homogeneously under the BMSW symmetry group. More precisely, in the abscence of matter sources, we have
\be\la{dcE}
\delta_{(\tau,Y)}\cE &\,\hat{=}\,   [\tau \pa_u+ {\cL}_Y +4\dot{\tau}] \cE\,,\\
\delta_{(\tau,Y)}\tilde\cE &=  [\tau \pa_u+ {\cL}_Y +4\dot{\tau}] \tilde\cE\,,\\
\delta_{(\tau,Y)}\cE_A &\,\doublehat{=}\,  [\tau \pa_u+ {\cL}_Y +3\dot{\tau}] \cE_A +
2\cE \pa_A \t + 2\tilde\cE \tilde{\pa}_A\t\,,\\
\delta_{(\tau,Y)}\cE_{AB} &\,\doublehat{=}\, [\tau \pa_u+ {\cL}_Y +2\dot{\tau}] \cE_{AB} +{3}\cE_{\langle A}  \p_{B\rangle}\t\,.
\ee
We  have seen as well that the asymptotic Einstein's equations constraining the metric functions $F, U, b$ transform as
\be
\d_{(\tau,Y)}  \textsf E_F&= \left[\cL_Y + 2 \dot{\tau} \right] \textsf E_F\,,\cr
 \delta_{(\tau,Y)} \textsf E_{U}^A &= [\tau \pa_u+ {\cL}_Y +3\dot{\tau}] \textsf E_{U}^A,\,\cr
  \delta_{(\tau,Y)} \textsf E_{\B}&=[\tau \pa_u+ {\cL}_Y +2\dot{\tau}] \textsf E_{\B} + \f14  \textsf E_{U}^A \p_A \t\,.
\ee

This is the first main result of this paper, namely the derivation of asymptotic Einstein's equations at null infinity from the unique demand of constructing tensorial operators starting from the time derivative of the  pseudo-tensors $(\cM,\tcM,\cP_A,\cT_{AB})$ in the abscence of matter sources.

 We now want to extend our symmetry argument also to the case where matter sources are present and use it to derive the asymptotic EEs coupled to matter.
Let us start by showing that a transformation structure similar to \eqref{dcE} for the asymptotic vacuum Einstein's equations is reproduced by the conservation equations of matter. 
We consider matter sources with stress-energy tensor (SET) $T_{\m\n}$.
We restrict our analysis to stress-energy tensor components that preserve the following asymptotic Einstein's equations
\be
\textsf E_F\, \hat{=}\,0\,\hat{=}\, \textsf E_{U}^A.
\ee
These conditions mean (see \cite{Freidel:2021yqe}, Section 8, for more general conditions) that the stress-energy tensor components have the  following expansions \cite{Flanagan:2015pxa}
  \begin{subequations}
 \be
 T_{AB}&=\f1{r} \hat T q_{AB}+\f1{r^2}\hat T_{ AB } + o(r^{-2}) \,,\\
 T_{uu}&=\f1{r^2} \hat T_{uu}+o(r^{-2})\,,\qquad 
   T_{uA}=\f1{r^2} \hat T_{uA}+o(r^{-2})\,,\\
    T_{rA}&=\f1{r^3} \hat T_{rA}+o(r^{-3})\,,\qquad
 T_{ru}= 
 o(r^{-3})\,,\\
 T_{rr}&=\f1{r^4} \hat T_{rr}
 +o(r^{-3})\,.
 \ee
   \end{subequations}
   The presence of the matter sources affects the equations of motion for $b$ and therefore it  changes the transformation property of the gravity evolution equations. We now have 
   \be\la{dcE}
\delta_{(\tau,Y)}\cE &\,\hat{=}\,   [\tau \pa_u+ {\cL}_Y +4\dot{\tau}] \cE\,,\\
\delta_{(\tau,Y)}\tilde\cE &=  [\tau \pa_u+ {\cL}_Y +4\dot{\tau}] \tilde\cE\,,\\
\delta_{(\tau,Y)}\cE_A &\,\hat{=}\,  [\tau \pa_u+ {\cL}_Y +3\dot{\tau}] \cE_A +
2\cE \pa_A \t + 2\tilde\cE \tilde{\pa}_A\t\, -2\pa_u (\dot {\textsf E}_\B\p_A\t) ,\\
\delta_{(\tau,Y)}\cE_{AB} &\,\hat{=}\, [\tau \pa_u+ {\cL}_Y +2\dot{\tau}] \cE_{AB} +{3}\cE_{\langle A}  \p_{B\rangle}\t\, \cr
& - {2}\dot {\textsf E}_\B \bD_{\langle A} \p_{B\rangle} \tau
 - {4} \textsf E_\B  \bD_{\langle A} \p_{B\rangle}\dot \tau
 +{2} \p_{\langle A} \dot {\textsf E}_\B  \p_{B\rangle} \tau.
\ee
In the presence of matter we no longer have $\textsf E_\B=0$, so the covariant combinations needs to be corrected by source terms.

In order to derive these source terms we nee to understand the  symmetry transformations of the stress tensor. 
   As shown in Appendix \ref{App:SETano}, the asymptotic components of the stress-energy tensor  transform as 
\begin{subequations}\la{SETtrans}
\be
\delta_{(\tau,Y)}\hat T_{uu} &=  [\tau \pa_u+ {\cL}_Y +4\dot{\tau}] \hat T_{uu}\,,\\
\delta_{(\tau,Y)}\hat T &=  [\tau \pa_u+ {\cL}_Y +3\dot{\tau}] \hat T\,,\\
\delta_{(\tau,Y)}\hat T_{rr} &=  [\tau \pa_u+ {\cL}_Y +2\dot{\tau}]\hat T_{rr}  \,,\\
\delta_{(\tau,Y)}\hat T_{uA} &=  [\tau \pa_u+ {\cL}_Y +3\dot{\tau}] \hat T_{uA} - \hat T  \p_A \dot \t  +\hat T_{uu} \pa_A \tau\,,\\
\delta_{(\tau,Y)}\hat T_{rA} &=  [\tau \pa_u+ {\cL}_Y +2\dot{\tau}]\hat T_{rA} -\hat T_{rr} \p_A \dot\t +\hat T \p_A \t\,,\\
\delta_{(\tau,Y)}\hat T_{\langle AB \rangle } &=  [\tau \pa_u+ {\cL}_Y +2 \dot{\tau}] \hat T_{\langle AB \rangle} \cr
& 
+ 2 \hat T_{u\langle A } \p_{B\rangle} \t 
  -2 \hat T_{r\langle A } \p_{B\rangle}\dot \t 
-  2 \hat T  D_{\langle A} \p_{B\rangle} \t\,,
\ee
\end{subequations}
where  the $r^{-2}$ component of $T_{AB}$ is given by $\hat T_{AB}:= \hat T_2 q_{AB}+\hat T_{\langle AB \rangle} $.
From these transformations we can identify the conservation equations as the combinations that 
transform as pseudo-tensors.\footnote{Strictly speaking its is the combination $\cC^1_A + \dot{\cC}^2_A$ which transforms as a pseudo tensor.}
These are  (see Appendix \ref{App:ConsEq} for their derivation)
\be\la{ConsEq}
\cC:= \pa_u \hat T_{r r}+2\hat T, \qquad
\cC^1_A:= \p_u \hat T_{rA}-\p_A\hat T,
\qquad
 \cC^2_A:=  \pa_A \hat{T}_{rr}+ 2\hat{T}_{rA}\,,
\ee
which transform as   
\be
\delta_{(\tau,Y)} \cC &= [\tau \pa_u+ {\cL}_Y +3\dot{\tau}]\cC,\cr
\delta_{(\tau,Y)} \cC^1_A&= [\tau \pa_u+ {\cL}_Y +3\dot{\tau}]\cC^1_A 
- \cC \pa_A\dot \tau \,,\cr
\delta_{(\tau,Y)} \cC^2_A&= [\tau \pa_u+ {\cL}_Y +3\dot{\tau}]\cC^2_A 
+ \cC \pa_A\tau.
\ee 
The proof follows from
\be
\delta_{(\tau,Y)} \pa_u\hat T_{rA} &=  [\tau \pa_u+ {\cL}_Y +3\dot{\tau}]\pa_u \hat T_{rA} - (\pa_u \hat T_{rr}- \hat T)  \p_A \dot\t+ \pa_u\hat T \p_A \t\,,\cr
\delta_{(\tau,Y)}\pa_A \hat T &=  [\tau \pa_u+ {\cL}_Y +3\dot{\tau}] \pa_A \hat T
+ 3\hat T \p_A  \dot{\tau} 
+ \pa_u \hat T \pa_A \tau\,,\cr
\delta_{(\tau,Y)}\pa_A \hat{T}_{rr} &= [\tau \pa_u+ {\cL}_Y +2\dot{\tau}]\pa_A \hat{T}_{rr} + \dot{\hat{T}}_{rr} \pa_A \tau +2 \hat{T}_{rr} \p_A \dot{\tau}\,.\label{Trr}
\ee 

Applying our symmetry argument, one then recovers the result that the conservation equations are given by 
$\cC=0, \cC^1_{A}=0, \cC^2_A=0$. 
Using that the leading order of the Einstein equation component $G_{rr}=T_{rr}$ (see, e.g., \cite{Freidel:2021yqe}) implies 
\be\la{TrrEb}
T_{rr} = -8{\textsf E}_\B\,,
\ee
 we see that the  conservation equations mean 
   \be\la{TEb}
\hat{T}=  4 \dot{\textsf E}_\B,
 \qquad \hat{T}_{rA}  = 4 \pa_A {\textsf E}_\B\,.
\qquad 
\ee

The asymptotic EEs  coupled to matter read\footnote{Given the topological nature of the dual mass charge, the corresponding EE does not acquire a source term. It is possible to have a matter source term for the dual mass if we relax the boundary condition and allow for $T_{uA} $ to contain a term proportional to $1/r$ \cite{Bieri:2020zki}.}
\be
\tilde\cE=0\,,\quad\cE+\cS=0\,,\quad \cE_A+\cS_A=0\,,\quad \cE_{AB}+\cS_{AB}=0\,,
\ee
 where the sources are
given by  
\be\la{sources}
\cS:= \f12 \hat T_{uu} ,
\qquad 
\cS_A:= \hat T_{uA} +\f12 \pa_A \hat T,\qquad
{\cS_{AB}:=\f32 \hat T_{\langle AB\rangle} -\f1{4} D_{\langle A} \pa_{B\rangle} \hat T_{rr} -\f32 \hat T C_{AB}}\,.
\ee
The source $\cS_A$ found by the symmetry arguments here agrees with the source found in the direct evaluation of the 
Einstein tensor components near null infinity  written in \cite{Freidel:2021yqe}.\footnote{The expansion  of the $\langle A B\rangle$-component of the Einstein tensor containing $\dot \cE_{AB}$ was not completed in \cite{Freidel:2021yqe}, thus we cannot compare the expression of the $\cS_{AB}$ source with that reference.}
Unfortunately, although the general form agrees, some of the detail numerical coefficients we found for the source terms $\cS_A$ and $\cS_{AB}$ disagree with the ones written in  \cite{Nichols:2018qac}.

This form of the  sources is derived in Appendix \ref{App:sources}, again applying our symmetry argument. More precisely, the expressions \eqref{sources} are obtained  by demanding that  EEs coupled to matter   transform   homogeneously under the BMSW symmetry group as
\be\la{dcES}
\delta_{(\tau,Y)}(\cE+\cS) &\,\hat{=}\,   [\tau \pa_u+ {\cL}_Y +4\dot{\tau}] (\cE+\cS)\,,\\
\delta_{(\tau,Y)}(\cE_A+\cS_A) &\,\hat{=}\,  [\tau \pa_u+ {\cL}_Y +3\dot{\tau}] (\cE_A+\cS_A)  +
2(\cE+\cS) \pa_A \t + 2\tilde\cE \tilde{\pa}_A\t\,,\\
\delta_{(\tau,Y)}(\cE_{AB}+\cS_{AB}) &\,\hat{=}\, [\tau \pa_u+ {\cL}_Y +2\dot{\tau}] (\cE_{AB}+\cS_{AB})  +{3}(\cE_{\langle A} +\cS_{\langle A} ) \p_{B\rangle}\t\,.
\ee
Notice that we have removed the double hat symbol over the equal signs as the $\textsf E_b$  asymptotic equation is sourced by the SET component $T_{rr}$ \eqref{TrrEb} and it thus contributes to the definition of the covariant tensors encoding the EEs in the presence of matter.

\section{ Properties of the covariant observables}\la{sec:CO}

Let us now summarize the covariance properties that we have revealed so far and the nested structure that organizes them.
The covariant observables are  the radiative 
observable 
\be
\cN^{AB}:= \dot N^{AB}\,,
\ee
{corresponding to the truly free data at $\scri^+$  encoding the two polarizations of the outgoing gravitational radiation}, 
and
the corner 
observables $(\cJ^A,\cM,\tcM,\cP_A,\cT_{AB})$ defined as 
\begin{subequations}\la{c}
\be
\cJ^A &:= \f12 \bD_B N^{AB} + \f14 \pa^A R(q)\,,\\
{\cal M}&:= M + \f18 N^{AB} C_{AB}\,, \la{Mdot}\\
\tcM&:= \f14  {D}_{B}(\bD\!\cdot\! {\tilde C})^{B}+  \f18   \tilde{N}^{AB}C_{AB}\,, \la{tMdot}\\
{\cal P}_A &:= P_A  
+\f1{16} \bD_A (C^{BC}C_{BC}) +  \f14   C_{AB} (\bD\!\cdot\! {C})^{B}\,,\la{Pdot}\\
{\cal T}_{AB}&:= 3\left(E_{AB}-\f1{16}  C_{AB} (C_{CD} C^{CD})\right)\,, \la{Tdot}
\ee
\end{subequations}
{and corresponding to initial data}.
They transform as follows  
\begin{subequations}\la{dc}
\be
\d_{(\tau,Y)} {\cN}^{AB}
&=\left[ \tau \pa_u  + \cL_Y + 5 \dot{\tau}\right]  {\cN}^{AB}, \\
\d_{(\tau,Y)} \cJ^A &= \left[\tau \pa_u +\cL_Y + 4\dot{\tau} \right] \cJ^A + 
 \f12 {\cN}^{AB} \pa_B\tau, \\
 \d_{(\tau,Y)} {\cal M} &\,\hat{=}\, 
\left[\tau \pa_u +\cL_Y + 3 \dot{\tau} \right]{\cal M}
+ \cJ^A \pa_A \tau, \\
\delta_{(\tau,Y)}  {\tcM} &=
[\tau \pa_u+ {\cL}_Y +3\dot{\tau}] {\tcM} 
+  \tilde{\cJ}^A \pa_A {\tau}, \\
\d_{(\tau,Y)} {\cal P}_A &\,\doublehat{=}\,  \left[\tau\pa_u + \cL_{Y} + 2 \dot\tau\right] {\cal P}_A + 3 \left({\cal M} q_{AB} +   \tcM \epsilon_{AB} \right) \pa^B\tau, \\
\d_{(\tau,Y)} \cT_{AB} &\,\doublehat{=}\,  \left[\tau\pa_u + \cL_{Y} +  \dot\tau  \right] \cT_{AB}  +
{4} \cP_{\langle A}  \p_{B \rangle} \tau 
\,.
\ee
\end{subequations}
These transformation properties show a clear pattern where the linear anomaly of each covariant quantity of conformal dimension $s$ is uniquely determined by the ones of  conformal dimension $s-1$.

The requirement of covariance under the full BMSW group  gives the equations of motion
\begin{subequations}\la{ceom}
\be
 \dot \cJ^A &= \f12  \bD_B {\cN}^{AB}\,,\la{ceom-J}\\ 
\dot \cM &= \f12  \bD_A \cJ^A  +  \f18  C_{AB} {\cN}^{AB}\,,\la{ceom-M}\\
\dot \tcM &= \f12  \bD_A \tilde \cJ^A  +  \f18  C_{AB} {\tilde \cN}^{AB}\,,\la{ceom-tM}\\
\dot\cP_A &=  D_A \cM + \tilde{D}_A\tcM  + C_{AB} \cJ^B\,, \la{ceom-P}\\
\dot \cT_{AB}&={ \bD_{\langle A} \cP_{B\rangle}+\f32\left( C_{AB}\cM+\tilde C_{AB} \tcM\right)}\,.\la{ceom-T}
\ee
\end{subequations}
We can now combine the action on the soft observable with the equations of motion to write down the action of the symmetry group on the {\it corner phase space} variables. That is, the phase space obtained after imposition of the constraint equations. This gives us the on-shell actions
\begin{subequations}\la{dc}
\be
\d_{(\tau,Y)} \cJ^A &= \left[\cL_Y + 4\dot{\tau} \right] \cJ^A + \f12   \cN^{AB}\pa_B\t
+ \f\tau 2  \bD_B \cN^{AB}, \\
 \d_{(\tau,Y)} {\cal M} &\,\hat{=}\, 
\left[\cL_Y + 3 \dot{\tau} \right]{\cal M}
+  \cJ^A \pa_A \tau   +  
\f{\tau}2 \left( \bD_A \cJ^A + \f14 C_{AB} \cN^{AB}\right), \\
\delta_{(\tau,Y)}  {\tcM} &=
[ {\cL}_Y +3\dot{\tau}] {\tcM} 
+  \tilde{\cJ}^A \pa_A {\tau} +  
\f{\tau}2 \left(\bD_A\tilde \cJ^A+   \f14 C_{AB} {\tilde \cN}^{AB}\right), \\
\d_{(\tau,Y)} {\cal P}_A &\,\doublehat{=}\,  \left[ \cL_{Y} + 2 \dot\tau\right] {\cal P}_A + 
3 ({\cal M} \pa_A \tau + \tcM \tilde\pa_A\tau)
 + \tau \left(\p_A \cM +\tilde{\p}_A \tcM +  C_{AB} \cJ^B \right), \\
\d_{(\tau,Y)} \cT_{AB} &\,\doublehat{=}\,  \left[ \cL_{Y} +  \dot\tau  \right] \cT_{AB}  
{ +4  \cP_{\langle A}  \p_{B \rangle} \tau 
 + \tau\left( \bD_{\langle A}\cP_{B\rangle} +   \f32 C_{AB}\cM+ \f32 \tilde C_{AB} \tcM\right)
 }
.
\ee
\end{subequations}
This information is enough to recover the equations of motion which are given by 
\be
\dot \cJ^A=   \d_{(1,0)}\cJ^A\,,\quad   \dot\cM= \d_{(1,0)}\cM\,, \qquad \dot\cP_A = \d_{(1,0)} \cP_A\,,\quad  \dot\cT_{AB}  = \d_{(1,0)} \cT_{AB} \,,
\ee
where $\tau=1$ denotes the constant function on the sphere.

\subsection{Non-radiative Phase space }
The non-radiative phase space is  obtained after imposing the no-radiation condition \footnote{The fact that this is a correct quantity to set to zero in the absence of radiation follows from the fact the it transforms covariantly under the BMSW transformations (as well as under the standard  BMS group). 
Our choice of no-radiation condition is motivated by the demand that the corner charges be conserved in time, as shown below, and it corresponds to a Weyl tensor of Petrov type I. 
However, one can argue that this choice is conventional. Another accepted choice is to call radiation  ${\rm Im} \Psi_2, \Psi_3, \Psi_4$
since they depends explicitely on the shear and its time derivative only and to include in  the charges 
$ {\rm Re }\Psi_2, \Psi_1,\Psi_0$, which represent independent data. This is the choice  adopted in \cite{Compere:2018ylh}, where it was demanded the vanishing of the shifted news tensor \eqref{bN},  also transforming with no anomaly, and of the dual mass. In this second perspective the conventional notion of no radiation would be to impose that ${\rm Im} \Psi_2, \Psi_3, \Psi_4$ vanish, with a Weyl tensor of type II. Ultimately understanding what is the right notion of no-radiation should be related to whether the charges not included in the radiation are not only conserved but also form a coadjoint orbit. So far it only has been proven that the even stronger non-radiative condition ${\rm Im} \Psi_2= \Psi_3=\Psi_4=\Psi_0=0$, which is of Petrov type D,  forms a coadjoint orbit  (see e.g. \cite{Barnich:2021dta}). Understanding whether the no-radiation condition we work with here qualifies under this criterion still needs to be investigated.
}
 \be
 \cN^{AB}=0\,.
\ee 
  This means that  we have 
  \be
C_{AB}= u n_{AB} + c_{AB}\,,
\ee
 where $n_{AB}$ and $c_{AB}$ are time independent.
Once the no-radiation condition is imposed, we can solve the evolution equations 
\eqref{ceom} explicitly as 
follows (we use the label ${\mathrm{NR}}$ to denote the non-radiative solutions)
\begin{subequations}\la{C-exp}
\be
\cJ_A^{\mathrm{NR}} &= j_A \,, \la{J-exp}\\
\cM^{\mathrm{NR}} &= m +  \f{u}{2}D_A j^A\, ,\la{M-exp}
\\
\tcM^{\mathrm{NR}} &= \tilde{m} +  \f{u}{2}D_A \tilde{j}^A \,,\la{tM-exp}
\\
\cP_A^{\mathrm{NR}}&= 2 p_A + u(D_A m + \tilde{D}_A \tilde{m} +c_{AB} j^B) 
+  \f{u^2}{2}(D_{\langle A} D_{B \rangle} +  n_{AB})j^B \,, \la{P-exp}
 \\
{\cT}_{AB}^{\mathrm{NR}} &=3t_{AB}+  u\left({2} D_{\langle A} p_{B \rangle} +{ \frac{3}2 }(m c_{AB}  +\tilde{m} \tilde{c}_{AB})\right)\cr
&+ 
{\frac{3 u^2}{2} }\left(\left[\frac13 D_{\langle A}D_{B \rangle} + \frac12 n_{AB}\right]m +
\left[\frac13  D_{\langle A}\tilde{D}_{B \rangle} + \frac12 \tilde{n}_{AB}\right]\tilde{m} +
\frac13 D_{\langle A}c_{{B \rangle}C} j^C + \frac56 c_{C \langle A } D_{ {B \rangle}}j^C\right) \cr
&
+  {{u^3}}\left(\frac16 D_{\langle A} D_{ B} D_{C \rangle}j^C + 
 \frac16 D_{\langle A}n_{B \rangle C} j^C + \frac23 n_{C \langle A } D_{ {B \rangle}}j^C\right)\,,\la{T-exp}
\ee
\end{subequations}
where $(j_A, m,\tilde{m},p_A,t_{AB})$ are constant tensors on the sphere. In fact, given the evolution equations \eqref{ceom-J},  \eqref{ceom-M},  \eqref{ceom-tM}, it is immediate to see that $\dot  j^A=\dot m=\dot {\tilde m}=0$.
For $ p_A$ and $t_{AB} $ the proof goes as follows.
From the evolution equation \eqref{ceom-P} and the expansions (\ref{M-exp},\ref{tM-exp}),
 we get 
\be
\dot{\cP}_A &= D_A m +\tilde{D}_A\tilde{m} + c_{AB} \cJ^B +
    \f{u}{2}(D_AD_B j^B - \tilde{D}_A\tilde{D}_B j^B  + 2 n_{AB} \cJ^B)\cr
    &= (D_A m +\tilde{D}_A\tilde{m} + c_{AB} \cJ^B) +
    u(D_{\langle A} D_{B \rangle} +  n_{AB})j^B\,,
\ee
where we used
the identity
\be
\bD_A\bD_B-\tilde \bD_A \tilde \bD_B=2\bD_{\langle A} \bD_{B\rangle}\,.
\ee

Next, from  the evolution equation \eqref{ceom-T} and also the expansion \eqref{P-exp}, we get
{
\be
\dot{\cT}_{AB} &= 2 D_{\langle A} p_{B \rangle} + 
{u} (D_{\langle A}D_{B \rangle} m + D_{\langle A}\tilde{D}_{B \rangle} \tilde{m} +
D_{\langle A}[c_{{B \rangle}C} j^C]) \cr
&+  \f{u^2}{2}(D_{\langle A} D_{ B} D_{C \rangle}j^C +  D_{\langle A}[n_{B \rangle C} j^C])\cr
&+ \frac32 m c_{AB}  +\f32\tilde{m} \tilde{c}_{AB} 
+ \frac{3u}{2} \left(\frac12 c_{AB} D_C j^C+\frac12 \tilde{c}_{AB} D_C \tilde{j}^C+ m n_{AB} +\tilde{m} \tilde{n}_{AB}\right)\cr
&+ \frac{3u^2}{4} (n_{AB} D_C j^C+\tilde{n}_{AB} D_C \tilde{j}^C)\,,
\ee
}
and we use that 
\be
 \frac14 c_{AB} D_C j^C+ \frac14 \tilde{c}_{AB} D_C \tilde{j}^C=\frac12 c_{C \langle A } D_{ {B \rangle}}j^C
\ee
to arrive at
{
\be
\dot{\cT}_{AB} &= 2 D_{\langle A} p_{B \rangle} + \frac32 (m c_{AB}  +\tilde{m} \tilde{c}_{AB})\cr
&+ 
u \left(\left[ D_{\langle A}D_{B \rangle} + \frac32 n_{AB}\right]m +
\left[  D_{\langle A}\tilde{D}_{B \rangle} + \frac32 \tilde{n}_{AB}\right]\tilde{m} +
 D_{\langle A}c_{{B \rangle}C} j^C + \frac52 c_{C \langle A } D_{ {B \rangle}}j^C\right) \cr
&
+  {u^2}\left(\frac12 D_{\langle A} D_{ B} D_{C \rangle}j^C + 
 \frac12 D_{\langle A}n_{B \rangle C} j^C + 2n_{C \langle A } D_{ {B \rangle}}j^C\right)\,.
\ee
}

The conserved quantities $(j^A, m,\tilde{m},p_A,t_{AB})$ represent the charges parametrizing the non-radiative corner phase space.
They encode on $\scri$ the physical content of the spacetime. We come back to this important point in a moment.

Let us first remark that, in the asymptotic analysis of symmetry, it is often customary to define a stronger version of the no radiation condition to be specified by 
\be
\cN^{AB}=0\,, \qquad \cJ^A=0\,,
\ee
and  the spacetime is said to be strongly non-radiative.
The second condition means that $n_{AB}+\frac12 R q_{AB}$ is the Liouville energy-momentum tensor.
In this case, the conservation equations look simpler,
with $\cM^{\mathrm{NR}}=m$ and $\tcM^{\mathrm{NR}}=\tilde{m}$  independent of time and
\be
\cP_A^{\mathrm{NR}}&= 2 p_A + u(D_A m + \tilde{D}_A \tilde{m} )\,, \\
{\cT}_{AB}^{\mathrm{NR}} &=3t_{AB}
+  {
u\left(2 D_{\langle A} p_{B \rangle} + \frac{3}2 (m c_{AB}  +\tilde{m} \tilde{c}_{AB})\right)
}\cr
&+ 
 {
\frac{u^2}{2} \left(\left[ D_{\langle A}D_{B \rangle} + \frac32 n_{AB}\right]m +
\left[  D_{\langle A}\tilde{D}_{B \rangle} + \frac32 \tilde{n}_{AB}\right]\tilde{m}\right).
}
\ee
\subsection{Symmetry transformations}

Under a symmetry transformation the shear components transform as
\be
\d_{(T,W,Y)} c_{AB} &= \left[\cL_Y -W \right] c_{AB}
- \left[2\bD_{\langle A} \bD_{B\rangle} -n_{AB}\right]T\, , \la{Dc}\\
\d_{(T,W,Y)} n_{AB} &=\cL_Y n_{AB}- {2\bD_{\langle A} \bD_{B\rangle} W}\, ,\la{Dn}
\ee
while the symmetry transformations of the conserved charge aspects are given by 
\begin{subequations}\la{ddc}
\be
\d_{(T,W,Y)} j^A &= \left[\cL_Y + 4W \right] j^A\,, \\
 \d_{(T,W,Y)} m &\,\hat{=}\, 
\left[\cL_Y + 3 W \right]m 
+  j^A \pa_AT   +  
\f{T}2 \bD_A j^A\, , \\
\delta_{(T,W,Y)}  \tilde{m} &=
[ {\cL}_Y +3W] \tilde{m} 
+  \tilde{j}^A \pa_A {T} +  
\f{T}2  \bD_A\tilde j^A\,, \\
\d_{(T,W,Y)} p_A &\,\doublehat{=}\, \left[ \cL_{Y} + 2 W \right] p_A + 
{\f32} (m \pa_A T + \tilde{m} \tilde\pa_AT)
 + {\f T2} \left(\p_A m +\tilde{\p}_A \tilde{m} +  c_{AB} j^B \right)\,, \\
\d_{(T,W,Y)} t_{AB} &\,\doublehat{=}\, { \left[ \cL_{Y} +  W \right] t_{AB}  +
\f83  p_{\langle A}  \p_{B \rangle} T 
 + T\left(  \f23\bD_{\langle A}p_{B\rangle} +   \f12 c_{AB}m+ \f12 \tilde c_{AB} \tilde{m}\right)\,
.
}
\ee
\end{subequations}
 These transformation properties represent the second main result of the paper. There are two  important aspects  related to them we now  highlight. 
 
First,  we see that  the  conserved charge aspects parametrizing the corner phase space transform homogeneously for the asymptotic corner symmetry group (see Section \ref{sec:gext})  where  $T=0$. 
Second, and most importantly, the transformations \eqref{ddc} are conjectured to define a moment map between the corner phase space at $\scri$ and the dual Lie algebra of the extended corner symmetry group of null infinity. This fundamental conjecture, that will be investigated  in \cite{FMP}, gives a precise meaning to our claim above that the conserved charges $(j^A, m,\tilde{m},p_A,t_{AB})$ parametrize the corner phase space at $\scri$.

\section{ An impulsive  wave solution}\la{sec:IW}

Now that we have found the non-radiative solutions $\cN_{AB}=0$ we investigate the nature of the non-linear impulsive solutions that describe the fundamental transitions among vacua.
An impulsive gravitational wave or gravitational impulse is an \emph{exact } solution of the 
vacuum Einstein's equations of motion. 
Their study goes back to the work of Aichelburg and Sexl \cite{Aichelburg:1966su} and of  Szekeres,   Khan and Penrose \cite{Szekeres:1970vg, Khan:1971vh}. Their mathematical study
through a cut and paste approach   started with the work of Penrose 
 \cite{Penrose-72}. The study  of spherical impulsive waves has continued and  followed many formal mathematical developments since then, see \cite{Hogan:1993xj,Aliev:2000jp, Podolsky:2002sa, Luk:2012hi,Luk:2013zr}, supplemented by the   study  of their  collisions  \cite{Nutku:1977wp,Chandrasekhar:1986jn, Luk:2013zr} and the relationship with the memory effects \cite{Zhang:2017jma, Bhattacharjee:2019jaf}.
It is important to appreciate that impulsive waves that are asymptotically flat have to be spherical; this excludes the extensively studied pp-waves, which  are planar.

A gravitational impulse is analogous to the gravitational shock wave studied thoroughly by Dray and 't Hooft \cite{Dray:1984ha,Dray:1985yt,Dray:1985ie}, in the sense that they are,  by definition, solutions of the Einstein's equations that produce radiation localized on a null hypersurface, the hypersurface $u=\mathrm{cste}$.
Gravitational impulses \footnote{ Here we defer from the  accepted nomenclature of  Penrose \cite{Penrose-72} who 
calls an impulsive gravitational wave a gravitational wave whose metric is continuous but not $C_1$ on some (null) hypersurface, while  shock waves refer, for him, to metrics which are $C_1$. The curvature tensor of an impulsive gravitational wave is proportional to a delta function while the curvature tensor of a shock wave is proportional to a step function. }  are fundamentally  different in nature from shock waves though,
in the sense that a shock wave needs a non-vanishing energy-momentum source while a gravitational  impulse does not need any energy-momentum source. 
Gravitational impulses are made of pure geometry.
It is interesting to realize that the gravitational impulse solution we are constructing here 
is a solution of full non-linear gravity.
Its linearization is related to the so called gravitational soft mode introduced by Strominger et al. in \cite{Strominger:2013lka, He:2015zea} and studied more thoroughly in \cite{Pasterski:2016qvg, Donnay:2018neh, Pate:2019lpp, Guevara:2021abz, Pasterski:2021fjn, Pasterski:2021dqe}. Let us also point out that  plane-fronted gravitational impulses were considered in \cite{Wieland:2016exy} as solutions to the gluing conditions between interfaces of bounded finite regions in a discrete gravitational context.
  
\subsection{Impulsive wave phase space} \la{sec:IW-sol}

A shock wave localized at $u=0$ describes the transition between an initial  vacuum labelled by 
$O^{-}=(c_{AB}^-,n_{AB}^-,\cJ^{A-}, \cM^-,\tilde{\cM}^-,\cP_A^-,\cT_{AB}^-)$ to the corresponding out vacuum labelled by
$O^+$.
By definition a non-expanding gravitational impulse solution satisfies \footnote{We used that $\theta(u)+\theta(-u)=1$.}
\be
C_{AB}&=  (c_{AB}^+ + u n_{AB}^+)\theta(u)+ (c_{AB}^- + u n_{AB}^-)\theta(-u), \cr
&= c_{AB} + u n_{AB} + (\mathring c_{AB} + u \mathring n_{AB}) \epsilon(u)\,,
\ee
where we denote $\mathring c_{AB}:=  (c_{AB}^+-c_{AB}^-)$ the jump across the impulse   and $c_{AB} := \frac12 ( c_{AB}^+ +c_{AB}^-)$ the average  value  (similarly for $\mathring n_{AB}$ and $n_{AB}$).
In the last line above, we have introduced the step function 
\be
\epsilon(u) := \frac12[\theta(u)-\theta(-u)],\qquad \dot{\epsilon}(u)=\delta(u)\,,
\ee 
where 
 $\dot\delta(u):=\pa_u\delta(u)$ is the derived delta function.

An impulsive wave corresponds to the choice where the induced metric is continuous. This means that we impose $\mathring c_{AB}=0$. This condition is necessary in order to ensure that the energy flux is finite.\footnote{When $\mathring c_{AB} \neq 0$,
we have that $ N_{AB}N^{AB}= \mathring c_{AB}\mathring c^{AB}\delta(u)^2 +\cdots$ which is ill defined.}
This continuity condition means that 
\be
N_{AB}&=   n_{AB} +  \mathring n_{AB} \epsilon(u) \,,\\
\cN_{AB}&=    \mathring n_{AB} \delta(u)\,,
\ee
 and we see that the Weyl tensor component $\cN_{AB}$ is proportional to a delta function. 
 
 We can now easily integrate out the evolution equations and express the 
 evolution of the covariant quantities $(\cJ^A,\cM,\tcM,\cP_A,\cT_{AB})$
 in terms of the conserved  quantities 
  $(j_A, m,\tilde{m},p_A,t_{AB})$,  which have the property that they are constant in time in the non-radiative zone $u<0$ and $u>0$ before and after the gravitational impulse, and the impulse strength $\mathring{n}_{AB}$.
 
  Let us start with the current $ \cJ^A$.  It is immediate to see that the solution to \eqref{ceom-J} can be written as
 \be
 \cJ^A(u) &=   \cJ_{A}^ {\mathrm{NR}}+\cJ_{A}^{ \mathrm{R}1}\cr
&= \frac14 D^A R(q)+   \f12  D_B n^{AB}  +\f12\epsilon(u)  D_B \mathring{n}^{AB}\,,
 \ee
 where we made explicit its structure as a sum of the non-radiative solution \eqref{J-exp} and a  distributional radiative component linear in $\mathring{n}_{AB}$.
 
 To get the mass evaluation one integrates the evolution equation \eqref{ceom-M}
 \be
 \dot \cM & =  \f12 \bD_A \cJ^A  +  \f18  C_{AB} \cN^{AB}.
 \ee
 The first term can be easily integrated if one uses that 
 \be
 \pa_u[u\epsilon(u)]= \epsilon(u) + u\delta(u) =\epsilon(u).
 \ee
 The product $ C_{AB}\cN^{AB}$  contains product of distribution 
 which are evaluated using
 \be
 u\delta(u)=0,\qquad \epsilon(u) \delta(u) =0,
 \qquad
 u \epsilon(u)  \dot\delta(u) =0,
 \qquad 
 \epsilon(u) \delta(u)= 0.
 \ee
 Explicitly, we get  \footnote{We use the following regularization 
 \be
 \delta(u)\theta(u)=\frac12 \delta(u) , 
 \qquad
  \delta(u)\theta(-u)=\frac12 \delta(-u).
 \ee
}
 \be
 C_{AB} \cN^{AB}&= 
 [ c_{AB}+ u {n}_{AB} +  u \epsilon(u) \mathring n_{AB} ] 
  \mathring n^{AB} \delta(u)=  c_{AB} \mathring n^{AB}   \delta(u)
  = \pa_u[ c_{AB} \mathring n^{AB}   \epsilon(u)].
   \ee
 This means that we obtain the solution 
 \be
 \cM(u) &= \cM^{\mathrm{NR}}+\cM^{\mathrm{R1}}\cr
 &=m + \f{u}2 D_A j^A +  \f18 \epsilon(u)\, c_{AB} \mathring n^{AB}  
 + \f14 u \epsilon(u) \,D_A D_B \mathring{n}^{AB}.
 \ee
 We see that the covariant mass is the sum of  a non-distributional component $\cM^{\mathrm{NR}}$, which agrees with the non-radiative expression
 \eqref{M-exp},
and a   distributional radiative component $\cM^{\mathrm{R1}}$,
 which is linear in the impulse radiative news $\mathring{n}_{AB}$. A similar analysis gives the expression for the dual mass in the presence of an impulse
 \be 
 \tcM(u) &=\tcM^{\mathrm{NR}}+\tcM^{\mathrm{R1}}\cr
 &= \tilde m + \f{u}2\, D_A \tilde{j}^A +  \f18 \epsilon(u)\, c_{AB} \mathring{\tilde{n}}^{AB}  
 + \f14 u \epsilon(u) \,D_A D_B \mathring{\tilde{n}}^{AB}.
 \ee
 The expression  for the covariant momentum can be obtained  by integrating out \eqref{ceom-P}
 \be 
 \dot\cP_A &=  D_A \cM + \tilde{D}_A\tcM  + C_{AB} \cJ^B\,.
 \ee
  To perform the integration one uses the expansion
 \be 
 C_{AB}\cJ^B 
 &= [c_{AB} + u n_{AB} ] j^B 
 +\f12   \epsilon(u)  c_{AB} D_C \mathring{n}^{CB}
 + u\epsilon(u)\left[ \f12n_{AB}D_C \mathring{n}^{CB}+  \mathring{n}_{AB} j^B\right]
 + {\f12}u \mathring{n}_{AB}D_C \mathring{n}^{CB}\,,
 \ee
 where we used that $\epsilon^2(u)=1$ as a distribution.
The final expression is given by
 \be 
 \cP_A&=
  \cP_{A}^{\mathrm{NR}}+\cP_{A}^{\mathrm{R}1}+\cP_{A}^{\mathrm{R}2}\cr
 &=2 p_A + u(D_A m + \tilde{D}_A \tilde{m} +c_{AB} j^B) 
+  \f{u^2}{2}(D_{\langle A} D_{B \rangle} +  n_{AB})j^B\cr
&+   \f18 u\epsilon(u)\, \left[D_A (c_{BC} \mathring n^{BC})  +D_A( c_{BC} \mathring{\tilde{n}}^{BC})+4 c_{AB} D_C \mathring{n}^{CB}
 \right]\cr
 &
 + \f18 u^2 \epsilon(u) \left[D_A D_B D_C \mathring{{n}}^{BC}+D_A D_B D_C \mathring{\tilde{n}}^{BC} + 2  n_{AB}D_C \mathring{n}^{CB}+  4 \mathring{n}_{AB} j^B \right]\cr
 & + \f{u^2}{4} \mathring{n}_{AB}D_C \mathring{n}^{CB}
 \,.
 \ee
 We see that the covariant momentum is the sum of the non-radiative expression $\cP_A^\mathrm{NR}$ given in \eqref{P-exp}, a distributional expression $\cP_A^\mathrm{R1}$ proportional to the impulse news $\mathring{n}_{AB}$ and a secular component $\cP_A^\mathrm{R2}$ quadratic in the impulse news. 
 
Finally, the expression for the stress tensor can be obtained  by integrating out \eqref{ceom-T}
{
\be
\dot \cT_{AB}&= \bD_{\langle A} \cP_{B\rangle}+\f32\left( C_{AB}\cM+\tilde C_{AB} \tcM\right)\,.
\ee
}
A similar analysis shows  that the solution 
can be written as 
\be
\cT_{AB}= \cT_{AB}^{\mathrm{NR}}+\cT_{AB}^{\mathrm{R}1}+
\cT_{AB}^{\mathrm{R}2}\,,
\ee
showing that the stress tensor  can be decomposed into a non-radiative component $\cT_{AB}^{\mathrm{NR}}$,   a distributional radiative component $\cT_{AB}^{\mathrm{R}1}$  linear in $\mathring{n}_{AB}$ and a secular radiative component $\cT_{AB}^{\mathrm{R}2}$ quadratic in $\mathring{n}_{AB}$.
The non-radiative component is already given in \eqref{T-exp}.   The distributional radiative 
component reads
\be 
\cT_{AB}^{\mathrm{R}1}&= 
{ \f3{16}} u\epsilon(u) \left[c_{AB}c_{CD} \mathring n^{CD}+  \tc_{AB}c_{CD} \mathring \tn^{CD}\right]\cr
&+{\f1{16}}u^2\epsilon(u) \bigg[
 D_{\langle A}\left(D_{ B\rangle} (c_{CD} \mathring n^{CD})  +D_{ B\rangle}( c_{CD} \mathring{\tilde{n}}^{CD})+4 c_{{ B\rangle}C} D_D \mathring{n}^{DC}
 \right)\cr
& +3(c_{AB} D_C D_D \mathring n^{CD}+\tc_{AB} D_C D_D \mathring \tn^{CD})
 +\f32 (n_{AB} c_{CD} \mathring n^{CD} +\tn_{AB} c_{CD} \mathring \tn^{CD} )
 +12 (m\mathring n_{AB}+\tilde m\mathring \tn_{AB})
\bigg]\cr
&
 +{ \f1{24}} u^3 \epsilon(u) \bigg[ D_{\langle A}\left( D_{ B\rangle}  D_C D_D \mathring{{n}}^{CD}+D_{ B\rangle}  D_C D_D \mathring{\tilde{n}}^{CD} + 2  n_{{ B\rangle} C}D_D\mathring{n}^{CD}+  4 \mathring{n}_{{ B\rangle} C} j^C \right)\cr
&+3(n_{AB} D_C D_D \mathring n^{CD}+\tn_{AB} D_C D_D \mathring \tn^{CD})
+6 (D_Cj^C  \mathring n_{AB}+D_C \tilde j^C\mathring \tn_{AB})
\bigg]\,,
\ee

while the secular radiative components is 
\be
\cT_{AB}^{\mathrm{R}2}&={ \f{3 u^2}{32}} \left[\mathring{n}_{AB} c_{CD} \mathring{n}^{CD} + \mathring{\tilde{n}}_{AB} c_{CD} \mathring{\tilde{n}}^{CD}  \right]\cr
& +{ \f{u^3}{8}} \left[\mathring{n}_{AB} D_C D_D \mathring{n}^{CD}+\mathring{\tilde{n}}_{AB} D_C D_D \mathring{\tilde{n}}^{CD}
 +\f23  D_{\langle A} (\mathring{n}_{B\rangle C}D_D \mathring{n}^{DC})\right]\,.
\ee

\subsection{Recovering Penrose's solution}

The solution first described by Penrose in \cite{Penrose-72}, and obtained by a holomorphic gluing along a null-cone 
 of  two portions of flat space, is a particular example of the construction we have just given.
  Penrose's solution can be revealed by imposing that 
  \be
  c_{AB}=0,\qquad
   D_B \mathring{n}^{AB}=0.\label{Dn}
  \ee 
  Under these conditions, we see that the radiative components of the current, mass, momentum and stress tensor all vanish
  \be
  \cJ_A^{\mathrm{R}}=0\,,\qquad \cM^{\mathrm{R}}=0=\tcM^{\mathrm{R}}\,,\qquad
  \cP_A^{\mathrm{R}}=0\,,\qquad \cT^{\mathrm{R}}_{AB}=0\,. 
  \ee
The Penrose's solution is characterized by demanding that the non-radiative component is also flat. This means that the only non-vanishing component is the radiative one $\cN^{AB}= \mathring{n}^{AB} \delta(u)$.
This solution  is integrable in the bulk exactly. 
It is obtained by patching up two flat space solutions
\be
\rd s^2 = -2\rd u \rd r + \rd u^2 + \frac{4r^2}{(1+|z|^2)^2} \rd z \rd \bar{z}, 
\ee
along the sphere at $u=0$.
The key element is to  recognize that the asymptotic news can be written as a  Schwarzian derivative
\be
\mathring{n}_{zz}= \{ h, z\}=\frac{h'''}{h'} - \frac32 \left(\frac{h''}{h'}\right)^2,\qquad
 \mathring{n}_{\bar{z}\bar{z}}= \{\bar{ h}, \bar{z}\}\,,
\ee
where $h$ is holomorphic.
The full solution can then be obtained by the following matching condition
at $u=0$
\be
(r,z,\bar{z})_+ =\left( \frac{r}{|h'|}\frac{1+|z|^2}{1+|h|^2}, h(z), \bar{h}(\bar{z})  \right)_-.
\ee

\section{Conclusions}\la{sec:conc}

Exploiting the BMSW extension \cite{Freidel:2021yqe} of the residual diffeomorphism symmetry of null infinity, we have constructed in Section \ref{sec:Ano} charges associated to all the Weyl scalars and that transform semi-covariantly, i.e. with only linear anomaly appearing, under the action of the BMSW group. The characterization of the full phase space of $\scri$ led us to the introduction of a duality transform and in particular to the definition of the dual covariant mass \eqref{dualM}. We have
 shown in Section \ref{sec:EOM} how the sole demand of anomaly freedom is enough to recover the asymptotic Einstein's equations coupled to matter, written as evolution equations for the covariant charges,  by identifying the quantities that transform homogeneously under the symmetry transformations. 
 
 {This derivation of the gravitational dynamics from purely a symmetry principle highlights  the central role of the extended corner symmetry algebra, revealed in \cite{Ciambelli:2021vnn, Freidel:2021cbc}, in providing a local holographic description of gravity. In particular,
 borrowing the terminology from representation theory,
  we have seen how the evolution equations can be understood as {\it intertwiners} for the BMSW group, as they imply that a given combination is left invariant by the action of the this group, whose Lie algebra represents a subalgebra of the  extended corner symmetry one \cite{Freidel:2021cbc}. }

More precisely, our derivation of the asymptotic evolution   Einstein's equations as the functionals of the gravitational phase space variables left invariant by the asymptotic symmetry group  opens a new way to think about the quantization of gravitational dynamics in terms of  representation theory structures associated to the quantization of this group.  Among these,  the  intertwiner space represents the  subspace of invariant tensors in the tensor product of a given set of irreducible representations of the quantum symmetry algebra. One can then envisage a regularization procedure where a notion of intertwiner can be used to fuse tensor products of   irreducible representations associated to corners at consecutive instants of time at $\scri$, so that a quantum version of constraint equations  is holographically implemented. In order for this strategy to correctly capture the gravitational dynamics at the quantum level it is crucial to identify a basis where the propagating degrees of freedom of the radiation for general spacetimes can be represented  explicitly and possibly in a non-perturbative manner.

Within this program of describing the gravitational dynamics starting from the representation of the corner symmetry group, we have taken here a further step in this direction in Section \ref{sec:CO} by identifying  the conserved charges that define the non-radiative corner phase space. 
We have shown that they transform under a representation of the extended corner symmetry group. This statement is supported by the transformation properties \eqref{ddc}. 
We have then studied a fundamental vacua transition process by solving the evolution equations in the presence of an impulsive gravitational wave, representing an exact solution of the vacuum Einstein's equations. Interestingly, we found that all the Weyl scalars in the asymptotic corner phase space are non-vanishing. The solutions  consist of a vacuum component, given by the conserved charges describing the non-radiative phase space, and a radiative component. The latter contains a distributional contribution linear in the gravitational impulse news and a secular contribution quadratic in it.
This opens the way towards a description of an arbitrary signal as a succession of gravitational impulses. The next step in the program is to ensure that the representation that we have identified for the conserved charges can be understood as a coadjoint representation. {Imposition of asymptotic dynamics at the quantum level can then be phrased  in terms of a notion of intertwiners between the irreducible representations of the asymptotic symmetry group and the quantum numbers associated to radiation in an impulsive wave basis}.

Let us conclude by pointing out an interesting implication of our strategy in recovering the asymptotic dynamics of gravity.
A natural question is whether our symmetry argument can implement any constraint on modifications of gravitational dynamics beyond Einstein's theory. 
 An answer to this question can be provided by relying on the relatively recent discovery of the equivalence between  soft graviton theorems and asymptotic symmetries
 (see \cite{Strominger:2017zoo,  Pasterski:2021rjz, Raclariu:2021zjz} for reviews). 
In particular, the leading, subleading and sub-subleading tree-level soft theorems have been shown to be equivalent to respectively the covariant mass and dual mass  EOM \eqref{Mdot}, \eqref{tMdot}, the covariant momentum EOM \eqref{Pdot}, and the spin-2 charge EOM \eqref{Tdot} \cite{Freidel:2021dfs}. 
Moreover, it was shown in  \cite{Elvang:2016qvq, Laddha:2017ygw} that tree-level soft graviton theorems  at leading and subleading orders do not receive higher derivative
corrections, while the sub-subleading soft graviton theorem  corrections vanish for pure gravity. One exception where we expect corrections to the sub-subleading soft theorem, which is beyond the scope of our analysis here, is when gravity is coupled to a dilaton field.
We can thus conclude that our strategy to derive  asymptotic evolution equations at leading order in the large-$r$ expansion around null infinity  uniquley determined by symmetry  is unaffected by 
 higher derivative corrections to vacuum general relativity at leading,  subleading  and sub-subleading orders.  


\section*{Acknowledgement}

We would like to thank Glenn Barnich, Geoffrey Comp\`ere, Roberto Oliveri, Simone Speziale for helpful discussions and insights. 
Research at Perimeter Institute is supported in part by the Government of Canada through the Department of Innovation, Science and Economic Development Canada and by the Province of Ontario through the Ministry of Colleges and Universities. This project has received funding from the European Union's Horizon 2020 research and innovation programme under the Marie Sklodowska-Curie grant agreement No. 841923. 

\appendix

\section{Action of the symmetry}\la{App:action}

We want to find $\d_{(\t,Y)} \Phi^i$ such that
\be
{\cal L}_{\xi_{(\t,Y)}} g_{\mu\nu}[\Phi^i]=\left. \pa_\epsilon g_{\mu\nu}[\Phi^i+ \epsilon \d_{(\tau,Y)} \Phi^i] \right|_{\epsilon=0}\,.
\ee
In this section we concentrate on the case where $\Phi^i=\{ b, F, M, U_A, \bP_A ,q_{AB}, C_{AB}, E_{AB}\}$.
It will prove convenient to express the \bmw vector fields $ \xi_{(\tau,Y)} $ \eqref{vu} in the form
\be\la{xi}
\xi_{(\tau,Y)}=\bar\xi_{(\tau,Y)}+\f1r\xi_1+\f1{r^2} \xi_2+\f1{r^3} \xi_3\,,
\ee
where $
\bxi_{(\tau,Y)}:=\tau \pa_u + Y^A\pa_A -\dot{\tau} r\pa_r$ is the asymptotic component given in \eqref{xib}.
The lower order vector fields $\xi_i= y^A_i \p_A+ \rho_i r \p_r$ only have tangential and radial components.
Their expression is derived from (\ref{vA},\ref{vr}) and   the expansion of 
\be
I^{AB}&= \int_r^\infty  \f{\rd r'}{r^{'2}} e^{2\beta}  \gamma^{AB}\cr
&=\f1r \bq^{AB} - \f1{2r^2} C^{AB} 
+\f{\bq^{AB}}{r^3} \left( \f23 b + \f1{12}C^{CD}C_{CD} \right)
+ o(r^{-4})\,.
\ee
The expansion of the tangential vector is
\be
y_1^A=-\p^A\tau, \qquad
y_2^A=\f1{2} C^{AB}\p_B \tau,
\qquad
y_3^A=\left(2\B-\f83\textsf E_{\B}\right)  \p^A \tau\,,
\ee
while  the radial components are given by
\be
\rho_1&= \f12   \Delta \tau \,,\\
\rho_2&= -\f12\left( {\bD}_AC^{AB}\p_B \tau +\f12C^{AB}{\bD}_A \p_B \tau {- \textsf E_U^A \pa_A\tau} \right) \,,\\
\rho_3&=
-\left( 
\f43 \p_A  \B\p^A \tau
-\f43\p_A\textsf E_{\B}\p^A \tau
+\left(\B-\f43\textsf E_{\B}\right)   \Delta \tau
+\f13 \bP^A\p_A\tau
+\f13C^{AC}\bU_C\p_A\tau
\right) \,.
\ee
The metric components read as
\be
g_{uu}&= -2\Phi e^{2\beta}+\f1{r^2} \gamma_{AB} \Upsilon^A \Upsilon^B \cr
&=-2 F +\f{2 M} r +\f1{r^2} \left( \bq_{AB}   U^A  U^B -4 \bF \B \right)
+o(r^{-2})\,,\\
g_{ur}&=-e^{2\beta} = -1-\f2{r^2} \B +o(r^{-2})\,,\\
g_{rr}&=0\,,\\
g_{Au}&= - \gamma_{AB} \Upsilon^B= -  U_A   +
\frac{2 }{ 3 r} \left(\bP_A-\f12 C_{AB} U^B +\pa_A\B \right) +o(r^{-1})\,,\\
g_{Ar}&=0\,,\\
g_{AB}&=r^2 \gamma_{AB} = r^2 \bq_{AB}+r C_{AB} +\f14\bq_{AB}C_{CD} C^{CD}+ \f1{r} E_{AB} +  o(r^{-1})\,.
\ee

In order to compute their symmetry transformations, we use the 
field expansion \eqref{xi} and  write a general field transformation as
\be
\d_{(\t, Y)}=\d_{\bar \xi}+\Delta_\t\,,
\ee
where
\be
\d_{\bar \xi} :=  \tau\pa_u +\cL_Y + s \dot\tau\,,
\ee
with $s$ the conformal weight of the given quantity and $\Delta_\t$ its anomaly.

\subsection{$g_{ur}$}

We start with 
\be
 {\cal L}_{\xi} g_{ur}&= 
 \xi^\nu \p_\nu g_{ur}
+ g_{u u} \pa_r \xi^u
+ g_{u r} \pa_r \xi^r
+ g_{u A} \pa_r \xi^A
+ g_{r u} \pa_u \xi^u\,.
\ee
The dominant contribution is obtained by replacing $\xi\to\bar\xi$ and we find 
 that $g_{ur}$ transforms as a scalar  of weight $[-r\p_r]$, since
\be \la{cor}
\d_{\bar{\xi}} g_{ur}= \tau \dot{g}_{ur} + \cL_Y g_{ur}-\dot{\tau} r\pa_r  g_{ur}\,.
\ee
Given the notation $y^A=\sum_{i} y^A_i/r^i, \rho=\sum_i \rho_i/r^i$, the anomaly is given by 
\be
 \Delta_\tau  g_{ur}&= 
 y^A \p_A g_{ur}+\rho r\p_r g_{ur}
+ g_{u r} \pa_r (r\rho)
+ g_{u A} \pa_r y^A\cr
&= 
-\p_r \left(\f1{r} \rho_2\right )-U_A \p_r \left(\f1{r} y^A_1\right )+o(r^{-2}) \cr
&=-\f{1}{r^2}\left[ 
\f12\left( {\bD}_AC^{AB}\p_B \tau +\f12C^{AB}{\bD}_A \p_B \tau {- \textsf E_U^A \pa_A\tau}\right)
+U^A\p_A\t
\right] +o(r^{-2})\cr
&=-\f{2}{r^2}\left[ \f18 C^{AB}{\bD}_A \p_B \tau +{\f14 \textsf E_U^A \pa_A\tau }\right] +o(r^{-2})\,.\la{Dgur}
\ee
Now, since $ g_{ur}= -2(1+ b/r^2) +o(r^{-2})$, this means that we have 
\be
\delta_{(\t, Y)} \B 
&=[\tau \pa_u +Y^A \pa_A + 2\dot{\tau} ] \B 
+\f18 C^{AB} D_A \pa_B\tau
+{\f14  \textsf E_{U}^A \p_A \t}\,,
\ee
where we used that 
\be
 \textsf E_{U}^A&:=  {U}^A+ \frac12{D}_B \bC^{AB}\,\hat=\,0\,.
 \ee
\subsection{$g_{uu}$}

Next, we look at
\be
{\cal L}_{\xi} g_{uu}
&=
\xi^u \p_u g_{uu}
+\xi^A \p_A g_{uu}
+\xi^r \p_r g_{uu}
+ 2g_{uu} \p_u \xi^u
+ 2g_{uA} \p_u \xi^A
+ 2g_{ur} \p_u \xi^r.
\ee
This means that $g_{uu}$ transforms as a scalar  of weight $[2-r\p_r]$, since
\be
\d_{\bar{\xi}} g_{uu}= \tau \dot{g}_{uu} + \cL_Y g_{uu} +\dot{\tau}(2- r\pa_r) g_{uu} .
\ee
The anomaly is given by 
\be
\Delta_\tau g_{uu} &=
y^A \p_A g_{uu}
+\rho\,r \p_r g_{uu}
+ 2g_{uA} \p_u y^A
+ 2 r g_{ur} \dot\rho \cr
&=
-\f2r\left( y_1[F]  + U_A \dot{y}^A_1 \right)
-2 \left(\dot{\rho}_1 +\f1r \dot\rho_2 \right)  +o(r^{-1})
\cr
&= 
- \Delta \dot\tau +\f2r\left( \pa^A\tau \pa_AF   + 
\f12 {\bD}_AN^{AB}\p_B \tau +\f14 \pa_u \left(C^{AB}{\bD}_A \p_B \tau \right) + {\f12 ({ \textsf E_U^A \pa_A \dot\tau}-  \dot{ \textsf E}_U^A \pa_A \tau)}\right) +o(r^{-1})\,.\cr
\ee
We can thus read off the field variations
\be
\d_{(\tau,Y)} \bF&=\left[\tau \pa_u +\cL_Y + 2 \dot{\tau} \right] \bF  + \f12  \Delta \dot{\tau},\\
\d_{(\tau,Y)} M&=\left[\tau \pa_u +\cL_Y + 3 \dot{\tau} \right]M
+ \left( \f12 \bD_B N^{AB} +\pa^A F \right)\pa_A \tau\cr
&+
\f14 N^{AB} \bD_A\pa_B\tau 
+\f14  C^{AB} \bD_A \p_B \dot\tau
+  {\f12 ({\textsf E_U^A \pa_A \dot\tau}-  \dot{ \textsf E}_U^A \pa_A \tau)} \,.
\ee

\subsection{$g_{uA}$}
Next, we rewrite
\be
g_{Au}=-U_A+\f1r V_A +o(r^{-1})\,,
\ee
where, following \eqref{PA} and \eqref{gamma}, we have 
\be
V_A:=\f23\left(\bP_A-\f12  C_{AB} U^B +\pa_A\B \right)\,,
\ee
and compute
\be
 {\cal L}_{\xi} g_{Au}&= 
 \xi^\nu \p_\nu g_{Au}
+ g_{A B} \pa_u \xi^B
+ g_{A u} \pa_u \xi^u
+ g_{u u} \pa_A \xi^u
+ g_{u B} \pa_A \xi^B
+ g_{u r} \pa_A \xi^r\,.
\ee
This means that $g_{Au}$ transforms as a vector  of weight $[1-r\p_r]$, since
\be
\delta_{\bar{\xi}} g_{Au}= \tau \dot{g}_{Au} + \cL_Y g_{Au} +\dot{\tau}(1- r\pa_r) g_{Au} .
\ee
The anomaly is given by 
\be
\Delta_\tau g_{Au}&= 
 \rho r\pa_r g_{Au}
+y^B\bD_B g_{Au}
+ g_{A B} \pa_u y^B
+ g_{u u} \pa_A \tau
+ g_{u B} \pa_A y^B
- g_{u r} \pa_A(r\dot\t)
+ r g_{u r} \pa_A \rho\cr
&= - \f1r y_1^B\bD_B U_A
+ \left(r^2 \bq_{AB}+r C_{AB} +\f14\bq_{AB}C_{CD} C^{CD}\right)\left( \f1r  \dot{y}^B_1 + \f1{r^2}  \dot{y}^B_2+  \f1{r^3}  \dot{y}^B_3\right)\cr
&+\left(-2 F +\f{2 M} r\right)\p_A \t
-\f1r\left(-U_B+\f1r V_B \right)D_A \p^B \t
-\left(1+\f2{r^2} \B\right) \pa_A\left(-r\dot\t + \rho_1+\f1{r} \rho_2\right)+o(r^{-1}),\cr
&= \dot{y}_{2A} +  C_{AB} \dot{y}_{1}^B-2 F \pa_A\tau -\pa_A \rho_1  \cr
&{+\f1r\left[ 2M \pa_A \tau + 2 b\pa_A \dot\tau  -\pa_A \rho_2 -y_1^B\bD_B U_A + U^BD_A\pa_B\tau+ \dot{y}_{3A}+ C_{AB} \dot{y}_2^B +\f14 C_{CD} C^{CD} \dot{y}_{1A} \right]}\cr
&=\f1{2} N^{AB}\p_B \tau-\f1{2} C^{AB}\p_B \dot \tau
-2 F\p_A \t
-\f12 \p_A\Delta \tau
\cr
&+\f1r\bigg[
 {\bD_B U_A\p^B\t}  + U^BD_A\pa_B\tau
+2\dot\B \p_A\t
+2 \B \p_A\dot\t
-\f83 \dot {\textsf E}_{\B}\p_A\t
-\f83\textsf E_{\B}\p_A\dot \t
\cr
&+\f1{2} C_{AB}  N^{BC}\p_C \tau
+\underbrace{\f1{2} C_{AB}  C^{BC}\p_C \dot \tau
-\f14 C_{CD} C^{CD} \p_A\dot \t}_{=0}
\cr
& +\f12\p_A\left( {\bD}_C C^{CB}\p_B \tau +\f12C^{CB}{\bD}_C \p_B \tau - \textsf E_U^B\pa_B\tau \right)
+2M \p_A \t + 2\B \p_A\dot\t
\bigg]\,,\la{DguA}
\ee
from which we can read off the transformations
\be
\d_{(\t,Y)} \bU_A &=   \left[\tau\pa_u + \cL_{Y} + \dot\tau  \right]\bU_A 
   + \f12 (4 \bF  \pa_A\tau +\pa_A \Delta\tau)  +\f12\left( C_{A}{}^{B} \pa_B\dot\tau - N_{A}{}^{B}\pa_B\tau\right)\,,\\
 \d_{(\t,Y)} V_A &=   \left[\tau\pa_u + \cL_{Y} + 2\dot\tau  \right]V_A 
 +\f1{2} C_{AB}  N^{BC}\p_C \tau+2M\p_A \t+2\dot\B \p_A\t+4\B \p_A \dot\t\cr
 &+\f12\left( \bD_A{\bD}^C C_{CB} -\bD_B \bD^C C_{CA} \right)\p^B \tau
  +\f14\p_A\left(C^{CB}{\bD}_C \p_B \tau \right)\cr
  &+{ \f12  \textsf E_{U}^B\bD_A \p_B \t
  + \left(\bD_B  \textsf E_{U A}-\f12 \bD_A \textsf E_{UB}\right)\p^B\t  }
  -\f83 \dot{ \textsf E}_{\B}\p_A\t
-\f83\textsf E_{\B}\p_A\dot \t
  \,.
\ee

We can now use 
\be
\d_{(\tau,Y)} (\pa_A \B) &= [\tau\pa_u + \cL_Y +2\dot\tau ]
(\pa_A \B) +\f18 \pa_A (C^{BC} D_B \pa_C\tau)
+  \dot{\B} \pa_A \tau + 2 \B \pa_A\dot\tau\,,\\
\d_{(\tau,Y)} ( C_{AB}U^B)&= [\tau\pa_u + \cL_Y +2\dot\tau ]
 (C_{AB}U^B )
+\f14C_{BC}C^{BC} \pa_A \dot\tau
-\f12 C_{AB} N^{BC}\pa_C \tau \cr
&+\f12C_{AB} (4F\pa^B\tau +\pa^B\D \tau) 
+ \bD_{\langle A}\bD_{B\rangle} \tau \bD_C C^{CB}\cr
&
-2 \bD_{\langle A}\bD_{B\rangle} \tau  \textsf E_{U}^B\,,
\ee
to finally compute, on-shell of $\textsf E_{U}^A\,\hat=\,0$, the momentum transformation
\be
\d_{(\tau,Y)}P_A&=\f32  \d_{(\t,Y)} V_A +\f12  \d_{(\tau,Y)} ( C_{AB}U^B)-\d_{(\tau,Y)} (\pa_A \B)\cr
&\,\hat{=}\, [\tau\pa_u + \cL_Y +2\dot\tau ]P_A  
+3M\p_A \t
-\f18 C_{BC} N^{BC} \p_A\t
+\f12 C_{AB}  N^{BC}\p_C \tau
\cr
&+F C_{AB} \pa^B\tau +\f14C_{AB}  \pa^B\D \tau
 \cr
 &+\f34\left( \bD_A{\bD}^C C_{CB} -\bD_B \bD^C C_{AC} \right)\p^B \tau
  +\f14\p_A\left(C^{CB}{\bD}_C \p_B \tau \right)\cr
   &+\f12 \bD_{\langle A}\bD_{B\rangle} \tau \bD_C C^{CB}-2\dot{\textsf E}_{\B} \pa_A \tau\,. 
   \ee
We see that the momentum transformation does not contain any anomaly term proportional to $\pa_A\dot{\tau}$.


\subsection{Sphere metric}

We compute here the anomaly of the sphere metric component $E_{AB}$ in the expansion \eqref{gamma}. 
The Lie derivative of the metric component $g_{AB}$  yields 
\be
\cL_\xi g_{AB}
&=\xi^u \p_u g_{AB}
+\xi^C \p_C g_{AB}
+\xi^r \p_r g_{AB}
+ 2 g_{(A u} \p_{B)} \xi^u
+ 2 g_{C(A} \p_{B)} \xi^C\,.
\ee
This means that $g_{AB}$ transforms as a tensor   of weight $-[r\p_r]$, since
\be
\d_{\bar{\xi}} g_{AB}= \tau \dot{g}_{AB} + \cL_Yg_{AB}  -\dot{\tau} r\pa_r  g_{AB} .
\ee
The anomaly is given by 
\be
\Delta_\tau g_{AB}&=
(\cL_{y}+ \rho r\pa_r)(r^2 \gamma_{AB})
+2g_{u(A}\p_{B)} \tau\cr
&= \f1r  \left( \cL_{ y_1} + \rho_1 r\pa_r\right) ( r^2\bq_{AB})  +2g_{u(A}\p_{B)} \tau\cr
&+\f1r (\cL_{ y_1}+ \rho_1 r\pa_r)\left( r C_{AB}\right) +
\f1{r^2}  (\cL_{ y_2}+\rho_2 r\pa_r) ( r^2\bq_{AB})
\cr
&+
\f1{r^3}(\cL_{  y_3}+\rho_3 r\pa_r) ( r^2\bq_{AB})
 +\f1{r^2} (\cL_{ y_2}+\rho_2 r\pa_r) \left( r C_{AB}\right)
 +\f1r(\cL_{ y_1}+\rho_1 r\pa_r) \left(   \frac{1}{4} q_{AB}C_{CD}C^{CD} \right)
\cr
&=  r  \left(  
\cL_{y_1} \bq_{AB}+2\rho_1 \bq_{AB}
\right)+ \left[ \left( 
 \cL_{y_2} \bq_{AB}+2 \rho_2 \bq_{AB}
+\cL_{y_1} C_{AB} +\rho_1 C_{AB}
\right)
-2\bU_{(A}\p_{B)} \tau \right] 
\cr
&+ \f1{r} \left( 
 \cL_{y_3} \bq_{AB}+2\rho_3 \bq_{AB}
+\cL_{y_2} C_{AB} +\rho_2 C_{AB}
 +\cL_{y_1}\left( \f14 q_{AB}C_{CD}C^{CD}  \right)
\right)\cr
&+
\frac{4 }{ 3 r} \left(\bP_{(A}-\f12  C_{C(A}\bU^C +\pa_{(A}\B\right )  \p_{B)} \tau
+o(r^{-1})
\,.
\ee
Therefore, we can write
\begin{subequations}\la{D}
\be
\Delta_\t  \bq_{AB} &=0\,, \la{Dq}\\
\Delta_\t  C_{AB} &=\cL_{y_1} \bq_{AB}+2\rho_1 \bq_{AB}
= -2 \bD_{\langle A} \p_{B\rangle} \tau
\,,\la{DCAB}\\
 \Delta_\t  \left(\f14 q_{AB}C_{CD}C^{CD}\right )&=
  \cL_{y_2} \bq_{AB}+2\rho_2 \bq_{AB}
+\cL_{y_1} C_{AB} +\rho_1 C_{AB}
-2\bU_{(A}\p_{B)} \tau
 \,,\la{DqCC}\\
 \Delta_\t E_{AB} &=\cL_{y_3} \bq_{AB}+2\rho_3 \bq_{AB}
+\cL_{y_2} C_{AB} +\rho_2 C_{AB}
 +\cL_{y_1}\left( \f14 q_{AB}C_{CD}C^{CD}  \right)\cr
 &+
\frac{4 }{ 3 } \left(\bP_{(A}-\f12 C_{C(A}\bU^C +\pa_{(A}\B \right)  \p_{B)} \tau
 \,.\la{DEAB}
\ee
\end{subequations}
The first two anomaly relations in \eqref{D} yield \eqref{dqAB}, \eqref{dC}. The third one can be evaluated to be 
\be
 \Delta_\t  \left(\f14 q_{AB}C_{CD}C^{CD}\right )
&\,\hat{=}\,
- \f12\bq_{AB} C^{CD}{\bD}_C \p_D \tau
-C_{C(A} \bD_{B)} \p^C \tau+\f12C_{AB}\Delta \tau\cr
&= - \bq_{AB} C^{CD}{\bD}_C \p_D \tau
\,,\la{DCC1}
\ee
where we have used\footnote{The relation \eqref{an4} is an application of the general property that, for any pair of $2\times 2$ symmetric and traceless matrices $A,B$,
the  identity
\be\la{key}
A_{\langle A}{}^C B_{B\rangle C}=0
\ee
holds.
Then  \eqref{an4} follows from
$A_{AB}= C_{AB}$ and $B_{AB} =D_{\langle A} \pa_{B\rangle} \tau$.} 
\be
C_{C\langle A} \bD_{B\rangle} \p^{C} \tau=
    \f12 C_{AB} \bD_C \p^C \tau \,,
    \la{an4}
\ee
and it is thus consistent with \eqref{Dq}, \eqref{DCAB}, as $ \Delta_\t  \left(\f14 q_{AB}C_{CD}C^{CD}\right )=
 \f12 q_{AB}C^{CD} \Delta_\t C_{CD}$.
 
 \subsubsection{Stress tensor anomaly}\la{App:Eano}

  We can  now compute our quantity of interest, namely
 \be
 \Delta_\t E_{AB} &=
4\left(\B -\f43  \textsf E_\B \right)\bD_{(A} \p_{B)} \tau 
+4\p_{(A}\left(\B -\f43  \textsf E_\B \right) \p_{B)} \tau \cr
& -2\left( 
\f43 \p_C \B\p^C \tau
-\f43\p_C\textsf E_{\B}\p^C \tau
+\left(\B-\f43\textsf E_{\B}\right)   \Delta \tau
+\f13 \bP^C\p_C\tau
+\f13C^{CD}\bU_C\p_D\tau
\right)\bq_{AB}\cr
&+\f1{2} C^{CD}\p_C \tau\bD_D C_{AB}
+C_{C(A} \bD_{B)} (C^{CD}\p_D \tau)
 -\f12 C_{AB} \left( {\bD}_CC^{CD}\p_D \tau +\f12C^{CD}{\bD}_C \p_D \tau \right) \cr
 &-\f12 C_{CD}C^{CD} \bD_{(A} \p_{B)}\tau  \underbrace{-\f14 \bq_{AB} \p^C \tau \p_C (C_{DE}C^{DE})}_{= -8 \bq_{AB}  \p_C  \textsf E_\B \p^C \tau +8 \bq_{AB}  \p_C  \B \p^C\tau }\cr
 &+\frac{4 }{ 3 } \left(\bP_{(A}-\f12  C_{C(A}\bU^C +\pa_{(A}\B \right)  \p_{B)} \tau\cr
&=4 \B \bD_{(A} \p_{B)} \tau -2\bq_{AB} \B  \Delta  \tau - \f14 C_{AB} C^{CD}{\bD}_C \p_D\t
\cr
&+\frac{4 }{ 3 } \left(\bP_{(A}-\f12  C_{C(A}\bU^C -8\pa_{(A}\B \right)  \p_{B)} \tau
-\f 23\bq_{AB} \left(  \bP^D
-\f12 C^{CD}\bU_C -8  \p^D \B
 \right)\p_D\tau\cr
&+\f1{2}\left( 
C^{CD}\bD_C C_{AB}
+C_{CA} \bD_{B} C^{CD} 
+C_{CB} \bD_{A} C^{CD} 
- C_{AB} {\bD}_CC^{CD}
\right)\p_D \tau\cr
& +16 \pa_{(A}\B \pa_{B)} \tau - q_{AB} C^{CD}\bU_C\pa_D\tau 
\cr
&-\f{16}3 \textsf E_\B \bD_{\langle A} \p_{B\rangle } \tau -\f{16}3 \p_{\langle A}  \textsf E_\B\p_{B\rangle } \tau\,.
\ee


On-shell of the asymptotic Einstein's equation $ \textsf E_U^A\,\hat{=}\,0$, we thus have
\be
\Delta_\t E_{AB} &\,\hat{=}\,
4 \B \bD_{\langle A} \p_{B\rangle } \tau  - \f14 C_{AB} C^{CD}{\bD}_C \p_D\tau
\cr
&+\frac{4 }{ 3 } \left(\bP_{\langle A}-\f12  \bU^CC_{C\langle A} -8\pa_{\langle A}\B \right)  \p_{B\rangle } \tau \cr
&+\f1{2}\left( 
C^{CD}\bD_C C_{AB}
- C_{AB} {\bD}_CC^{CD}
\right)\p_D \tau
  \cr
& + C_{C\langle A} \bD_{B \rangle } C^{CD}  \p_D \tau
+ 16 \pa_{\langle A}\B \pa_{B\rangle} \tau \cr
&-\f{16}3 \textsf E_\B \bD_{\langle A} \p_{B\rangle } \tau -\f{16}3 \p_{\langle A}  \textsf E_\B\p_{B\rangle } \tau\,.
\ee
The last line can be simplified since
\be
C_{C\langle A} \bD_{B \rangle } C^{CD}   \p_D \tau+  16 \pa_{\langle A}\B \pa_{B\rangle} \tau 
&=\bD_{ \langle A} ( C_{ B \rangle C } C^{CD} )  \p_D \tau-
\bD_{\langle A} C_{B \rangle  C}  C^{CD}  \p_D \tau +  16 \pa_{\langle A}\B \pa_{B\rangle} \tau 
\cr
&= 16 \p_{ \langle A}  \textsf E_\B  \p_{B \rangle} \tau -
\bD_{\langle A} C_{B \rangle  C}  C^{CD}  \p_D \tau
\,.
\ee
This means that the spin 2 anomaly is given, on-shell of $\textsf E_U^A\,\hat{=}\, 0$, by 
\be
\Delta_\t E_{AB} &\,\hat{=}\, 
\frac{4 }{ 3 } \left( \bP_{\langle A}-\f12 \bU^C  C_{C  \langle A} -8\pa_{ \langle A}\B\right) \p_{B\rangle } \tau 
\cr
&+\f1{2}\left( 
C^{CD}\bD_C C_{AB}
- C_{AB} {\bD}_CC^{CD}
\right)\p_D \tau  - \bD_{\langle A} C_{B \rangle  C}  C^{CD}  \p_D \tau\cr
&+ 4 \B \bD_{\langle A} \p_{B\rangle } \tau  - \f14 C_{AB} C^{CD}{\bD}_C \p_D\tau\cr
&-\f{16}3\textsf E_\B  \bD_{\langle A} \p_{B\rangle} \tau+\f{32}3\p_{ \langle A}  \textsf E_\B  \p_{B \rangle} \tau\,.
\ee
We could use the  definition of the covariant momentum
$\cP_A= \bP_{A}-\f12  C_{C  A}\bU^C +\f1{16}\pa_{ A}(C_{BC}C^{BC})$ to rewrite
\be
\Delta_\t E_{AB} &\,\hat{=}\,
\f 43 \cP_{\langle A}  \p_{B \rangle} \tau 
\cr
&+\f1{2}\left( 
C^{CD}\bD_C C_{AB}
- C_{AB} {\bD}_CC^{CD}
\right)\p_D \tau\cr
&+\f14\pa_{ \langle A}(C_{CD}C^{CD}) \p_{B\rangle } \tau
 - \bD_{\langle A} C_{B \rangle  C}  C^{CD}  \p_D \tau\cr
&-\f18 C_{CD}C^{CD} \bD_{\langle A} \p_{B\rangle } \tau  - \f14 C_{AB} C^{CD}{\bD}_C \p_D\tau\cr
&-\f{4}3\textsf E_\B  \bD_{\langle A} \p_{B\rangle} \tau\,.\la{EABano-off}
\ee


\section{Variations}\la{sec:Variations}
In this section we compute the behavior  under symmetry transformation of different quantities.

\begin{itemize}
\item{\bf Connection}

One establishes that 
\bea
\d_{(\tau,Y)} \Gamma_{AB}^C &=& \frac12 q^{CD} \left(D_A\d_{(\tau,Y)} q_{BD} + D_B \d_{(\tau,Y)}q_{AD} -  D_D \d_{(\tau,Y)}q_{AB}  \right)\cr
&=& \frac12 q^{CD} \left(D_AD_B Y_D   
+ D_B D_A Y_D + [D_A,D_D ]Y_B+ [D_B,D_D]Y_A  \right)\cr
&-& q^{CD} \left(D_A\dot{\tau} q_{BD} + D_B\dot{\tau} q_{AD} -  D_D\dot{\tau} q_{AB}  \right)\cr
&=&  D_{(A}D_{B)} Y^C   
+\f12  \left(R_{BDA}{}^C+ R_{ADB}{}^C  \right) Y^D -  2 D_{\langle A} \dot{\tau} \delta^C_{B\rangle} \,,
\eea
where we used that $[D_A,D_B] V_C= R_{ABC}{}^D V_D$ and $[D_A,D_B] V^C=R^{C}{}_{DAB}V^D$.
This means that the  contribution to  the anomaly due to $\tau$ is due to the presence of Weyl rescaling and  given by 
\be
\Delta_\tau   \Gamma_{AB}^C  &= -D_A\dot{\tau} \delta^C_B - D_B\dot{\tau} \delta^C_A + D^C\dot{\tau} q_{AB}\,,\la{DG1}\\
\Delta_\tau   \Gamma_{AB}^A  &= -2 D_B\dot{\tau}\,.\la{DG2}
\ee
Given a vectorial section $V^A$  of scale $s$ it can be checked that  the anomaly only depends on $\tau$
\be
\d_{(\tau,Y)} D_A V^C &= D_{A} \d_{(\tau,Y)}{V}^C + \d_{(\tau,Y)} \Gamma_{AB}^C  V^B \cr
&=\left[\tau \pa_u +\cL_Y + s \dot{\tau} \right]  (D_A V^C) + \Delta_\tau (D_A V^C)\,.
\ee

To evaluate the anomaly one establishes that 
\bea
\delta_\tau D_{A}{V}^C &=&
D_{A} \d_{\tau}{V}^C + (\d_{\tau} \Gamma_{AB}^C)  V^B\cr
&=&  D_{A}[\tau \dot{V}^C + s \dot{\tau}{V}^C] - \left( D_A\dot{\tau} \delta^C_B + D_B\dot{\tau} \delta^C_A -  D^C\dot{\tau} q_{AB}  \right) V^B \cr
&=&[\tau  D_{A}\dot{V}^C + s \dot{\tau}{ D_{A}V}^C] \cr
&+&   D_{A}\tau \dot{V}^C + s (D_{A}\dot{\tau}) {V}^C -  \left( D_A\dot{\tau} V^C  +  \delta^C_A V^B D_B\dot{\tau}  -  D^C\dot{\tau} V_A   \right)\,,
\eea
which means that even if $V^A$ is a section of weight $s$ its spatial derivative contains an anomaly given by 
\bea 
\Delta_\tau (D_{A}{V}^C)&=&   D_{A}\tau \dot{V}^C + s (D_{A}\dot{\tau}) {V}^C +\Delta_\tau\Gamma_{AB}^C V^B \cr
 &=&D_{A}\tau \dot{V}^C + s (D_{A}\dot{\tau}) {V}^C -  \left( D_A\dot{\tau} V^C  +   \delta^C_A V^B D_B\dot{\tau}  -  D^C\dot{\tau} V_A   \right)\,.
\eea
Similarly, the anomaly for the derivative of a form  of weight $s$ is 
\bea\label{Anomaly}
\Delta_\tau (D_{A}V_B) &=&
 D_{A}\tau \dot{V}_B + s (D_{A}\dot{\tau}) {V}_B +  \left( D_A\dot{\tau} V_B  +  V_A D_B\dot{\tau}  -  V^CD_C\dot{\tau} q_{AB}   \right).
\eea

\item{\bf $D\cdot C$ vector}

One evaluates 
\bea
\Delta_\tau (D_B C^{CB}) &=& N^{AB} \pa_B \tau + 3 C^{CB} \pa_B \dot{\tau} +
D_B \Delta_\tau C^{CB} + 
\Delta_\tau\Gamma_{B A}^C C^{AB} + 
\Delta_\tau \Gamma_{B A}^B C^{AC}\cr
&=&N^{AB} \pa_B \tau + 3 C^{CB} \pa_B \dot{\tau} -2 
D_B D^{\langle B}\pa^{C\rangle} \tau 
 - 4  C^{AC} \pa_A\dot\tau.\cr
&=&N^{CB} \pa_B \tau - C^{CB} \pa_B \dot{\tau} - 
2 \Delta D^{C} \tau
+ D^C  \Delta \tau\cr
&=&N^{CB} \pa_B \tau - C^{CB} \pa_B \dot{\tau} - 
R(q)  D^{C} \tau
- D^C  \Delta \tau\,,
\eea
where we used that 
\be
[\Delta, D_{C}] \tau
=
[D^B, D_C] D^B \tau = R^B{}_{DBC} D^D\tau= \f12  R(q) D_C \tau.
\ee

\item{\bf $\cJ^A$ vector}

Given the definition \eqref{cJ} of the vector  $\cJ^A$,
and by means of \eqref{dDC} and 
\be 
\delta_{(\tau,Y)} (\pa^A \bF)&=\bq^{AC}\delta_{(\tau,Y)} (\pa_C \bF)
+ \pa_C \bF \delta_{(\tau,Y)} \bq^{AC}
\cr
&=\bq^{AC}\delta_{(\tau,Y)} (\pa_C \bF)
- (\bD^A Y^C+\bD^C Y^A) \pa_C \bF
+2\dot\tau \pa^A \bF\cr
&=\pa^A (\delta_{(\tau,Y)}  \bF)
- (\bD^A Y^C+\bD^C Y^A) \pa_C \bF
+2\dot\tau \pa^A \bF\cr
&=\left[\tau \pa_u +\cL_Y + 4 \dot{\tau} \right] (\pa^A \bF) 
+ 2 \bF \pa^A  \dot{\tau} 
+ \f12 \pa^A  \Delta \dot{\tau}\,,
\ee
we compute
\be
 \d_{(\tau,Y)} \cJ^A &=
\left[\tau \pa_u +\cL_Y + 4\dot{\tau} \right] \cJ^A +
\f12 \dot{N}^{AB} \pa_B\tau\,.
\ee

A similar calculation shows that
\be
\d_{(\tau,Y)} \tilde \cJ^A &= \left[\tau \pa_u +\cL_Y + 4\dot{\tau} \right]\tilde \cJ^A + 
 \f12 \dot{\tilde N}^{AB} \pa_B\tau\,.
 \ee

Next, we can write
\be
\delta_{(\tau,Y)} (\bD_A \cJ^A )=
 \bD_A(\delta_{(\tau,Y)} \cJ^A ) +(\delta_{(\tau,Y)} \Gamma^A_{AB}) \cJ^B\,,
\ee
from which, by means of \eqref{DG2}, we get
\be
\delta_{(\tau,Y)} (\bD_A \cJ^A )=
 \left[\tau \pa_u +\cL_Y + 4\dot{\tau} \right] \bD_A \cJ^A
 +2\p_u (\cJ^A \bD_A \tau  ) 
+\f12 \dot{N}^{AB}  \bD_A \pa_B\tau\,,\la{DM}\nn\\
\ee
where we have used 
\be
  \f12\bD_A\dot{N}^{AB} = \dot \cJ^B 
\ee
and
\be
\bD_A ( \cL_Y \cJ^A)+\cJ^B \bD_B \bD_A Y^A = \cL_Y (\bD_A \cJ^A)\,.
\ee

We also have
\be
\delta_{(\t, Y)} (C_{AB} \cJ^B)&=
\delta_{(\t, Y)} C_{AB} \cJ^B+C_{AB}  \delta_{(\t, Y)} \cJ^B\cr
&=  \left[ \tau \pa_u  + \cL_Y +3 \dot{\tau} \right]  C_{AB} \cJ^B\cr
& - {2 \cJ^B \bD_{\langle A} \bD_{B\rangle} \tau }
+\f12C_{AB}  \dot{N}^{BC} \pa_C\tau\cr
&=  \left[ \tau \pa_u  + \cL_Y +3 \dot{\tau} \right]  C_{AB} \cJ^B\cr
& - {  \bD_C N^{BC}  \bD_{\langle A} \bD_{B\rangle} \tau 
-2 \pa^B \bF  \bD_{\langle A} \bD_{B\rangle} \tau 
}
+ \f12 C_{AB}  \dot{N}^{BC} \pa_C\tau
\cr
&=  \left[ \tau \pa_u  + \cL_Y +3 \dot{\tau} \right]  C_{AB} \cJ^B\cr
& + {  \f12 C_{AB}  \dot{N}^{BC} \pa_C\tau}
- \left( \bD_C N^{BC}  +
2 \pa^B \bF \right) \bD_{\langle  A} \p_{B\rangle} \tau 
\,.
\ee

\item{\bf Covariant mass $\cM$}

Given the covariant mass transformation \eqref{dcM}, we have
\be
\delta_{(\t, Y)} (\p_A{\cal M})&=\p_A(\delta_{(\t, Y)} {\cal M})\cr
&\,\hat{=}\, \left[ \tau \pa_u  + \cL_Y+3\dot \tau\right]\p_A {\cal M}\cr
&+\p_A\tau  \dot {\cal M}+3\p_A\dot \tau {\cal M}\cr
&+\bD_A \cJ^B \p_B\t
+ \cJ^B \bD_A \p_B\t
\,.
\la{dDM}
\ee

\item{\bf Derivative  $\bD_{[A} ( \bD\!\cdot\!C)_{B]}$}

From \eqref{dDC} we see that $( \bD\!\cdot\!C)_{B}$ is a generalized tensor of dimension $1$ and hence 
its anomaly is 
\be
\Delta_{\t}(\bD_{[A} ( \bD\!\cdot\!C)_{B]})) &= 
D_{[A}\tau ( \bD^C N_{B]C})  +  (D_{[A}\dot{\tau}) ( \bD^C C_{B]C})+\bD_{[A} (\Delta_{\t} \bD\!\cdot\!C)_{B]} \cr
&=D_{[A}\tau ( \bD^C N_{B]C})  +  (D_{[A}\dot{\tau}) ( \bD^C C_{B]C}) \cr
&+  
D_{[A}( N_{B]}{}^{C} \pa_C \tau - C_{B]}{}^{C} \pa_C \dot{\tau} - 
R(q)  D_{B]} \tau
- D_{B]} \Delta \t)\cr
&= D_{[A}\tau ( 2 \bD^C N_{B]C} + D_{B]} R )
- ( N_{[A}{}^{C} D_{B]}\pa_C \tau + C_{[B}{}^{C} D_{A]}\pa_C \dot{\tau} )\cr
&= 4 D_{[A}\tau \cJ_{B]} 
- 
 ( N_{[A}{}^{C} D_{B]}\pa_C \tau + C_{[B}{}^{C} D_{A]}\pa_C \dot{\tau} )\,.
\ee
Contracting this identity with $\epsilon^{AB}$ gives the identity
\be \la{DDDC}
\Delta_{\t}(\bD_{A} ( \bD\!\cdot\!\tilde{C})^{A}) &=4 \tilde{\cJ}^A \pa_A\tau - ( N^{AB} \tilde{D}_{A}\pa_B \tau + \tilde{C}^{AB} D_{A}\pa_B \dot{\tau} )\,.
\ee

\item{\bf Dual covariant mass $\tcM$}

From the definition \eqref{dualM} and the relation   \eqref{DDDC},
 we can derive
\be
\Delta_{\t}\tcM &= \f14 \Delta_{\t}(\bD_{A} ( \bD\!\cdot\!\tilde{C})^{A})  
+\f18\left( \Delta_{\t}C_{AB} \tilde{N}^{AB}
+C_{AB}  \Delta_{\t}\tilde{N}^{AB} \right)\cr
&= \tilde{\cJ}^A \pa_A\tau - \f14( N^{AB} \tilde{D}_{A}\pa_B \tau + \tilde{C}^{AB} D_{A}\pa_B \dot{\tau} \cr
&-\f14   D_{\langle A}\pa_{B\rangle} \tau  \tilde{N}^{AB}-\f14  C_{AB}  \tilde D_{\langle A}\pa_{B\rangle} \dot \tau
\cr
&= \tilde{\cJ}^A \pa_A\tau\,.
\ee

From this anomaly and the definition \eqref{cJ}, we can further compute
\be\la{dDcM}
\delta_{\tau,Y}  (\tilde\p_A\tilde {\cal M})&=
\tilde\p_A( \delta_{\tau,Y} \tilde {\cal M}) \cr
&=\left[ \tau \pa_u  + \cL_Y+3\dot \tau\right]\tilde\p_A {\tcM}\cr
&+\tilde\p_A\tau  \dot {\tcM}+3\tilde\p_A\dot \tau {\tcM}\cr
&+\tilde{D}_A \tcJ^B \pa_B\t
+ \tcJ^B\tilde{D}_A \pa_B\t
\,.
\ee

\item{\bf Covariant momentum $\cP_A$}

Given the covariant momentum transformation \eqref{dcP}, we want to compute the anomaly of the quantity $\bD_A \cP_B$. 
From the  relation
\bea
 \Delta_\t( \bD_A \cP_B)=
  \bD_A ( \Delta_\t \cP_B)
  -(  \Delta_\t \Gamma^D_{AB})  \cP_D
 \,,\la{DDP1}
 \eea
and the anomaly \eqref{DG1}, we see that the last term in \eqref{DDP1} yields the following contribution to the anomaly
\bea
 ( \d^D_B \p_A \dot \tau + \d^D_A \p_B \dot \tau
-\bq_{BA}  \p^D \dot \tau)\cP_D
&=&2\cP_{(A}\p_{B)} \dot \tau -\bq_{BA}  \cP_D \p^D \dot \tau\cr
&=&2\cP_{\langle A}\p_{B\rangle} \dot \tau\,.
\eea

Therefore, we have
\be
 \Delta_\t( \bD_A \cP_B)
&\,\doublehat{=}\, 
 \p_A\t \dot \cP_B
 +2 \p_A \dot\t\cP_B+2\cP_{\langle A}\p_{B\rangle} \dot \tau\cr
&+3  \bD_A  \tcM \tilde  \pa_B\tau +3\bD_A  {\cal M} \pa_B\tau
 +3  \tcM \bD_A\tilde \pa_B\tau + 3 {\cal M}  \bD_A\pa_B\tau
\,.
\la{DDP}
 \ee
 
 \item{\bf Moebius derivative}
 
 We now want to analyse the transformations of ``Moebius derivative operator''
 $[D_{\langle A} D_{B\rangle} + \frac{s}2 n_{AB}] \phi$ for a section of conformal weight $s$,
 and the transformation of $[D_{\langle A} D_{B\rangle} + \frac{s}2 n_{AB}] j^B$ for a vector of weight $s+1$.
 One starts with the computation of the conformal anomaly 
 \be
 \delta_W D_{\langle A} D_{B\rangle}  \phi &=
 {-} \delta_W\Gamma_{AB}^C D_C\phi +   D_{\langle A} \delta_W D_{B\rangle}  \phi \cr
 &= 2 D_{\langle A}W D_{B\rangle}\phi + s D_{\langle A} D_{B\rangle}(W \phi)\cr
 &= 2(s+1) D_{\langle A}W D_{B\rangle}\phi + s W D_{\langle A} D_{B\rangle} \phi
 + s \phi D_{\langle A} D_{B\rangle}W.
 \ee
 Combining this with the fact that $\delta_W n_{AB}= -2 D_{\langle A} D_{B\rangle}W$, we find that 
 \be
 \delta_W  \left[D_{\langle A} D_{B\rangle} + \frac{s}2 n_{AB}\right] \phi =
 sW \left[D_{\langle A} D_{B\rangle} + \frac{s}2 n_{AB}\right] \phi + 2(s+1)  D_{\langle A}W D_{B\rangle}\phi\,,
 \ee
 which shows that the Moebius combination $\left[D_{\langle A} D_{B\rangle} + \frac{s}2 n_{AB}\right]$ possesses no quadratic anomaly and that it is tensorial for sections of conformal weight $s=-1$.

Similarly, one evaluates
\be
 \delta_W D_{\langle A} D_{B\rangle} V^B &=
 \delta_W\Gamma_{AB}^C D_CV^B +   D_{\langle A}( \delta_W \Gamma_{B\rangle C}^B V^C) + 
  D_{\langle A} D_{B\rangle} \delta_W V^B \cr
 &= -2 D_{\langle A}W D_{B\rangle}V^B  -2 D_{\langle A}( D_{B\rangle}W V^B) )  
 + s D_{\langle A} D_{B\rangle}(W V^B)\cr
 &= 2(s-2) D_{\langle A}W D_{B\rangle}V^B + s W D_{\langle A} D_{B\rangle} \phi
 + (s-2) \phi D_{\langle A} D_{B\rangle}W.
 \ee
 Combining this with the fact that $\delta_W n_{AB}= -2 D_{\langle A} D_{B\rangle}W$, we find that 
 \be
 \delta_W  \left[D_{\langle A} D_{B\rangle} + \frac{(s-2)}2 n_{AB}\right] V^B =
 sW \left[D_{\langle A} D_{B\rangle} + \frac{(s-2)}2 n_{AB}\right] V^B 
 + 2(s-2)  D_{\langle A}W D_{B\rangle}V^B \,,
 \ee
which shows that the Moebius combination $\left[D_{\langle A} D_{B\rangle} + \frac{(s-2)}2 n_{AB}\right]V^B$ possesses no quadratic anomaly and that it is tensorial for vectorial sections of conformal weight $s=2$.

\end{itemize}

\section{Derivation of the momentum evolution equation}\la{App:Peom}

Given the metric parametrization \eqref{eq:FallOff}, the asymptotic Einstein's equation for the momentum $P_A$ is given by \cite{Barnich:2011mi, Compere:2018ylh, Freidel:2021yqe}
\be
 \dot \bP_A&=
 \bD_AM
 + \f 1{8}  \bD_A( C^{BC} N_{BC})\cr
  &+ C_{AB} \bD^B  \bF\cr
  &-\frac14  N^{CB}\bD_A C_{CB}
 \cr
 &-\f14\left( \bD_B D^B\bD^C C_{AC}- \bD_B D_A \bD_C C^{BC}
 \right)\cr
&-
     \f14 \bD_B  (N_{AC}C^{CB}) 
+ \frac14  \bD_{B} ( N^{CB}  C_{AC} )
 \,.
\ee

Recalling the definition \eqref{cP} of the covariant momentum,
we can write
\be
 \dot{\cal P}_A&=
 \bD_AM
 + \f 1{8}  \bD_A( C^{BC} N_{BC}) + C_{AB}  \bD^B  \bF \cr
  &-\frac14  N^{CB}\bD_A C_{CB} +{ \f14 N_{AB}( \bD\!\cdot\! C)^B + \f14 C_{AB}( \bD\!\cdot\! N)^B +\f1{8} \pa_{A}(C^{CB}N_{CB}) }
 \cr
 &-\f14 \bD^B \left(  D_B( \bD\!\cdot\! C)_{A}-  D_A (\bD\!\cdot\!C)_{B}
 \right)\cr
&-
     \f14 \bD^B  (N_{A}{}^CC_{CB}-  N_B{}^{C}  C_{CA} )
    \\
&=
 \bD_A
 \left( M
 + \f 1{8}  C^{BC} N_{BC}\right) + C_{AB} \left(\f12 D_B N_A{}^B + \bD^B  \bF\right) 
 + D^B J_{[AB]} \cr
  &
  {
  -\frac14  N^{CB}\bD_A C_{CB} +\f14 N_{AB}( \bD\!\cdot\! C)^B - \f14 C_{AB}( \bD\!\cdot\! N)^B + \f1{8} \pa_{A}(C^{CB}N_{CB})
} \cr
 &{
 -  \f18 \bD^B  (N_{A}{}^CC_{CB}-  N_B{}^{C}  C_{CA} )
 }
 \\
 &= D_A \cM + \tilde{D}_A\tcM  + C_{AB} \cJ^B
  \cr
  &
  { 
  -\frac14  N^{CB}\bD_A C_{CB} +\f14 N_{AB}( \bD\!\cdot\! C)^B - \f14 C_{AB}( \bD\!\cdot\! N)^B + \f1{8} \pa_{A}(C^{CB}N_{CB})
} \cr
 &{
 -  \f18 \bD^B  (N_{A}{}^CC_{CB}-  N_B{}^{C}  C_{CA} )
 }
 \la{Pdot1}
 \,,
\ee
where we have used the definitions \eqref{cJ},  \eqref{JABanti}, \eqref{dualM}.
The terms in the second and third lines of \eqref{Pdot1} can be expanded and  simplified as
\bea
&-&\frac14  N^{CB}\bD_A C_{CB} +\f14 N_{AB}( \bD\!\cdot\! C)^B - \f14 C_{AB}( \bD\!\cdot\! N)^B + \f1{8} \pa_{A}(C^{CB}N_{CB})
 \cr
 &-&
     \f18 (\bD_BN_{AC}C^{CB}-  N^{BC}  \bD_BC_{CA} ) 
     -\f18  (N_{AB} ( \bD\!\cdot\! C)^B-  ( \bD\!\cdot\! N)^B  C_{BA} )\cr
 &+&
     \f18 (N^{CB}\bD_B C_{CA} - C^{CB} \bD_BN_{AC})\cr
     &=& \f18 N_{AB}( \bD\!\cdot\! C)^B - \f18 C_{AB}( \bD\!\cdot\! N)^B 
     +  \f1{4}(   C^{CB} \bD_{[A}N_{B]C}-N^{CB}\bD_{[A} C_{B]C} ).\la{mess2}
\eea
Now 
we use that 
\be
\epsilon_{AB} \epsilon^{CD}= \delta _A^C \delta_B^D -\delta_A^D\delta_B^C 
\ee
to massage terms like 
\be
C^{CB} \bD_{[A}N_{B]C}= \f12 \epsilon_{AB} C^{BC}  ( \bD\!\cdot\! \tilde{N})_C 
= \f12 \tilde{C}_{AB} ( \bD\!\cdot\! \tilde{N})^B\,,
\ee
where $\tilde{N}_{BC}= \epsilon_B{}^A N_{AC}$.
This means that the   contributions in the second and third lines of \eqref{Pdot1} can be written as
\be
\f18 (N_{AB}( \bD\!\cdot\! C)^B - \tilde{N}_{AB} ( \bD\!\cdot\! \tilde{C})^B]
 - \f18 [C_{AB}( \bD\!\cdot\! N)^B - \tilde{C}_{AB} ( \bD\!\cdot\! \tilde{N})^B]\,.
\ee
Finally, 
since $N$ is symmetric and traceless, we can derive the relation
\be
N_{AB}( \bD\!\cdot\! C)^B = \tilde{N}_{AB} ( \bD\!\cdot\! \tilde{C})^B\,,
\ee
and similarly
\be
C_{AB}( \bD\!\cdot\! N)^B = \tilde{C}_{AB} ( \bD\!\cdot\! \tilde{N})^B\,.
\ee

We thus arrive at the sought after expression
\be
 \dot{\cal P}_A&=D_A \cM + \tilde{D}_A\tcM  + C_{AB} \cJ^B 
 \,.
\ee
Note that, since $C_{AB} \cJ^B = \tilde C_{AB} \tilde\cJ^B$, we can write the momentum evolution equation in a completely self-dual manner as
\be
\dot{\cal P}_A&=D_A \cM + \tilde{D}_A\tcM  + \f12\left( C_{AB} \cJ^B +\tC_{AB} \tcJ^B \right)   
  \,.
\ee

\section{Stress-energy tensor}\la{sec:SET}


\subsection{SET anomaly proof}\la{App:SETano}

 We give here the explicit derivation of the transformation properties of all the SET components \eqref{SETtrans}. 
From the analysis of Appendix \ref{App:action}, we can see immediately that the component $\hat T_{uu}$ transforms as a scalar  of weight $4$, since
\be
\d_{\bar{\xi}} T_{uu}= \tau \dot{T}_{uu} + \cL_YT_{uu} +\dot{\tau}(2- r\pa_r) T_{uu} .
\ee
The anomaly is given by 
\be
\Delta_\tau T_{uu} &=
y^A \p_A T_{uu}
+\rho\,r \p_r T_{uu}
+ 2T_{uA} \p_u y^A
+ 2 r T_{ur} \dot\rho 
=o(r^{-2})\,,
\ee
which implies 
\be
\Delta_\tau \hat T_{uu} =0\,.
\ee


The component $\hat T_{uA}$ transforms as a vector  of weight $3$ since from
\be
 {\cal L}_{\xi} T_{Au}= \xi^\nu \p_\nu T_{Au} +T_{Au} \p_u  \xi^u+T_{Ar} \p_u  \xi^r+T_{AB} \p_u  \xi^B
 + T_{uu} \p_A \xi^u+ T_{ur} \p_A \xi^r+ T_{uB} \p_A \xi^B
\ee
we have that
\be
\d_{\bar{\xi}} T_{Au}= \tau \dot{T}_{Au} + \cL_YT_{Au} +\dot{\tau}(1- r\pa_r) T_{Au}\,.
\ee
The anomaly is given by
\be
\Delta_\tau T_{Au}&= 
y^B\bD_B T_{Au}
+\rho\,r \p_r T_{Au}
+T_{Ar} \p_u  \xi^r
+ T_{A B} \pa_u y^B
+ T_{u u} \pa_A \tau
+ T_{u r} \pa_A \xi^r
+T_{uB}\p_A y^B
\cr
&=-\f1{r^2} \hat T  \p_A \dot \t
+\f1{r^2} \hat T_{uu} \pa_A \tau
+o(r^{-2})\,,
\ee
from which 
\be
\Delta_\tau \hat T_{Au} =- \hat T  \p_A \dot \t  +\hat T_{uu} \pa_A \tau\,.
\ee

From the Lie derivative of the $T_{AB}$ component 
\be
\cL_\xi T_{AB}
&=\xi^u \p_u T_{AB}
+\xi^C \p_C T_{AB}
+\xi^r \p_r T_{AB}
+ 2 T_{u(A } \p_{B)} \xi^u
+ 2 T_{r(A } \p_{B)} \xi^r
+ 2 T_{C(A} \p_{B)} \xi^C\,,
\ee
it is immediate to see that $\hat T$ transforms as a scalar  of weight $3$, since
\be
\d_{\bar{\xi}} T_{AB}= \tau \dot{T}_{AB} + \cL_Y T_{AB}  -\dot{\tau} r\pa_r  T_{AB}\,,
\ee
and $q_{AB}$ has conformal dimension $-2$, while $\hat T_{ AB } $ transforms as a tensor of weight 2.
 It is also easy to see that $\hat T$ has no anomaly, since both  the leading terms in  $T_{AB}$ and $q_{AB}$ have no anomaly. 
 At the same time, the anomaly of  the $r^{-2}$ component of $T_{AB}$ given by $\hat T_{AB}:= \hat T_2 q_{AB}+\hat T_{\langle AB \rangle} $ can be read off of
 \be
 \Delta_\tau  T_{ AB }&=
  y^C \p_C T_{AB}
+r\rho \p_r T_{AB}
+ 2 T_{u(A } \p_{B)} \t
-2 T_{r(A } \p_{B)}( \dot \t r)
+ 2 T_{C(A} \p_{B)} y^C
 \ee
 and it is given by
 \be
  \Delta_\tau  \hat T_{ AB }&=
  -q_{AB} \p^C \t  \p_C \hat T
  -\f12 q_{AB} \hat T  \Delta \t
  + 2 \hat T_{u(A } \p_{B)} \t
  -2 \hat T_{r(A } \p_{B)}\dot \t 
-  2 \hat T  D_{(A} \p_{B)} \t\cr
&=-q_{AB} \p^C \t  \p_C \hat T
  -\f32 q_{AB} \hat T  \Delta \t
   +q_{AB} \hat T_{u C} \p^C \t 
   -q_{AB} \hat T_{r C } \p_{C}\dot \t 
  \cr
&  + 2 \hat T_{u\langle A } \p_{B\rangle} \t 
  -2 \hat T_{r\langle A } \p_{B\rangle}\dot \t 
-  2 \hat T  D_{\langle A} \p_{B\rangle} \t\,.
 \ee
 This  means that 
 \be
\Delta_\tau  \hat T_2&=  \p_C \hat T \p^C \t 
  -\f32\hat T  \Delta \t
   + \hat T_{u C} \p^C \t 
   -\hat T_{r C } \p_{C}\dot \t 
 \ee
 and
 \be\la{DTAB}
 \Delta_\tau  \hat T_{\langle A B\rangle }&=2 \hat T_{u\langle A } \p_{B\rangle} \t 
  -2 \hat T_{r\langle A } \p_{B\rangle}\dot \t 
-  2 \hat T  D_{\langle A} \p_{B\rangle} \t\,.
 \ee

The component $\hat T_{rr}$   transforms as a vector  of weight $2$ since
from
\be
 {\cal L}_{\xi} T_{rr}&= 
 \xi^\nu \p_\nu T_{rr}
+ 2T_{r u} \pa_r \xi^u
+2 T_{r r} \pa_r \xi^r
+2 T_{r A} \pa_r \xi^A 
\ee
we have that
\be
\cL_{\bar{\xi}} T_{rr}= \tau \dot{T}_{rr} + \cL_Y T_{rr} -\dot{\tau}(2+r\pa_r) T_{rr}\,.
\ee
The anomaly is given by
\be
\Delta_\tau T_{rr}&= 
y^A\bD_A T_{rr}
+\rho\,r \p_r T_{rr}
+2T_{rr} \p_r(r\rho)
+2 T_{r A} \pa_r y^A\,,
\ee
from which
\be
\Delta_\tau \hat T_{rr} =0\,.
\ee

The component $T_{rA}$   transforms as a vector  of weight 2
since from
\be
 {\cal L}_{\xi} T_{rA}&= 
 \xi^\nu \p_\nu T_{rA}
+ T_{r u} \pa_A \xi^u
+ T_{r r} \pa_A \xi^r
+ T_{r B} \pa_A \xi^B
+ T_{A r} \pa_r \xi^r
+ T_{AB} \pa_r \xi^B
\ee
we have that
\be
\cL_{\bar{\xi}} T_{rA}=\tau \dot{T}_{rA} + \cL_Y T_{rA} -\dot{\tau}(1+r\pa_r) T_{rA}\,.
\ee
The anomaly is given by
\be
\Delta_\tau T_{rA}&= 
y^A\bD_A T_{rA}
+\rho\,r \p_r T_{rA}
+ T_{r u} \pa_A \t
+T_{r r} \pa_A \xi^r
+ T_{r B} \pa_A y^B
+ T_{A r} \pa_r (r\rho)
+ T_{AB} \pa_r y^B \,,
\ee
from which
\be
\Delta_\tau \hat T_{rA} =-\hat T_{rr} \p_A \dot\t +\hat T \p_A \t\,.
\ee
Notice that this anomaly is consistent with the  conservation equations  $\cC, \cC^2_A$ in \eqref{ConsEq} as 
\be
\Delta_\tau  \pa_A \hat{T}_{rr} =  \pa_u  \hat T_{rr} \pa_A\tau+2 \hat T_{rr} \pa_A \dot{\tau}= 
-2 \hat T\pa_A\tau+2 \hat T_{rr} \pa_A \dot{\tau}\,.
\ee

It also follows that the vector $\dot{\hat T}_{rA}$ transforms with weight 3 and anomaly
\be
\Delta_\tau \dot{\hat T}_{rA}&= -\dot{\hat T}_{rr} \p_A \dot\t +\dot{\hat T} \p_A \t +\hat T \p_A\dot \t\cr
&=\dot{\hat T} \p_A \t +3 \hat T \p_A\dot \t\,,
\ee
consistently with the conservation equation  $\cC^1_A$ in \eqref{ConsEq} as
\be
\Delta_\tau \p_A \hat T=
 \pa_u \hat T \pa_A \tau+3\hat T \p_A  \dot{\tau} \,.
\ee

%

\subsection{Conservation equations proof}\la{App:ConsEq}

Here we derive the SET conservation equations \eqref{ConsEq}. These follow  
 from 
 \be
 \nabla^\mu T_{\mu\nu}=g^{\mu\rho}\left(\p_\rho T_{\m\n} - \Gamma^\sigma_{\m\rho} T_{\sigma\n} - \Gamma^\sigma_{\n\rho} T_{\sigma\m}\right)=0\,,
 \ee
 and  the inverse metric components given by
 \begin{subequations}
 \be
g^{uu}&= 0\,,\\
g^{ur}&=-e^{-2\beta}\,,\\
g^{rr}&=2\Phi e^{-2\beta}=2\bF -\f{2 M} r +o(r^{-1})\,\\
g^{Au}&= 0\,,\\
g^{Ar}&= -e^{-2\beta} \f{\Upsilon^A}{r^2}=-\f {\bU^A}{r^2}+o(r^{-2})\,,\\
g^{AB}&= \f 1{r^2} \gamma^{AB} =  \f 1{r^2} \bq^{AB}- \f 1{r^3} C^{AB} +o(r)\,.
 \ee
  \end{subequations}

Let us consider first the component $\n=r$.  We have 
 \be
 0&=g^{u\rho}\left(\p_\rho T_{u r} - \Gamma^\sigma_{u\rho} T_{\sigma r}  - \Gamma^\sigma_{r \rho} T_{\sigma u}\right)\cr
 &+g^{r \rho}\left(\p_\rho T_{r r} - \Gamma^\sigma_{r\rho} T_{\sigma r}-   \Gamma^\sigma_{r \rho} T_{\sigma r}\right)\cr
 &+g^{A \rho}\left(\p_\rho T_{A r} - \Gamma^\sigma_{A\rho} T_{\sigma r} - \Gamma^\sigma_{r \rho} T_{\sigma A}\right)\cr
  &=\f1{r^4}\left[
  -\p_u \hat T_{r r}-q^{AB} \hat T q_{AB} 
  \right]+o(r^{-4})\,,
 \ee
 where only the spin connection component $ \Gamma^C_{r B}=\f1r \d^C_B$ contributes at the leading order. We thus 
 obtain the conservation equation
 \be\la{Ceom}
\cC:= \pa_u \hat T_{r r}+2\hat T=0\,.
 \ee

Next, we consider the component $\n=u$ and we obtain
 \be
 0&=g^{u\rho}\left(\p_\rho T_{u u} - \Gamma^\sigma_{u\rho} T_{\sigma u}  - \Gamma^\sigma_{u \rho} T_{\sigma u}\right)\cr
 &+g^{r \rho}\left(\p_\rho T_{r u} - \Gamma^\sigma_{r\rho} T_{\sigma u}-   \Gamma^\sigma_{u\rho} T_{\sigma r}\right)\cr
 &+g^{A \rho}\left(\p_\rho T_{A u} - \Gamma^\sigma_{A\rho} T_{\sigma u}  - \Gamma^\sigma_{u \rho} T_{\sigma A}\right)\cr
 &=\f1{r^3}\left[2 \hat T_{uu}-2 \hat T_{uu}
 \right]+o(r^{-3})\,,
 \ee
 where   only the spin connection component $ \Gamma^u_{AB}=r q_{AB}$ contributes at the leading order.
We thus see that this component yields a trivial relation. 

Finally, the component $\n=A$ yields
 \be
 0&=g^{u\rho}\left(\p_\rho T_{u A} - \Gamma^\sigma_{u\rho} T_{\sigma A}  - \Gamma^\sigma_{A \rho} T_{\sigma u}\right)\cr
 &+g^{r \rho}\left(\p_\rho T_{r A} - \Gamma^\sigma_{r\rho} T_{\sigma A}-   \Gamma^\sigma_{A \rho} T_{\sigma r}\right)\cr
 &+g^{B\rho}\left(\p_\rho T_{B A} - \Gamma^\sigma_{B\rho} T_{\sigma A} - \Gamma^\sigma_{A \rho} T_{\sigma B}\right)\cr
 &=\f1{r^3}\left[2\hat  T_{u A}  -\p_u \hat T_{rA} +q^{BC}D_C\hat T_{AB}-2\hat  T_{u A}
 \right]+o(r^{-3})\,,
\ee
where again only the spin connection component $ \Gamma^u_{AB}=r q_{AB}$ contributes at the leading order.
We thus 
 obtain the conservation equation
 \be\la{C1eom}
\cC^1_A:= \p_u \hat T_{rA}-\p_A\hat T=0\,.
 \ee

Combining the two conservation equations \eqref{Ceom}, \eqref{C1eom} one gets the third conservation equation
\be
\cC^2_A:=  \pa_A \hat{T}_{rr}+ 2\hat{T}_{rA}=0\,.
\ee

\subsection{Sources}\la{App:sources}

Let us derive the explicit expressions \eqref{sources} for the matter sources by applying our symmetry argument. 
We start with the covariant mass and momentum equations. By inspection of the conformal weights under the BMSW group action, we can consider the general ansatz 
\be
\cS:=  \f12 \left( \hat T_{uu}+\beta \dot{\hat T}\right)\,  ,
\qquad 
\cS_A:= \hat T_{uA} +\alpha \pa_A \hat T\,,
\ee
with $\alpha, \beta$ two free numerical coefficients to be determined. It follows that
\be 
\delta_{(\tau,Y)}\cS_A= [\tau \pa_u+ {\cL}_Y +3\dot{\tau}] \cS_A +\hat T_{uu} \pa_A \tau+\alpha\dot{ \hat T} \p_A \t +(3\alpha-1) \hat T \pa_A \dot \tau\,.
\ee

Next, given the definition \eqref{cP} of the covariant momentum and  the transformation \eqref{dP}, one gets that
\be
\d_{(\t,Y)} {\cal P}_A\,\hat{=}\,  [\tau\pa_u + \cL_{Y} + 2 \dot\tau  ] {\cal P}_A 
 + 3 \left({\cal M}\pa_A\tau  +   \tcM \tilde\pa_A\tau\right) 
-2\dot{ \textsf E}_{\B} \pa_A \tau\,.\la{dcP-off}
\ee
Using that in the presence of matter we have $4\dot{\textsf E}_\B =\hat{T}$, this means 
\be
\Delta_\tau {\cal E}_A&\,\hat{=}\,  
-\f12 \dot {\hat T} \p_A\t -\f12 \hat T  \p_A\dot \t \,,
\ee
and
\be
\delta_{(\tau,Y)} \left(\cE_A+\cS_A\right)&= [\tau \pa_u+ {\cL}_Y +3\dot{\tau}] \left(\cE_A+\cS_A\right)+ 2\tilde\cE \tilde{\pa}_A\t\cr
&+2 \left(\cE+ \f12 \left(\hat T_{uu}+ \left(\alpha-\f12\right)\dot{\hat T}\right)\right) \pa_A\tau 
+\hat T \left(  3\alpha-\f32\right)\p_A\dot \t \,.
\ee
We thus see that we need $\alpha=1/2$ to remove the anomaly and then the sources read
\be
\cS:=  \f12 \hat T_{uu}\,  ,
\qquad 
\cS_A:= \hat T_{uA}+\f12 \pa_A \hat T\,,
\ee
so that
\be
\delta_{(\tau,Y)} \left(\cE_A+\cS_A\right)&= [\tau \pa_u+ {\cL}_Y +3\dot{\tau}] \left(\cE_A+\cS_A\right)
+2 \left(\cE+\cS\right)\pa_A\t
+ 2\tilde\cE \tilde{\pa}_A\t\,.
\ee


For the spin-2 asymptotic EE, compatibility of the conformal weights suggests that we start with the 
 ansatz
\be
\cS_{AB}:= \gamma \hat T_{\langle AB\rangle} +\beta D_{\langle A} \pa_{B\rangle} \hat T_{rr} +\zeta \hat T C_{AB}\,.
\ee
By means of  the transformations \eqref{DTAB} and
\be
\delta_{(\tau,Y)}D_{\langle A} \p_{B\rangle} \hat T_{rr} 
&=  [\tau \pa_u+ {\cL}_Y +2\dot{\tau}] D_{\langle A} \p_{B\rangle} \hat T_{rr}
+  2\pa_u D_{\langle A} T_{rr} \pa_{B\rangle} \tau +6  D_{\langle A}\hat T_{rr} \pa_{B\rangle} \dot{\tau}\cr
&+ \pa_uT_{rr} D_{\langle A} \p_{B\rangle} \tau + 2 \hat T_{rr} D_{\langle A} \p_{B\rangle} \dot{\tau},
\ee
where note that the coefficient $6$ above involves using \eqref{Trr} and a contribution from   $-\delta_{(\tau,Y)} \Gamma^C_{\langle AB\rangle} \p_{C} \hat T_{rr}$,
we have
\be
\delta_{(\tau,Y)} \cS_{AB}
     &= [\tau \pa_u+ {\cL}_Y +2\dot{\tau}]\cS_{AB}\cr
&+\left[  2 \gamma\hat T_{u\langle A }-4\beta D_{\langle A} \hat T \right] \p_{B\rangle} \t \cr
&+\left[   -2 \gamma\hat T_{r\langle A } + 6 \beta  D_{\langle A} \hat T_{rr} \right] \p_{B\rangle}\dot \t \cr
 &-2\left[     \gamma  +\beta+\zeta \right] \hat T D_{\langle A} \p_{B\rangle} \t \cr
  &+ 2 \beta\hat T_{rr}D_{\langle A} \p_{B\rangle} \dot\t\,.
\ee

 Next, we can use  the definition \eqref{Vireom} to compute the off-shell of the $\textsf E_b $ equation of motion anomaly
 \be
 \Delta_\t \cE_{AB}\,\hat{=}\,
 - {2}\dot {\textsf E}_\B \bD_{\langle A} \p_{B\rangle} \tau
 - {4} \textsf E_\B  \bD_{\langle A} \p_{B\rangle}\dot \tau
 +{2} \p_{\langle A} \dot {\textsf E}_\B  \p_{B\rangle} \tau\,,
 \ee
 which follows from \eqref{dcP-off}  and the covariant spin-2 pseudo-tensor transformation (see \eqref{EABano-off})
\be 
\d_{(\tau,Y)} \cT_{AB} &\,\hat {=}\, \left[\tau\pa_u + \cL_{Y} +  \dot\tau  \right] \cT_{AB}  +
{4} \cP_{\langle A}  \p_{B \rangle} \tau-{4} \textsf E_\B  \bD_{\langle A} \p_{B\rangle} \tau\,.
\ee 

By demanding
\be
\delta_{(\tau,Y)} (\cE_{AB}+\cS_{AB}) &= [\tau \pa_u+ {\cL}_Y +2\dot{\tau}] (\cE_{AB}+\cS_{AB})+{3} \left(\cE_{\langle A}+\cS_{\langle A}\right)\p_{B\rangle}\t\,,
\ee
we can  fix the coefficients $\beta, \gamma,\zeta$ from the conditions
\be
2\beta+{\f12}&=0\,,\\
-2(    \gamma  +\beta+\zeta)-{\f12}&=0\,,\\
-2 \gamma \hat T_{r A }+ 6 \beta  D_{ A} \hat T_{rr}  &=0\quad \rightarrow \quad  -2 \gamma=12  \beta \,,\\
2 \gamma\hat T_{u A }+\left({\f12}-4\beta\right) \p_{ A} \hat T&= {3}\left(\hat T_{uA} +\f12 \pa_A \hat T\right)\,.
\ee
It is immediate to see that the system of equations is solved by
{
\be
\beta=-\f1{4}\,,\quad \gamma=\f32\,,\quad \zeta=-\f32\,,
\ee
}
from which
\be
\cS_{AB}:= {\f32} \hat T_{\langle AB\rangle} - {\f1{4}} D_{\langle A} \pa_{B\rangle} \hat T_{rr} - {\f32} \hat T C_{AB}\,.\la{SAB}
\ee
\bibliographystyle{bib-style2}
\bibliography{biblio-fluxes}

\providecommand{\href}[2]{#2}\begingroup\raggedright\begin{thebibliography}{10}

\bibitem{DonnellyFreidel}
W.~Donnelly and L.~Freidel, \emph{{Local subsystems in gauge theory and
  gravity}}, \href{http://dx.doi.org/10.1007/JHEP09(2016)102}{\emph{JHEP}
  {\bfseries 09} (2016) 102},
  [\href{https://arxiv.org/abs/1601.04744}{{\ttfamily 1601.04744}}].

\bibitem{Freidel:2015gpa}
L.~Freidel and A.~Perez, \emph{{Quantum gravity at the corner}},
  \href{http://dx.doi.org/10.3390/universe4100107}{\emph{Universe} {\bfseries
  4} (2018) 107}, [\href{https://arxiv.org/abs/1507.02573}{{\ttfamily
  1507.02573}}].

\bibitem{Freidel:2016bxd}
L.~Freidel, A.~Perez and D.~Pranzetti, \emph{{Loop gravity string}},
  \href{http://dx.doi.org/10.1103/PhysRevD.95.106002}{\emph{Phys. Rev.}
  {\bfseries D95} (2017) 106002},
  [\href{https://arxiv.org/abs/1611.03668}{{\ttfamily 1611.03668}}].

\bibitem{Freidel:2018pvm}
L.~Freidel and E.~R. Livine, \emph{{Bubble networks: framed discrete geometry
  for quantum gravity}},
  \href{http://dx.doi.org/10.1007/s10714-018-2493-y}{\emph{Gen. Rel. Grav.}
  {\bfseries 51} (2019) 9}, [\href{https://arxiv.org/abs/1810.09364}{{\ttfamily
  1810.09364}}].

\bibitem{Freidel:2019ees}
L.~Freidel, E.~R. Livine and D.~Pranzetti, \emph{{Gravitational edge modes:
  from Kac-Moody charges to Poincar{\'e} networks}},
  \href{http://dx.doi.org/10.1088/1361-6382/ab40fe}{\emph{Class. Quant. Grav.}
  {\bfseries 36} (2019) 195014},
  [\href{https://arxiv.org/abs/1906.07876}{{\ttfamily 1906.07876}}].

\bibitem{Freidel:2019ofr}
L.~Freidel, E.~R. Livine and D.~Pranzetti, \emph{{Kinematical Gravitational
  Charge Algebra}},
  \href{http://dx.doi.org/10.1103/PhysRevD.101.024012}{\emph{Phys. Rev. D}
  {\bfseries 101} (2020) 024012},
  [\href{https://arxiv.org/abs/1910.05642}{{\ttfamily 1910.05642}}].

\bibitem{Freidel:2020xyx}
L.~Freidel, M.~Geiller and D.~Pranzetti, \emph{{Edge modes of gravity - I:
  Corner potentials and charges}},
  [\href{https://arxiv.org/abs/2006.12527}{{\ttfamily 2006.12527}}].

\bibitem{Freidel:2020svx}
L.~Freidel, M.~Geiller and D.~Pranzetti, \emph{{Edge modes of gravity - II:
  Corner metric and Lorentz charges}},
  [\href{https://arxiv.org/abs/2007.03563}{{\ttfamily 2007.03563}}].

\bibitem{Freidel:2020ayo}
L.~Freidel, M.~Geiller and D.~Pranzetti, \emph{{Edge modes of gravity. Part
  III. Corner simplicity constraints}},
  \href{http://dx.doi.org/10.1007/JHEP01(2021)100}{\emph{JHEP} {\bfseries 01}
  (2021) 100}, [\href{https://arxiv.org/abs/2007.12635}{{\ttfamily
  2007.12635}}].

\bibitem{Donnelly:2020xgu}
W.~Donnelly, L.~Freidel, S.~F. Moosavian and A.~J. Speranza,
  \emph{{Gravitational Edge Modes, Coadjoint Orbits, and Hydrodynamics}},
  [\href{https://arxiv.org/abs/2012.10367}{{\ttfamily 2012.10367}}].

\bibitem{Noether:1918zz}
E.~Noether, \emph{{Invariant Variation Problems}},
  \href{http://dx.doi.org/10.1080/00411457108231446}{\emph{Gott. Nachr.}
  {\bfseries 1918} (1918) 235--257},
  [\href{https://arxiv.org/abs/physics/0503066}{{\ttfamily physics/0503066}}].

\bibitem{Freidel:2021dxw}
L.~Freidel, \emph{{A canonical bracket for open gravitational system}},
  [\href{https://arxiv.org/abs/2111.14747}{{\ttfamily 2111.14747}}].

\bibitem{Ciambelli:2021nmv}
L.~Ciambelli, R.~G. Leigh and P.-C. Pai, \emph{{Embeddings and Integrable
  Charges for Extended Corner Symmetry}},
  [\href{https://arxiv.org/abs/2111.13181}{{\ttfamily 2111.13181}}].

\bibitem{Ciambelli:2021vnn}
L.~Ciambelli and R.~G. Leigh, \emph{{Isolated Surfaces and Symmetries of
  Gravity}},  [\href{https://arxiv.org/abs/2104.07643}{{\ttfamily
  2104.07643}}].

\bibitem{Freidel:2021cbc}
L.~Freidel, R.~Oliveri, D.~Pranzetti and S.~Speziale, \emph{{Extended corner
  symmetry, charge bracket and Einstein's equations}},
  [\href{https://arxiv.org/abs/2104.12881}{{\ttfamily 2104.12881}}].

\bibitem{Chandrasekaran:2018aop}
V.~Chandrasekaran, E.~E. Flanagan and K.~Prabhu, \emph{{Symmetries and charges
  of general relativity at null boundaries}},
  \href{http://dx.doi.org/10.1007/JHEP11(2018)125}{\emph{JHEP} {\bfseries 11}
  (2018) 125}, [\href{https://arxiv.org/abs/1807.11499}{{\ttfamily
  1807.11499}}].

\bibitem{Freidel:2021yqe}
L.~Freidel, R.~Oliveri, D.~Pranzetti and S.~Speziale, \emph{{The Weyl BMS group
  and Einstein's equations}},
  \href{http://dx.doi.org/10.1007/JHEP07(2021)170}{\emph{JHEP} {\bfseries 07}
  (4, 2021) 170}, [\href{https://arxiv.org/abs/2104.05793}{{\ttfamily
  2104.05793}}].

\bibitem{Bondi:1960jsa}
H.~Bondi, \emph{{Gravitational Waves in General Relativity}},
  \href{http://dx.doi.org/10.1038/186535a0}{\emph{Nature} {\bfseries 186}
  (1960) 535--535}.

\bibitem{BMS}
H.~Bondi, M.~G.~J. van~der Burg and A.~W.~K. Metzner, \emph{{Gravitational
  waves in general relativity. 7. Waves from axisymmetric isolated systems}},
  \href{http://dx.doi.org/10.1098/rspa.1962.0161}{\emph{Proc. Roy. Soc. Lond.}
  {\bfseries A269} (1962) 21--52}.

\bibitem{Sachs62}
R.~Sachs, \emph{{On the characteristic initial value problem in gravitational
  theory}}, {\emph{J.Math.Phys.} {\bfseries 3} (1962) 908--914}.

\bibitem{Barnich:2009se}
G.~Barnich and C.~Troessaert, \emph{{Symmetries of asymptotically flat 4
  dimensional spacetimes at null infinity revisited}},
  \href{http://dx.doi.org/10.1103/PhysRevLett.105.111103}{\emph{Phys. Rev.
  Lett.} {\bfseries 105} (2010) 111103},
  [\href{https://arxiv.org/abs/0909.2617}{{\ttfamily 0909.2617}}].

\bibitem{Barnich:2011mi}
G.~Barnich and C.~Troessaert, \emph{{BMS charge algebra}},
  \href{http://dx.doi.org/10.1007/JHEP12(2011)105}{\emph{JHEP} {\bfseries 12}
  (2011) 105}, [\href{https://arxiv.org/abs/1106.0213}{{\ttfamily 1106.0213}}].

\bibitem{Campiglia:2014yka}
M.~Campiglia and A.~Laddha, \emph{{Asymptotic symmetries and subleading soft
  graviton theorem}},
  \href{http://dx.doi.org/10.1103/PhysRevD.90.124028}{\emph{Phys. Rev. D}
  {\bfseries 90} (2014) 124028},
  [\href{https://arxiv.org/abs/1408.2228}{{\ttfamily 1408.2228}}].

\bibitem{Flanagan:2015pxa}
E.~E. Flanagan and D.~A. Nichols, \emph{{Conserved charges of the extended
  Bondi-Metzner-Sachs algebra}},
  \href{http://dx.doi.org/10.1103/PhysRevD.95.044002}{\emph{Phys. Rev. D}
  {\bfseries 95} (2017) 044002},
  [\href{https://arxiv.org/abs/1510.03386}{{\ttfamily 1510.03386}}].

\bibitem{Compere:2018ylh}
G.~Comp\`{e}re, A.~Fiorucci and R.~Ruzziconi, \emph{{Superboost transitions,
  refraction memory and super-Lorentz charge algebra}},
  \href{http://dx.doi.org/10.1007/JHEP11(2018)200}{\emph{JHEP} {\bfseries 11}
  (2018) 200}, [\href{https://arxiv.org/abs/1810.00377}{{\ttfamily
  1810.00377}}].

\bibitem{Banerjee:2020zlg}
S.~Banerjee, S.~Ghosh and P.~Paul, \emph{{MHV graviton scattering amplitudes
  and current algebra on the celestial sphere}},
  \href{http://dx.doi.org/10.1007/JHEP02(2021)176}{\emph{JHEP} {\bfseries 02}
  (2021) 176}, [\href{https://arxiv.org/abs/2008.04330}{{\ttfamily
  2008.04330}}].

\bibitem{Banerjee:2020vnt}
S.~Banerjee and S.~Ghosh, \emph{{MHV gluon scattering amplitudes from celestial
  current algebras}},
  \href{http://dx.doi.org/10.1007/JHEP10(2021)111}{\emph{JHEP} {\bfseries 10}
  (2021) 111}, [\href{https://arxiv.org/abs/2011.00017}{{\ttfamily
  2011.00017}}].

\bibitem{Banerjee:2021cly}
S.~Banerjee, S.~Ghosh and S.~S. Samal, \emph{{Subsubleading soft graviton
  symmetry and MHV graviton scattering amplitudes}},
  [\href{https://arxiv.org/abs/2104.02546}{{\ttfamily 2104.02546}}].

\bibitem{Banerjee:2021dlm}
S.~Banerjee, S.~Ghosh and P.~Paul, \emph{{(Chiral) Virasoro invariance of the
  tree-level MHV graviton scattering amplitudes}},
  [\href{https://arxiv.org/abs/2108.04262}{{\ttfamily 2108.04262}}].

\bibitem{Strominger:2017zoo}
A.~Strominger, \emph{{Lectures on the Infrared Structure of Gravity and Gauge
  Theory}},  [\href{https://arxiv.org/abs/1703.05448}{{\ttfamily 1703.05448}}].

\bibitem{Pasterski:2021rjz}
S.~Pasterski, \emph{{Lectures on Celestial Amplitudes}},
  [\href{https://arxiv.org/abs/2108.04801}{{\ttfamily 2108.04801}}].

\bibitem{Raclariu:2021zjz}
A.-M. Raclariu, \emph{{Lectures on Celestial Holography}},
  [\href{https://arxiv.org/abs/2107.02075}{{\ttfamily 2107.02075}}].

\bibitem{Ciambelli:2018ojf}
L.~Ciambelli and C.~Marteau, \emph{{Carrollian conservation laws and Ricci-flat
  gravity}}, \href{http://dx.doi.org/10.1088/1361-6382/ab0d37}{\emph{Class.
  Quant. Grav.} {\bfseries 36} (2019) 085004},
  [\href{https://arxiv.org/abs/1810.11037}{{\ttfamily 1810.11037}}].

\bibitem{NP62}
E.~Newman and R.~Penrose, \emph{An approach to gravitational radiation by a
  method of spin coefficients}, {\emph{J.Math.Phys.} {\bfseries 3} (1962)
  566--578}.

\bibitem{Newman:1962cia}
E.~T. Newman and T.~W.~J. Unti, \emph{{Behavior of Asymptotically Flat Empty
  Spaces}}, \href{http://dx.doi.org/10.1063/1.1724303}{\emph{J. Math. Phys.}
  {\bfseries 3} (1962) 891}.

\bibitem{Adamo:2009vu}
T.~M. Adamo, C.~N. Kozameh and E.~T. Newman, \emph{Null geodesic congruences,
  asymptotically flat space-times and their physical interpretation},
  \href{http://dx.doi.org/10.12942/lrr-2009-6}{\emph{Living Rev. Rel.}
  {\bfseries 12} (2009) 6}, [\href{https://arxiv.org/abs/0906.2155}{{\ttfamily
  0906.2155}}].

\bibitem{Godazgar:2018qpq}
H.~Godazgar, M.~Godazgar and C.~Pope, \emph{{New dual gravitational charges}},
  \href{http://dx.doi.org/10.1103/PhysRevD.99.024013}{\emph{Phys. Rev. D}
  {\bfseries 99} (2019) 024013},
  [\href{https://arxiv.org/abs/1812.01641}{{\ttfamily 1812.01641}}].

\bibitem{Godazgar:2018dvh}
H.~Godazgar, M.~Godazgar and C.~Pope, \emph{{Tower of subleading dual BMS
  charges}}, \href{http://dx.doi.org/10.1007/JHEP03(2019)057}{\emph{JHEP}
  {\bfseries 03} (2019) 057},
  [\href{https://arxiv.org/abs/1812.06935}{{\ttfamily 1812.06935}}].

\bibitem{Godazgar:2019dkh}
H.~Godazgar, M.~Godazgar and C.~Pope, \emph{{Dual gravitational charges and
  soft theorems}}, \href{http://dx.doi.org/10.1007/JHEP10(2019)123}{\emph{JHEP}
  {\bfseries 10} (2019) 123},
  [\href{https://arxiv.org/abs/1908.01164}{{\ttfamily 1908.01164}}].

\bibitem{Godazgar:2020kqd}
H.~Godazgar, M.~Godazgar and M.~J. Perry, \emph{{Hamiltonian derivation of dual
  gravitational charges}},
  \href{http://dx.doi.org/10.1007/JHEP09(2020)084}{\emph{JHEP} {\bfseries 20}
  (2020) 084}, [\href{https://arxiv.org/abs/2007.07144}{{\ttfamily
  2007.07144}}].

\bibitem{Godazgar:2020gqd}
H.~Godazgar, M.~Godazgar and M.~J. Perry, \emph{{Asymptotic gravitational
  charges}},
  \href{http://dx.doi.org/10.1103/PhysRevLett.125.101301}{\emph{Phys. Rev.
  Lett.} {\bfseries 125} (2020) 101301},
  [\href{https://arxiv.org/abs/2007.01257}{{\ttfamily 2007.01257}}].

\bibitem{Kol:2019nkc}
U.~Kol and M.~Porrati, \emph{{Properties of Dual Supertranslation Charges in
  Asymptotically Flat Spacetimes}},
  \href{http://dx.doi.org/10.1103/PhysRevD.100.046019}{\emph{Phys. Rev. D}
  {\bfseries 100} (2019) 046019},
  [\href{https://arxiv.org/abs/1907.00990}{{\ttfamily 1907.00990}}].

\bibitem{Kol:2020vet}
U.~Kol, \emph{{Subleading BMS Charges and The Lorentz Group}},
  [\href{https://arxiv.org/abs/2011.06008}{{\ttfamily 2011.06008}}].

\bibitem{Oliveri:2020xls}
R.~Oliveri and S.~Speziale, \emph{{A note on dual gravitational charges}},
  \href{http://dx.doi.org/10.1007/JHEP12(2020)079}{\emph{JHEP} {\bfseries 12}
  (2020) 079}, [\href{https://arxiv.org/abs/2010.01111}{{\ttfamily
  2010.01111}}].

\bibitem{Grant:2021sxk}
A.~M. Grant, K.~Prabhu and I.~Shehzad, \emph{{The Wald-Zoupas prescription for
  asymptotic charges at null infinity in general relativity}},
  [\href{https://arxiv.org/abs/2105.05919}{{\ttfamily 2105.05919}}].

\bibitem{Barnich:2011ty}
G.~Barnich and P.-H. Lambert, \emph{{A Note on the Newman-Unti group and the
  BMS charge algebra in terms of Newman-Penrose coefficients}},
  \href{http://dx.doi.org/10.1155/2012/197385}{\emph{Adv. Math. Phys.}
  {\bfseries 2012} (2012) 197385},
  [\href{https://arxiv.org/abs/1102.0589}{{\ttfamily 1102.0589}}].

\bibitem{Barnich:2016lyg}
G.~Barnich and C.~Troessaert, \emph{{Finite BMS transformations}},
  \href{http://dx.doi.org/10.1007/JHEP03(2016)167}{\emph{JHEP} {\bfseries 03}
  (2016) 167}, [\href{https://arxiv.org/abs/1601.04090}{{\ttfamily
  1601.04090}}].

\bibitem{Barnich:2019vzx}
G.~Barnich, P.~Mao and R.~Ruzziconi, \emph{{BMS current algebra in the context
  of the Newman\textendash{}Penrose formalism}},
  \href{http://dx.doi.org/10.1088/1361-6382/ab7c01}{\emph{Class. Quant. Grav.}
  {\bfseries 37} (2020) 095010},
  [\href{https://arxiv.org/abs/1910.14588}{{\ttfamily 1910.14588}}].

\bibitem{Barnich:2021dta}
G.~Barnich and R.~Ruzziconi, \emph{Coadjoint representation of the bms group on
  celestial riemann surfaces},
  [\href{https://arxiv.org/abs/2103.11253}{{\ttfamily 2103.11253}}].

\bibitem{Freidel:2021dfs}
L.~Freidel, D.~Pranzetti and A.-M. Raclariu, \emph{{Sub-subleading Soft
  Graviton Theorem from Asymptotic Einstein's Equations}},
  [\href{https://arxiv.org/abs/2111.15607}{{\ttfamily 2111.15607}}].

\bibitem{Freidel:2021ytz}
L.~Freidel, D.~Pranzetti and A.-M. Raclariu, \emph{{Higher spin dynamics in
  gravity and $w_{1 + \infty}$ celestial symmetries}},
  [\href{https://arxiv.org/abs/2112.15573}{{\ttfamily 2112.15573}}].

\bibitem{Aichelburg:1966su}
P.~C. Aichelburg and H.~Balasin, \emph{{Symmetries of impulsive gravitational
  waves}}, {\emph{Helv. Phys. Acta} {\bfseries 69} (1966) 337--340}.

\bibitem{Szekeres:1970vg}
P.~Szekeres, \emph{{Colliding gravitational waves}},
  \href{http://dx.doi.org/10.1038/2281183a0}{\emph{Nature} {\bfseries 228}
  (1970) 1183--1184}.

\bibitem{Khan:1971vh}
K.~A. Khan and R.~Penrose, \emph{{Scattering of two impulsive gravitational
  plane waves}}, \href{http://dx.doi.org/10.1038/229185a0}{\emph{Nature}
  {\bfseries 229} (1971) 185--186}.

\bibitem{Penrose-72}
R.~Penrose, \emph{The geometry of impulsive gravitational waves}, {\emph{Part
  of General relativity : Papers in honour of J.L. Synge,} (1972) 101--115}.

\bibitem{Hogan:1993xj}
P.~A. Hogan, \emph{{A Spherical impulse gravity wave}},
  \href{http://dx.doi.org/10.1103/PhysRevLett.70.117}{\emph{Phys. Rev. Lett.}
  {\bfseries 70} (1993) 117--118}.

\bibitem{Aliev:2000jp}
A.~N. Aliev and Y.~Nutku, \emph{{Impulsive spherical gravitational waves}},
  \href{http://dx.doi.org/10.1088/0264-9381/18/5/308}{\emph{Class. Quant.
  Grav.} {\bfseries 18} (2001) 891--906},
  [\href{https://arxiv.org/abs/gr-qc/0011016}{{\ttfamily gr-qc/0011016}}].

\bibitem{Podolsky:2002sa}
J.~Podolsky and R.~Steinbauer, \emph{{Geodesics in space-times with expanding
  impulsive gravitational waves}},
  \href{http://dx.doi.org/10.1103/PhysRevD.67.064013}{\emph{Phys. Rev. D}
  {\bfseries 67} (2003) 064013},
  [\href{https://arxiv.org/abs/gr-qc/0210007}{{\ttfamily gr-qc/0210007}}].

\bibitem{Luk:2012hi}
J.~Luk and I.~Rodnianski, \emph{{Local Propagation of Impulsive
  GravitationalWaves}},
  \href{http://dx.doi.org/10.1002/cpa.21531}{\emph{Commun. Pure Appl. Math.}
  {\bfseries 68} (2015) 511--624},
  [\href{https://arxiv.org/abs/1209.1130}{{\ttfamily 1209.1130}}].

\bibitem{Luk:2013zr}
J.~Luk and I.~Rodnianski, \emph{{Nonlinear interaction of impulsive
  gravitational waves for the vacuum Einstein equations}},
  [\href{https://arxiv.org/abs/1301.1072}{{\ttfamily 1301.1072}}].

\bibitem{Sachs:1962wk}
R.~K. Sachs, \emph{{Gravitational waves in general relativity. 8. Waves in
  asymptotically flat space-times}},
  \href{http://dx.doi.org/10.1098/rspa.1962.0206}{\emph{Proc. Roy. Soc. Lond.}
  {\bfseries A270} (1962) 103--126}.

\bibitem{Barnich:2010eb}
G.~Barnich and C.~Troessaert, \emph{{Aspects of the BMS/CFT correspondence}},
  \href{http://dx.doi.org/10.1007/JHEP05(2010)062}{\emph{JHEP} {\bfseries 05}
  (2010) 062}, [\href{https://arxiv.org/abs/1001.1541}{{\ttfamily 1001.1541}}].

\bibitem{Winicour16}
T.~M{\"a}dler and J.~Winicour, \emph{{Bondi-Sachs Formalism}},
  \href{http://dx.doi.org/10.4249/scholarpedia.33528}{\emph{Scholarpedia}
  {\bfseries 11} (2016) 33528},
  [\href{https://arxiv.org/abs/1609.01731}{{\ttfamily 1609.01731}}].

\bibitem{Nichols:2018qac}
D.~A. Nichols, \emph{{Center-of-mass angular momentum and memory effect in
  asymptotically flat spacetimes}},
  \href{http://dx.doi.org/10.1103/PhysRevD.98.064032}{\emph{Phys. Rev. D}
  {\bfseries 98} (2018) 064032},
  [\href{https://arxiv.org/abs/1807.08767}{{\ttfamily 1807.08767}}].

\bibitem{Geroch:1977jn}
R.~Geroch, \emph{Asymptotic structure of space-time},  in \emph{Asymptotic
  Structure of Space-Time} (F.~P. Esposito and L.~Witten, eds.), (Boston, MA),
  Springer US, 1977.
\newblock \href{http://dx.doi.org/10.1007/978-1-4684-2343-3_1}{DOI}.

\bibitem{Compere:2016jwb}
G.~Comp\`ere and J.~Long, \emph{{Vacua of the gravitational field}},
  \href{http://dx.doi.org/10.1007/JHEP07(2016)137}{\emph{JHEP} {\bfseries 07}
  (2016) 137}, [\href{https://arxiv.org/abs/1601.04958}{{\ttfamily
  1601.04958}}].

\bibitem{Einstein:1916vd}
A.~Einstein, \emph{{The Foundation of the General Theory of Relativity}},
  \href{http://dx.doi.org/10.1002/andp.19163540702}{\emph{Annalen Phys.}
  {\bfseries 49} (1916) 769--822}.

\bibitem{Bieri:2020zki}
L.~Bieri, \emph{{New Effects in Gravitational Waves and Memory}},
  \href{http://dx.doi.org/10.1103/PhysRevD.103.024043}{\emph{Phys. Rev. D}
  {\bfseries 103} (2021) 024043},
  [\href{https://arxiv.org/abs/2010.09207}{{\ttfamily 2010.09207}}].

\bibitem{FMP}
L.~Freidel, S.~F. Moosavian and D.~Pranzetti, ``{\it Coadjoint Orbits of null
  infinity}.'' 2021, to appear.

\bibitem{Nutku:1977wp}
Y.~Nutku and M.~Halil, \emph{{Colliding Impulsive Gravitational Waves}},
  \href{http://dx.doi.org/10.1103/PhysRevLett.39.1379}{\emph{Phys. Rev. Lett.}
  {\bfseries 39} (1977) 1379--1382}.

\bibitem{Chandrasekhar:1986jn}
S.~Chandrasekhar and B.~C. Xanthopoulos, \emph{{A New Type of Singularity
  Created by Colliding Gravitational Waves}},
  \href{http://dx.doi.org/10.1098/rspa.1986.0116}{\emph{Proc. Roy. Soc. Lond.
  A} {\bfseries 408} (1986) 175--208}.

\bibitem{Zhang:2017jma}
P.~M. Zhang, C.~Duval and P.~A. Horvathy, \emph{{Memory Effect for Impulsive
  Gravitational Waves}},
  \href{http://dx.doi.org/10.1088/1361-6382/aaa987}{\emph{Class. Quant. Grav.}
  {\bfseries 35} (2018) 065011},
  [\href{https://arxiv.org/abs/1709.02299}{{\ttfamily 1709.02299}}].

\bibitem{Bhattacharjee:2019jaf}
S.~Bhattacharjee, S.~Kumar and A.~Bhattacharyya, \emph{{Memory Effect and
  BMS-like Symmetries for Impulsive Gravitational Waves}},
  \href{http://dx.doi.org/10.1103/PhysRevD.100.084010}{\emph{Phys. Rev. D}
  {\bfseries 100} (2019) 084010},
  [\href{https://arxiv.org/abs/1905.12905}{{\ttfamily 1905.12905}}].

\bibitem{Dray:1984ha}
T.~Dray and G.~'t~Hooft, \emph{{The Gravitational Shock Wave of a Massless
  Particle}}, \href{http://dx.doi.org/10.1016/0550-3213(85)90525-5}{\emph{Nucl.
  Phys. B} {\bfseries 253} (1985) 173--188}.

\bibitem{Dray:1985yt}
T.~Dray and G.~'t~Hooft, \emph{{The Effect of Spherical Shells of Matter on the
  Schwarzschild Black Hole}},
  \href{http://dx.doi.org/10.1007/BF01215912}{\emph{Commun. Math. Phys.}
  {\bfseries 99} (1985) 613--625}.

\bibitem{Dray:1985ie}
T.~Dray and G.~'t~Hooft, \emph{{The Gravitational Effect of Colliding Planar
  Shells of Matter}},
  \href{http://dx.doi.org/10.1088/0264-9381/3/5/013}{\emph{Class. Quant. Grav.}
  {\bfseries 3} (1986) 825--840}.

\bibitem{Strominger:2013lka}
A.~Strominger, \emph{{Asymptotic Symmetries of Yang-Mills Theory}},
  \href{http://dx.doi.org/10.1007/JHEP07(2014)151}{\emph{JHEP} {\bfseries 07}
  (2014) 151}, [\href{https://arxiv.org/abs/1308.0589}{{\ttfamily 1308.0589}}].

\bibitem{He:2015zea}
T.~He, P.~Mitra and A.~Strominger, \emph{{2D Kac-Moody Symmetry of 4D
  Yang-Mills Theory}},
  \href{http://dx.doi.org/10.1007/JHEP10(2016)137}{\emph{JHEP} {\bfseries 10}
  (2016) 137}, [\href{https://arxiv.org/abs/1503.02663}{{\ttfamily
  1503.02663}}].

\bibitem{Pasterski:2016qvg}
S.~Pasterski, S.-H. Shao and A.~Strominger, \emph{{Flat Space Amplitudes and
  Conformal Symmetry of the Celestial Sphere}},
  \href{http://dx.doi.org/10.1103/PhysRevD.96.065026}{\emph{Phys. Rev. D}
  {\bfseries 96} (2017) 065026},
  [\href{https://arxiv.org/abs/1701.00049}{{\ttfamily 1701.00049}}].

\bibitem{Donnay:2018neh}
L.~Donnay, A.~Puhm and A.~Strominger, \emph{{Conformally Soft Photons and
  Gravitons}}, \href{http://dx.doi.org/10.1007/JHEP01(2019)184}{\emph{JHEP}
  {\bfseries 01} (2019) 184},
  [\href{https://arxiv.org/abs/1810.05219}{{\ttfamily 1810.05219}}].

\bibitem{Pate:2019lpp}
M.~Pate, A.-M. Raclariu, A.~Strominger and E.~Y. Yuan, \emph{{Celestial
  operator products of gluons and gravitons}},
  \href{http://dx.doi.org/10.1142/S0129055X21400031}{\emph{Rev. Math. Phys.}
  {\bfseries 33} (2021) 2140003},
  [\href{https://arxiv.org/abs/1910.07424}{{\ttfamily 1910.07424}}].

\bibitem{Guevara:2021abz}
A.~Guevara, E.~Himwich, M.~Pate and A.~Strominger, \emph{{Holographic Symmetry
  Algebras for Gauge Theory and Gravity}},
  [\href{https://arxiv.org/abs/2103.03961}{{\ttfamily 2103.03961}}].

\bibitem{Pasterski:2021fjn}
S.~Pasterski, A.~Puhm and E.~Trevisani, \emph{{Celestial Diamonds: Conformal
  Multiplets in Celestial CFT}},
  [\href{https://arxiv.org/abs/2105.03516}{{\ttfamily 2105.03516}}].

\bibitem{Pasterski:2021dqe}
S.~Pasterski, A.~Puhm and E.~Trevisani, \emph{{Revisiting the Conformally Soft
  Sector with Celestial Diamonds}},
  [\href{https://arxiv.org/abs/2105.09792}{{\ttfamily 2105.09792}}].

\bibitem{Wieland:2016exy}
W.~Wieland, \emph{{Discrete gravity as a topological field theory with
  light-like curvature defects}},
  \href{http://dx.doi.org/10.1007/JHEP05(2017)142}{\emph{JHEP} {\bfseries 05}
  (2017) 142}, [\href{https://arxiv.org/abs/1611.02784}{{\ttfamily
  1611.02784}}].

\bibitem{Elvang:2016qvq}
H.~Elvang, C.~R.~T. Jones and S.~G. Naculich, \emph{{Soft Photon and Graviton
  Theorems in Effective Field Theory}},
  \href{http://dx.doi.org/10.1103/PhysRevLett.118.231601}{\emph{Phys. Rev.
  Lett.} {\bfseries 118} (2017) 231601},
  [\href{https://arxiv.org/abs/1611.07534}{{\ttfamily 1611.07534}}].

\bibitem{Laddha:2017ygw}
A.~Laddha and A.~Sen, \emph{{Sub-subleading Soft Graviton Theorem in Generic
  Theories of Quantum Gravity}},
  \href{http://dx.doi.org/10.1007/JHEP10(2017)065}{\emph{JHEP} {\bfseries 10}
  (2017) 065}, [\href{https://arxiv.org/abs/1706.00759}{{\ttfamily
  1706.00759}}].

\end{thebibliography}\endgroup

\end{document}